\documentclass[preprint,12pt]{elsarticle}

\usepackage{xcolor, soul}
\usepackage{float}
\usepackage{times}
\usepackage{hyperref}
\usepackage{natbib}
\usepackage{hypernat} 
\usepackage{amsmath}
\usepackage{amssymb}
\usepackage{bm}
\usepackage{setspace}
\usepackage{mathtools}
\usepackage{graphicx}
\usepackage{caption}
\usepackage{subcaption}
\usepackage{epstopdf}
\usepackage{grffile}
\setcitestyle{square}
\usepackage{booktabs}
\usepackage{multirow}
\usepackage{overpic} 
\usepackage{subcaption}
\usepackage[margin=2.5cm]{geometry}
\begin{document}
\begin{frontmatter}
  \title{An adaptive strategy based on conforming quadtree meshes\\ for kinematic limit analysis}
  \cortext[ca]{Corresponding author}
  \author[aa]{H. Nguyen-Xuan\corref{ca}}
  \ead{ngx.hung@hutech.edu.vn}
  \author[ab,ac]{Hien V. Do}
  \ead{hiendv@hcmute.edu.vn}
  \author[aa,ee]{Khanh N. Chau}
  \ead{cn.khanh@hutech.edu.vn}
  \address[aa]{Center for Interdisciplinary Research in Technology, Ho Chi Minh City University of Technology (HUTECH), Ho Chi Minh City, Vietnam}
  \address[ab]{Faculty of Mechanical Engineering, HCMC University of Technology and Education, Ho Chi Minh City, Vietnam}
  \address[ac]{Faculty of Civil Engineering, HCMC University of Technology and Education, Ho Chi Minh City, Vietnam}
  \address[ee]{Computational Engineering Department, Vietnamese--German University, Ho Chi Minh City, Vietnam}
  \begin{abstract}
    We propose a simple and efficient scheme based on adaptive finite elements over conforming quadtree meshes for collapse plastic analysis of structures. Our main interest in kinematic limit analysis is concerned with both purely cohesive-frictional and cohesive materials. It is shown that the most computational efficiency for collapse plastic problems is to employ an adaptive mesh strategy on quadtree meshes. However, a major difficulty in finite element formulations is the appearance of hanging nodes during adaptive process. This can be resolved by a definition of conforming quadtree meshes in the context of polygonal elements. Piecewise-linear shape functions in barycentric coordinates are used to approximate the velocity field. Numerical results prove the reliability and benefit of the present approach.
  \end{abstract}

\end{frontmatter}

\graphicspath{{./figures/}}

\section{Introduction}\label{Sec1}

Limit analysis has been known as a power tool to directly obtain the ultimate load bearing capacity and plastic collapse path of structures without any requirement of iterative or incremental analysis. The iteratively elastic-plastic analysis that might be undertaken with numerical methods is capable of providing a collapse load factor. However, the computational cost and convergence of the nonlinear solution is still questionable for large-scale structures. As an alternative methodology, limit analysis allows to show the most important features of the limit state of structures in the plastic regime. Until now, the slip-line field (SLF) theory is an analytical approach for evaluating the load bearing capacity of structures \cite{06Hill,07Ewing,08Miller}. Practically, SLF is well suited to the given problem with simple geometry and loading conditions \cite{09Prager}. The extensive development of different numerical methods has become more attractive beyond analytical approaches. Various numerical methods basically rely on the Koiter’s kinematic (upper bound) theorem \cite{11Koiter} or the Melan’s static (or lower bound) theorem \cite{12Melan}. The upper bound limit analysis is to find a load factor that results in minimizing the plastic dissipation problem (without the work of any additional loads) for any kinematically admissible displacement field, while the lower bound limit analysis determines statically and plastically admissible stress field that maximizes the plastic load factor. Among them, finite element methods have become popular in limit analysis \cite{13Nagtegaal,Bottero,15Capsoni,16Sloan,17Christiansen,18Zouain,19Krabben,20Vicente,21Lyamin,22Makrodimopoulos,23Makrodimopoulos,24Zouain}. Moreover, lower-order finite elements based on triangular or quadrilateral types are largely used due to its simplicity and efficiency, and specially, the flexibility in adaptive mesh refinements. Nevertheless, they do not perform well with unstructured meshes \cite{23Makrodimopoulos} and are sensitive to volumetric locking in the incompressibility limit. Several advanced technologies have therefore been devised in the literature \cite{22Makrodimopoulos,23Makrodimopoulos,24Zouain,29Nguyen}.

In attempts to enhance in the accuracy of the limit analysis solution, adaptive mesh refinement becomes rather important. We know that localized plastic deformations cause the slow convergence of the numerical approaches \cite{Borges}. Therewith, the mesh should be automatically refined along plastic zones. Theoretically, the error-based indicator has to be known to conduct adaptive mesh refinement\textcolor{blue}{\cite{Nhon01,Nhon02,Nhon03,Nhon04}}. However, finding a priori error estimate for the plastic problem is not easy. Hence, a posteriori error scheme is the most suitable choice. Initially, a posteriori indicator based on the recovery technique of the Hessian matrix was studied in \cite{Borges,47Lyamin}. Then, several alternative indicators related to the plastic dissipation and both the static and kinematic bound problems were studied in \cite{48Ciria,49Ciria,50Munoz,51Martin}. The plastic dissipation indicator in association with adaptive bubble-enhanced triangular finite element formulation was presented in \cite{29Nguyen}. On the other hand, we observe that plastic strain rates can be used to measure localized plastic deformation at limit state. In addition, there exist some regions in the structure represented by highly plastic strain rates. It requires that these zones must be refined to obtain the solution that converges quickly to the actual value with the lowest computational cost.

\begin{figure}[H]
  \centering
  \begin{subfigure}{1\textwidth}
    \centering
    \includegraphics[]{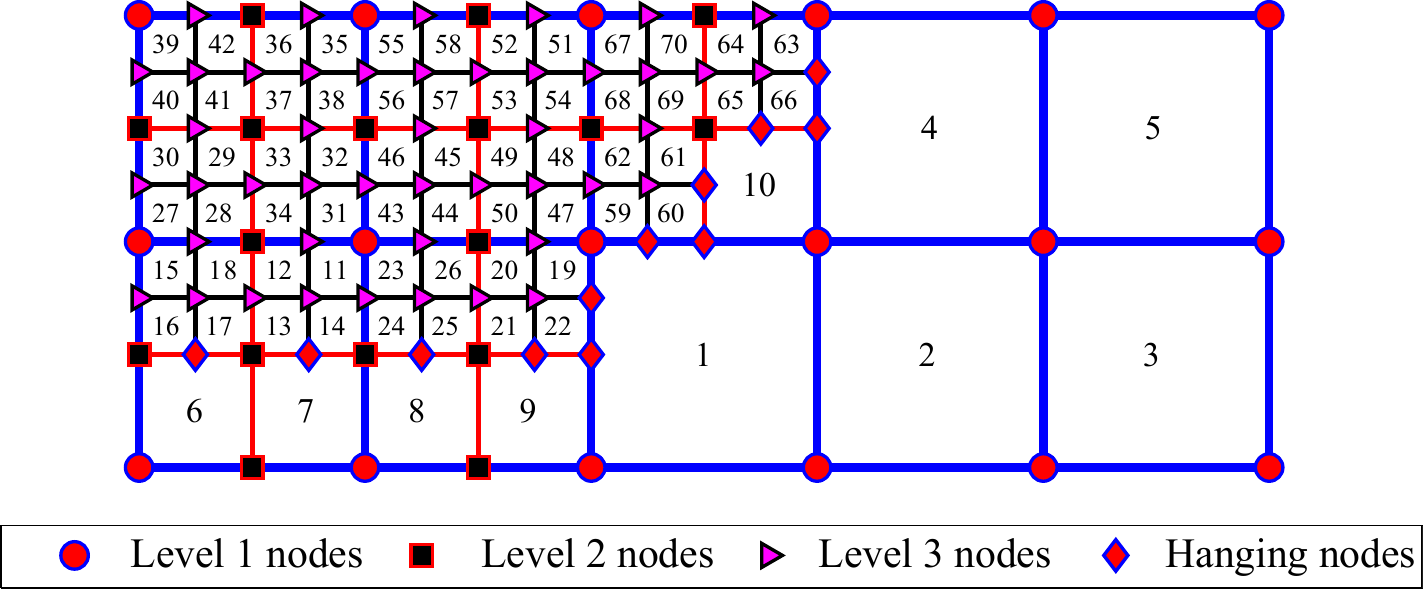}
    \caption{}
  \end{subfigure}
  \begin{subfigure}{1\textwidth}
    \centering
    \includegraphics[scale=0.8]{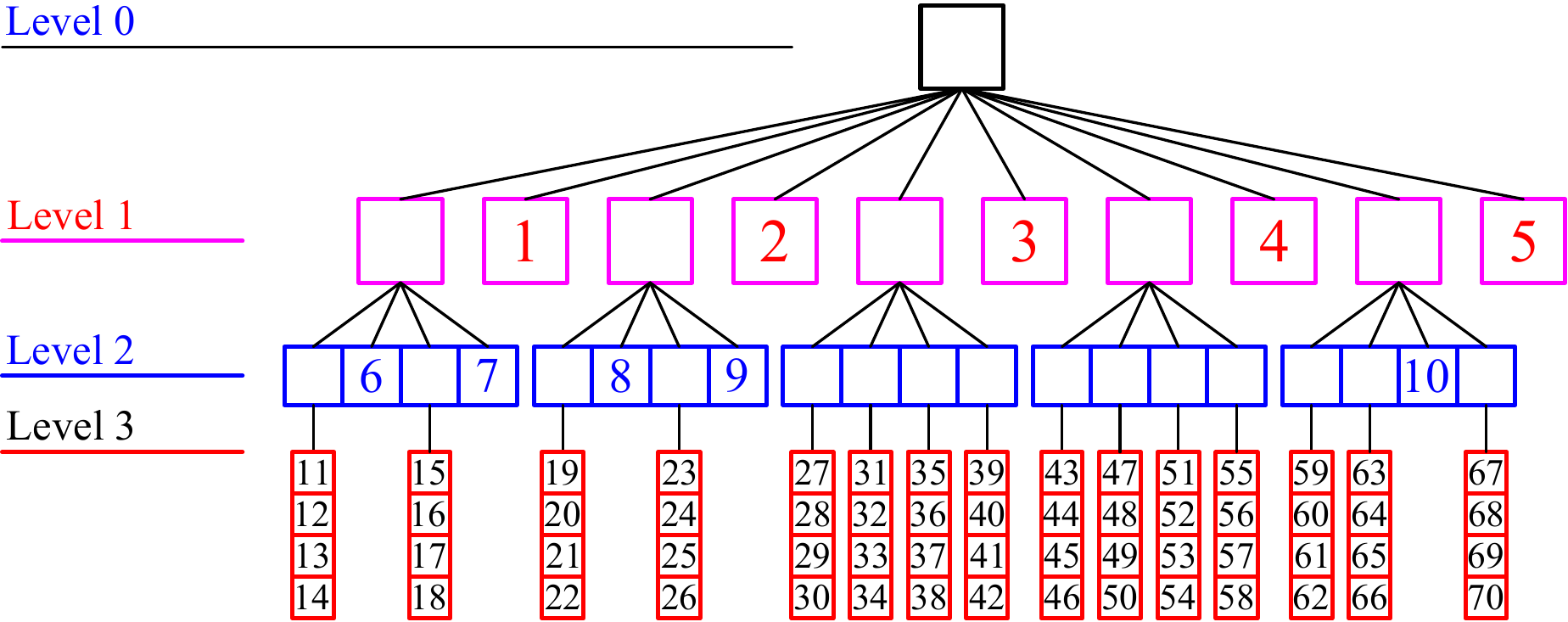}
    \caption{}
  \end{subfigure}
  \caption{An adaptive quadtree strategy: a) quadtree mesh and b) its representative tree.}
  \label{Fig: Fig7}
\end{figure}

Regarding advances in mesh generation, quadtree meshes \cite{01Samet} have been well known and applied largely to a wide range of various engineering multi-disciplines, e.g, computer graphics and image processing \cite{01SametImage}, moving fluid interface\cite{02Greaves}, large-scale earthquake ground motion simulation \cite{03Bielak}. They have been successfully incorporated into finite elements in the framework of adaptive mesh strategy to enhance the accuracy of approximate solution, especially for singular problems with high demand of computational resource or solution formulation \cite{04Liang,05Popinet}. Recently, quadtree meshes in combination with the scaled boundary finite element method (SBFEM) \cite{OoiManNatarajanSong,SaputraSong2017} showed a robust computational tool for mechanics problems. However, in the context of the standard FEM, after each refinement, hanging nodes are generated unless new elements and their neighbours are at same levels of refinement. For instance, Fig. \ref{Fig: Fig7} (a) illustrates three different levels of refinements. We see that hanging nodes are vertices of the child (smaller) element that is located on the side of the father (larger) element, i.e, \textcolor{blue}{diamond-shape nodes} displayed in Fig. \ref{Fig: Fig7} (a). Consequently, hanging nodes leads to the incompatibilities in finite element approximations. To overcome these difficulties, several special treatments have been studied such as Lagrange multipliers or penalty method \cite{82Hansbo}, adding temporary elements \cite{83Palle}, constraining hanging nodes to corner nodes \cite{84Ainsworth}, hierarchical enrichment \cite{85Krysl}, B-splines \cite{86Kagan} \cite{DO2017149}, and natural neighbor basis functions \cite{68Tabarraei}. In \cite{68Tabarraei}, natural neighbor (Laplace) basis functions were utilized to obtain $\mathcal{C}^{0}$-- continuous approximations along sides involving hanging nodes. The shape functions are constructed from the polygonal reference elements through the affine map. By this approach, we do not require the number of hanging nodes lied on each side and the computations are, moreover, performed over the framework of polygonal elements. We recall that polygonal finite elements have been applied to an extensive class of mechanics problems such as nonlinear constitutive modeling of polycrystalline materials \cite{31Sze,32Simone,33Menk}, solid mechanics problems \textcolor{blue}{\cite{34Dai,Timon01,Timon03,Timon04,Timon05,Timon06}}, topology optimization \cite{35Talischi,36Talischi,37Gain}, incompressible materials \cite{38Talischi}, fracture modeling \cite{39Tabarraei,40Bishop,41Ooi} and so on. However, we emphasize that, in the context of polygonal finite elements, the numerical solutions of limit analysis are less accurate for cohesive–-frictional materials, especially for purely cohesive material or incompressible material due to the presence of the flow rule constraints.

Last but not least, another important issue is how to efficiently solve a minimization problem existing in limit analysis models. Traditionally, linear and non-linear programming is totally possible, but linear programming problems lead to a large number of additional variables. It is interesting that most of the plastic yield criteria can be represented as an intersection of cones where the primal–dual interior point method \cite{43Andersen,44Andersen} implemented in optimization packages, e.g. MOSEK software, is really suitable to evaluate the limit analysis problem efficiently. For example, this optimization tool used for the limit analysis problems has been implemented in \cite{22Makrodimopoulos,23Makrodimopoulos}.

The main contribution of this study focuses on some following aspects:
\begin{itemize}
  \item A new bubble-enhanced FE formulation with piecewise-linear shape functions is presented to eliminate the volumetric locking in fully plastic regime.
  \item Quadrilateral elements with the presence of hanging nodes are resolved as conforming polygonal elements.
  \item Second-order cone programming (SOCP) is exploited to solve the large-scale optimization problems.
  \item A quadtree-based adaptive mesh generator guided by the $\mathcal{L}^{2}$ - norm of plastic strain rates is presented.
\end{itemize}

The paper is outlined as follows: A brief review on the upper bound theorem at hand is described in the next section. Section \ref{Sec3} presents a polygonal finite element formulation implemented with bubble functions through barycentric coordinates and a quadrature scheme based on an area-averaged edge-based projection technique. Section \ref{Sec4} states a solution procedure of the discrete problem. An adaptive quadtree mesh procedure is presented in Section \ref{Sec5}. Section \ref{Sec6} summarizes a numerical implementation. Numerical validations are given in Section \ref{Sec7}. Section \ref{Sec8} closes the paper with the concluding remarks.

\section{A short review of the kinematic theorem}\label{Sec2}
Let us consider a problem domain $\Omega$ bounded by a continuous boundary $\Gamma=\Gamma_{\dot{u}}\cup\Gamma_{t},\Gamma_{\dot{u}}\cap\Gamma_{t}=\emptyset $. The rigid-perfectly plastic body is acted by body forces $ \textbf{f} $ and external tractions $ \textbf{g} $ on $\Gamma_{t}$, and the boundary $\Gamma_{\dot{u}}$ is prescribed by the displacement velocity vector $\mathbf{\dot{u}}$. The equilibrium equation can be delineated as follows
\begin{align}
  W_{int}(\boldsymbol\sigma, \dot{\mathbf{u}}) = W_{ext}(\dot{\mathbf{u}}), \forall \dot{\mathbf{u}} \in \mathcal{V}
\end{align}
in which the internal work rate for stress tensor $\boldsymbol\sigma$ and velocity vector $ \dot{\mathbf{u}} $ is described by
\begin{align}
  W_{int}(\boldsymbol\sigma, \dot{\mathbf{u}})=\int_{\Omega} \boldsymbol\sigma \colon \boldsymbol\varepsilon(\dot{\mathbf{u}}) \mathrm{d}\Omega
\end{align}
and the external work rate is defined by
\begin{align}
  W_{ext}(\dot{\mathbf{u}})=\int_{\Omega} \mathbf{f} \cdot \dot{\mathbf{u}} \mathrm{d}\Omega + \int_{\Gamma_{t}}\mathbf{g} \cdot \dot{\mathbf{u}} \mathrm{d}\Gamma
\end{align}
in which $\mathcal{V}$ is a space of kinematically admissible velocity field denoted by
\begin{align}
  \mathcal{V} = \{\dot{\mathbf{u}}\in\mathcal{H}^{1}(\Omega)^{2}, \dot{\mathbf{u}} = \dot{\bar{\mathbf{u}}} \quad \text{on} \quad \Gamma_{\dot{u}}\}
\end{align}
We also define $\mathcal{L}^2(\Omega)$ -norm for strain rates which is regarded as a mesh-control indicator in adaptive mesh refinements as
\begin{equation}
  \eta = \| \dot{\boldsymbol\varepsilon} \|_{\mathcal{L}^{2}(\Omega)} = \Big(\int_{\Omega} \mathit{\dot{\varepsilon}}_{ij}\mathit{\dot{\varepsilon}}_{ij}\text{d}\Omega\Big)^{1/2}
  \label{eqn:L2}
\end{equation}
\textcolor{blue}{where plastic strain rates $\dot{\boldsymbol\varepsilon}$ adhere to the associated flow rule defined by
\begin{equation}
  \dot{\boldsymbol\varepsilon} = \dot{\mu} \dfrac{\partial \psi (\boldsymbol\sigma)}{\partial \boldsymbol\sigma}
  \label{eqn:associatedflow}
\end{equation}
and $\dot{\mu}$ is a non-negative plastic multiplier}. In addition, we define a convex set $\mathcal{B}$ which contains statically admissible stresses
\begin{align}
  \mathcal{B}=\{\mathbf{\boldsymbol\sigma \in \sum \big|\psi(\boldsymbol\sigma) \leq 0}\}
\end{align}
where $\sum$ is a space of symmetric stress tensors,  $ \boldsymbol\sigma $ satisfies the yield condition for assumed material and $ \psi(\boldsymbol\sigma) $ denotes the convex yield function.

Defining $ \mathcal{C}=\{\mathbf{\dot{u}}\in\mathcal{V}|W_{int}(\mathbf{\dot{u}})=1 \} $, the limit analysis problem is generally speaking to seek the limit load factor $ \lambda^{*} $ yielding the following optimization problem \cite{23Makrodimopoulos}
\begin{subequations}
  \begin{align}
    \textcolor{blue}{\lambda^{*}} & = \textcolor{blue}{\max \{\exists \boldsymbol{\sigma} \in \mathcal{B}|W_{int}(\boldsymbol\sigma, \dot{\mathbf{u}})=\lambda W_{ext}(\dot{\mathbf{u}}), \forall \dot{\mathbf{u}} \in \mathcal{V}\}} \\
                & = \max_{\boldsymbol\sigma \in \mathcal{B}} \min_{\dot{\mathbf{u}} \in \mathcal{C}} W_{int}(\boldsymbol\sigma, \dot{\mathbf{u}})                                         \\
                & = \min_{\dot{\mathbf{u}} \in \mathcal{C}} \max_{\boldsymbol\sigma \in \mathcal{B} } W_{int}(\boldsymbol\sigma, \dot{\mathbf{u}})                                        \\
                & = \min_{\dot{\mathbf{u}} \in \mathcal{C}} D (\dot{\mathbf{u}})
  \end{align}
\end{subequations}
in which
\begin{align}
  D(\dot{\mathbf{u}}) = \max_{\boldsymbol\sigma \in \mathcal{B}} W_{int}(\boldsymbol\sigma, \dot{\mathbf{u}})
\end{align}
For plane strain problems, von Mises and Mohr--Coulomb models \footnote{\textcolor{blue}{Our approach is also available for the Drucker--Prager model. Under plane strain conditions, the Drucker--Prager yield criterion can reduce to the Mohr--Coulomb yield criterion.}} can be coined in the following compact form \cite{23Makrodimopoulos}
\begin{align}
  \psi(\boldsymbol\sigma) & =\sqrt{J_{2}(\mathbf{s})} + \varsigma\sigma_{m} - \kappa\leq0
\end{align}
where
\begin{align}
  \sigma_{m} & =\dfrac{1}{2}tr(\boldsymbol\sigma), \ \mathbf{s} = \boldsymbol\sigma - \sigma_{m} \mathbf{I}, J_{2}(\mathbf{s})=\dfrac{1}{2}\mathbf{s} \colon \mathbf{s}
\end{align}
and $ (\varsigma=\sin\varphi,~\kappa=c\cos\varphi) $ is available for the Mohr--Coulomb criteria, \textit{c} denotes the cohesion coefficient, \textbf{I} is the identity matrix, and $\varphi$ is the internal friction angle. The Mohr-Coulomb criteria reduces to the von Mises criteria as $\varsigma=0, \kappa=\sigma_{y}/\sqrt{3} ,$ in which $\sigma_{y}$ is the yield stress.\\
Under in the associated flow rule, the plastic dissipation can be expressed as a function of plastically admissible strains
\begin{align}
  D(\dot{\mathbf{u}}) & =\int_{\Omega}\kappa\lambda \text{d}\Omega
\end{align}
where
\begin{align}
  \lambda\geq2\sqrt{J_{2}(\dot{\mathbf{e}})}\ \ \text{and}\ \  \dot{e}_{m}=\varsigma\lambda,\ \forall\varsigma\geq0
\end{align}
and $\dot{e}_{m} = tr(\dot{\bm{\varepsilon}}) = \boldsymbol\nabla \cdot \dot{\mathbf{u}} $ denotes the volume expansion rate with $\dot{e}_{ij}=\dot{\varepsilon}_{ij}-\dfrac{1}{2}\dot{e}_{m}\delta_{ij}$, and $J_{2}(\dot{\mathbf{e}}) = \dfrac{1}{2} \dot{\mathbf{e}} \colon \dot{\mathbf{e}}$.
As $\varsigma=0$, the plastic dissipation \cite{17Christiansen} is subjected to the incompressibility condition $\boldsymbol\nabla \cdot \dot{\mathbf{u}} = 0$. Although lower-order finite element formulations are extremely convenient for computation, they are sensitive to the volumetric locking \cite{13Nagtegaal,16Sloan,18Zouain}. In the current study, we introduce a simple formulation based on quadrilateral elements with piecewise-linear shape functions defined over barycentric coordinates \cite{FloaterMS} and employ an area-averaged edge-based projection technique \cite{42Liu} to achieve the numerical stability in the plastic regime.

\section{A finite element limit analysis formulation}\label{Sec3}
\subsection{A finite element space enriched with bubble piecewise-linear basis functions}
We consider a bounded domain $ \Omega $ discretized into a set $\mathcal{T}$(at primal mesh) of $ \mathit{n_{e}} $ non-overlapping polygonal elements \footnote{We use the concept of consider polygonal elements in order to emphasize that quadrilateral elements with presence of hanging nodes are defined as polygonal elements.} involving a set $ \partial\mathcal{T}$ of edges $ \mathit{n_{ed}} $  and $ \mathit{n_{n}} $  nodes such that  $ \Omega\approx\Omega^{h}=\bigcup^{n_{e}}_{e=1}\mathbf{\Omega^{e}}$. Let $ \mathcal{V}^{h}\subset\mathcal{V}$ be a finite element approximation space of kinematically admissible velocity fields. The discrete form of limit analysis is to seek an approximately collapse load factor $ \alpha^{+} $ as
\begin{subequations}
  \begin{align}
    \alpha^{+} & = \min\  D^{h}(\dot{\mathbf{u}}^{h})\nonumber \\
    s.t.       &
    \begin{cases}
      \lambda^{h}      & \geq2\sqrt{J_{2}(\dot{\mathbf{e}}^{h})} \\
      \textcolor{blue}{\varsigma\lambda^{h}} & \textcolor{blue}{=\dot{e}^{h}_{m}}              \\
      \dot{\mathbf{u}}^{h}\in \mathcal{V}^h
    \end{cases}
  \end{align}
  \label{eqn:LA}
\end{subequations}
It turns out that lower-order finite elements are not stable for purely cohesive materials. They perform poorly under the incompressibility condition. Mathematically, the numerical solution must satisfy the following condition
\begin{align}
  \textcolor{blue}{\dot{e}^{h}_{m} = 0} \quad \text{or} \quad \boldsymbol\nabla\cdot \dot{\mathbf{u}}=0
\end{align}
It is clear that such a constraint requires a special treatment to obtain a stable element formulation in plastic regime. Roughly speaking, the total number of degrees of freedom is not great enough to dominate the number of incompressible constraints in the FE discretization \cite{13Nagtegaal,Bottero,15Capsoni,16Sloan}. Several advanced remedies have been devised to provide locking-free finite elements for limit analysis such as mixed finite elements \cite{15Capsoni,18Zouain}, discontinuity velocity fields \cite{16Sloan,19Krabben,21Lyamin}, and linear strain elements \cite{23Makrodimopoulos}.
We now describe the finite dimension space $\mathcal{V}^h$ as
\begin{align}
  \mathcal{V}^{h}=\Big\{\dot{\mathbf{u}}^{h} \in \mathcal{H}^{1}(\Omega)^{2}, \dot{\mathbf{u}}^{h}\Big|_{\Omega^{e}} \in [\mathcal{Q}(\Omega^{e})]^{2}, \dot{\mathbf{u}}^{h}\Big|_{\Omega^{e}}\in \mathcal{C}^{\infty}, \dot{\mathbf{u}}^{h} \Big|_{\partial\Omega^{e}} \in \mathcal{C}^{0}, \Omega^{e} \in \mathcal{T}\Big\}
\end{align}
where $ \mathcal{Q}(\Omega^{e}) $ contains barycentric basis functions over the convex polygon $ \Omega^{e} $. As shown in \cite{13Nagtegaal}, the space $ \mathcal{V} $  needs be enriched to overcome volumetric locking in the fully plasticity regime. To address this, we introduce a piecewise-linear bubble space $ \mathcal{V}^{h}_{b} $ such that
\begin{align}
  \mathcal{V}^{h}_{b} = \Big\{\dot{\mathbf{u}}^{h}_{b} \in \mathcal{H}^{1}(\Omega)^{2}, \dot{\mathbf{u}}^{h}_{b} \Big|_{\partial\Omega^{e}}=0 \ \text{and}\  \dot{\mathbf{u}}^{h}_{b}|_{\Omega^{e}}\in \mathcal{C}^{0}\Big\}
\end{align}
The finite element space for the velocity field which is enriched with such bubble functions is defined as
\begin{align}
  \mathcal{W}^{h}=\mathcal{V}^{h}\oplus\mathcal{V}^{h}_{b}
\end{align}
It will be shown in numerical examples of plane strain problems that a following displacement-based FE formulation in combination with bubble functions and dual-mesh integration scheme can avoid volumetric locking in the plastic collapse analysis.
We write explicitly the velocity $\mathbf{\dot{u}}^{h}\in\mathcal{W}^{h}$ over $\Omega^{e}\in \mathcal{T}$ as follows
\begin{equation}
  \dot{\mathbf{u}}^{h}(\mathbf{x})\Big|_{\Omega^{e}} = \sum_{i=1}^{n_{nod}}(N^{e}_{i}(\mathbf{x})\mathbf{I}_{2}) \dot{\mathbf{d}}^{e}_{i}+(N^{e}_{b}(\mathbf{x})\mathbf{I}_{2}) \dot{\mathbf{d}}^{e}_{b}
  \label{eqn:uh}
\end{equation}
where $n_{nod}$ is the number of vertices of element, $ \mathbf{I}_{2} $ denotes the unit matrix of $ 2^{nd} $ rank, $ \dot{\mathbf{d}}^{e}_{i} $ is the vector of nodal degrees of freedom of $\dot{\mathbf{u}}^{h}(\mathbf{x})$ associated to the $ \textcolor{blue}{i^{th}} $ vertex of the element, $N^{e}_{i}$ is the standard nodal basis shape function at the $\textcolor{blue}{ i^{th} }$ vertex of element, $\dot{\mathbf{d}}^{e}_{b}$ is unknowns associated with the centroid node, and $N^{e}_{b}$ is the shape function at a centroid node of the element. In next section, we construct piecewise-linear shape functions for polygonal elements, which are used to define quadrilateral elements with the presence of hanging nodes.

\subsection{Basis functions on arbitrary polygonal elements}\label{Sec3_2}
\begin{figure}[H]
  \centering
  \includegraphics[]{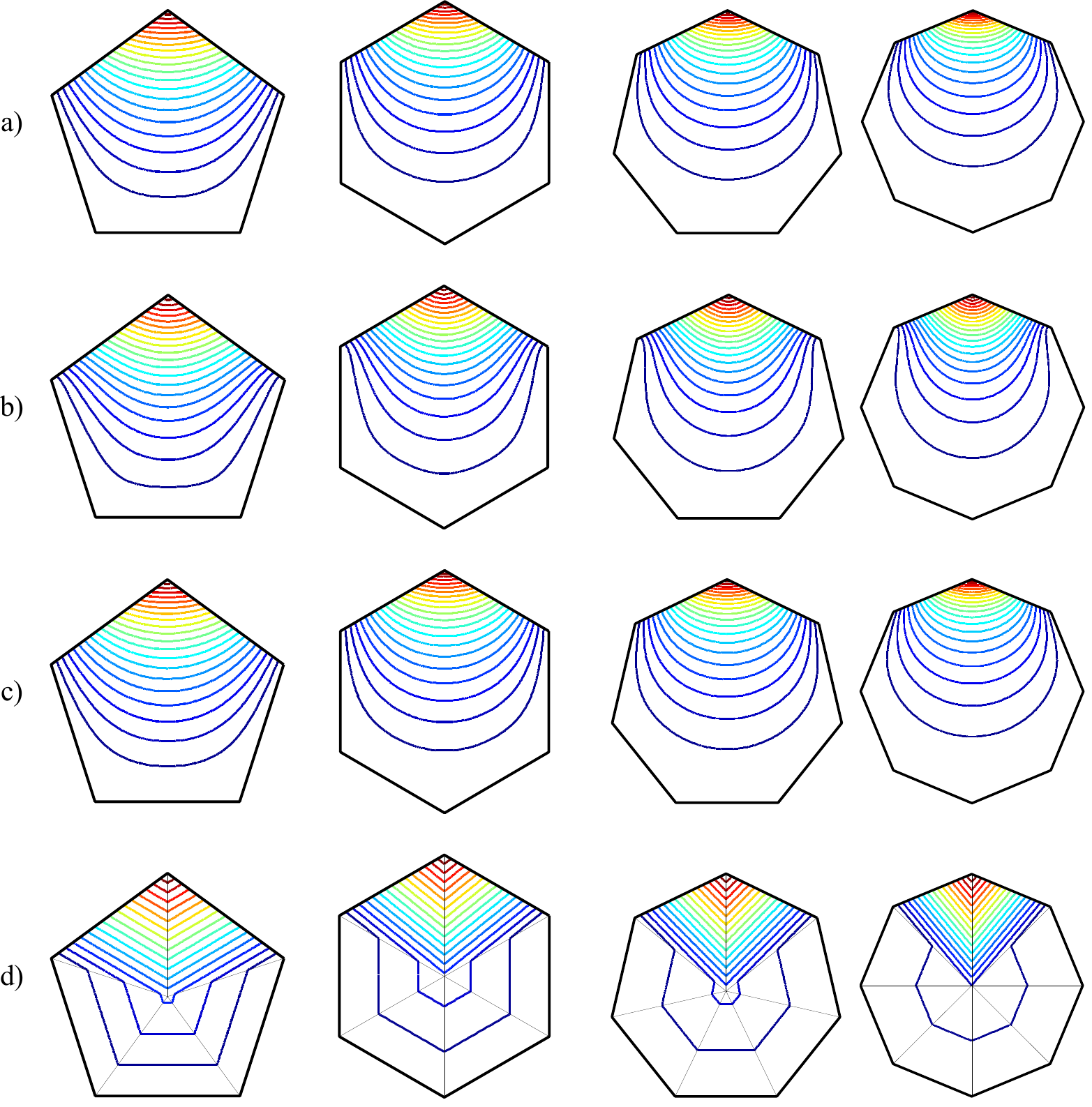}
  \caption{The shape functions defined over regular polygonal elements: a) Wachspress; b) Mean-value; c) Laplace and d) Piecewise-linear.}
  \label{Fig: Fig1}
\end{figure}

\begin{figure}[H]
  \centering
  \includegraphics[]{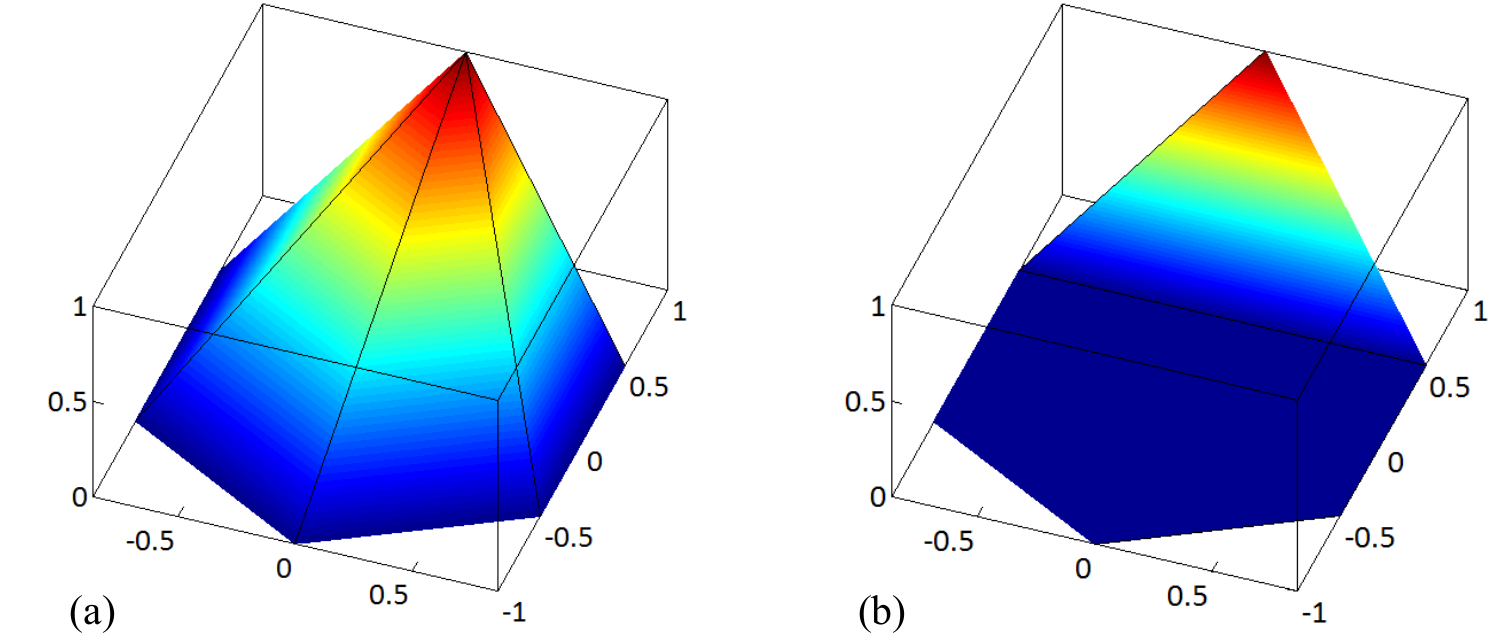}
  \caption{Piecewise-linear shape functions for polygonal elements: a) Upper bound shape function; b) Lower bound shape function.}
  \label{Fig: Fig2}
\end{figure}

\begin{figure}[H]
  \centering
  \includegraphics[]{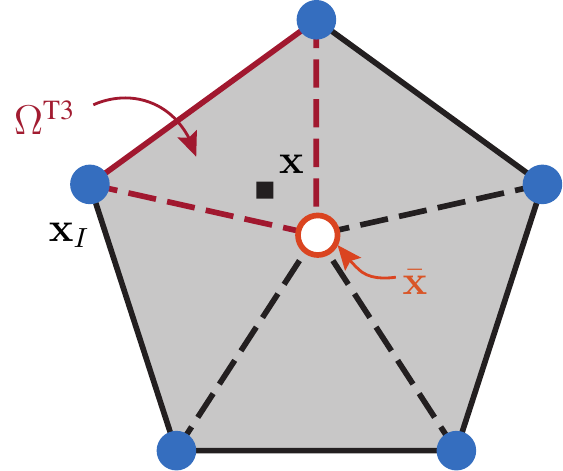}
  \caption{Definition for piecewise-linear shape functions.}
  \label{Fig: Fig3}
\end{figure}

It is recalled that construction of shape functions $\textcolor{blue}{N_{i}(\mathbf{x})}$ over arbitrary polygonal elements
cannot be defined in a similar way as those for the standard triangular or quadrilateral elements. Some approaches to basis functions have been developed such as Wachspress \cite{Wachspress,54Warren}, Mean-value, Laplace \cite{57Christ} and piecewise-linear shape functions. A comparison of different shape functions of four regular polygonal elements is shown in Fig. \ref{Fig: Fig1}.
More simply, Floater \textit{et al.} \cite{FloaterMS} proposed sharp upper and lower bound piecewise-linear functions (cf.  Fig. \ref{Fig: Fig2}(a) and (b)), which can be used for basis functions. These shape functions have the essential feature of interpolants such as non-negative, partition of unity, Kronecker delta, and linear precision.
On the other hand, the Gaussian integration rule is valid for triangular and quadrilateral finite elements. For arbitrary polygonal elements, the Gaussian integration is often performed over sub-triangles. However, this integration technique leads to the huge increase of incompatibility constraints resulting in the poor accuracy of limit analysis. Moreover, we learn in limit analysis that no significant difference in solution using Wachspress, mean-value, Laplace and piecewise-linear shape functions, while rational basis functions of Wachspress, mean-value and Laplace coordinates raise high computational cost. Therefore, piecewise-linear shape functions are recommended. They are defined over the sub-triangles which are created by connecting the centroid of the polygonal element to the vertices of the element as illustrated in Fig.\ref{Fig: Fig3}.

We recall that the shape functions at the vertices of polygonal elements fulfill the Kronecker delta property:
\begin{equation}
  \phi_{i}^{e}(\mathbf{x}_{j})=\delta_{ij}=
  \begin{cases}
    1\quad \mathbf{x}_{i} = \mathbf{x}_{j} \\
    0\quad \mathbf{x}_{i} \neq \mathbf{x}_{j}
  \end{cases}
\end{equation}
The shape functions at the centroid $\mathbf{x}_c$ are then evaluated by
\begin{equation}
  \phi_{i}^{e}(\mathbf{x}_c)=1/n_{nod}
\end{equation}

Now shape functions and their derivatives are accomplished by using linear shape functions over sub-triangles:
\begin{align}
  N_{i}^{e}(\mathbf{x})        & =\sum_{l=1}^{3}N_{\triangle}^{l}(\mathbf{x})\phi_{i}^{e}(\mathbf{x}_{l})\qquad \text{for}\quad \mathbf{x}\in \Omega_{\triangle} \label{Eq: ShapeI} \\
  \boldsymbol\nabla N_{i}^{e}(\mathbf{x}) & =\sum_{l=1}^{3}\boldsymbol\nabla N_{\triangle}^{l}(\mathbf{x})\phi_{i}^{e}(\mathbf{x}_{l})\quad\  \text{for}\quad \mathbf{x}\in  \Omega_{\triangle}
\end{align}
where $ N_{\triangle}^{l}(\mathbf{x}) $ and $\boldsymbol\nabla N_{\triangle}^{l}(\mathbf{x})$ are linear shape functions and their derivatives on each sub-triangle $\Omega_{\triangle}$, $\phi_{i}^e(\mathbf{x}_{l})$ denotes the shape function of node \textit{i} at the node \textit{l} of $\Omega_{\triangle}$. In Eq. \ref{Eq: ShapeI}, shape functions are the piecewise-linear continuous functions over sub-triangles and have the following basic properties:
(1) Kronecker delta $N_{i}^{e}(\mathbf{x}_{j})=\delta_{ij}$; (2) partition of unity $\sum_{i=1}^{n}N_{i}^{e}(\mathbf{x})=1$; (3) non-negative $N_{i}^{e}(\bar{\mathbf{x}})\geq0$ and linear compatibility $\sum_{i=1}^{n}N_{i}^{e}(\mathbf{x})\mathbf{x}_{i}=\mathbf{x}$.

\begin{figure}[H]
  \centering
  \includegraphics[]{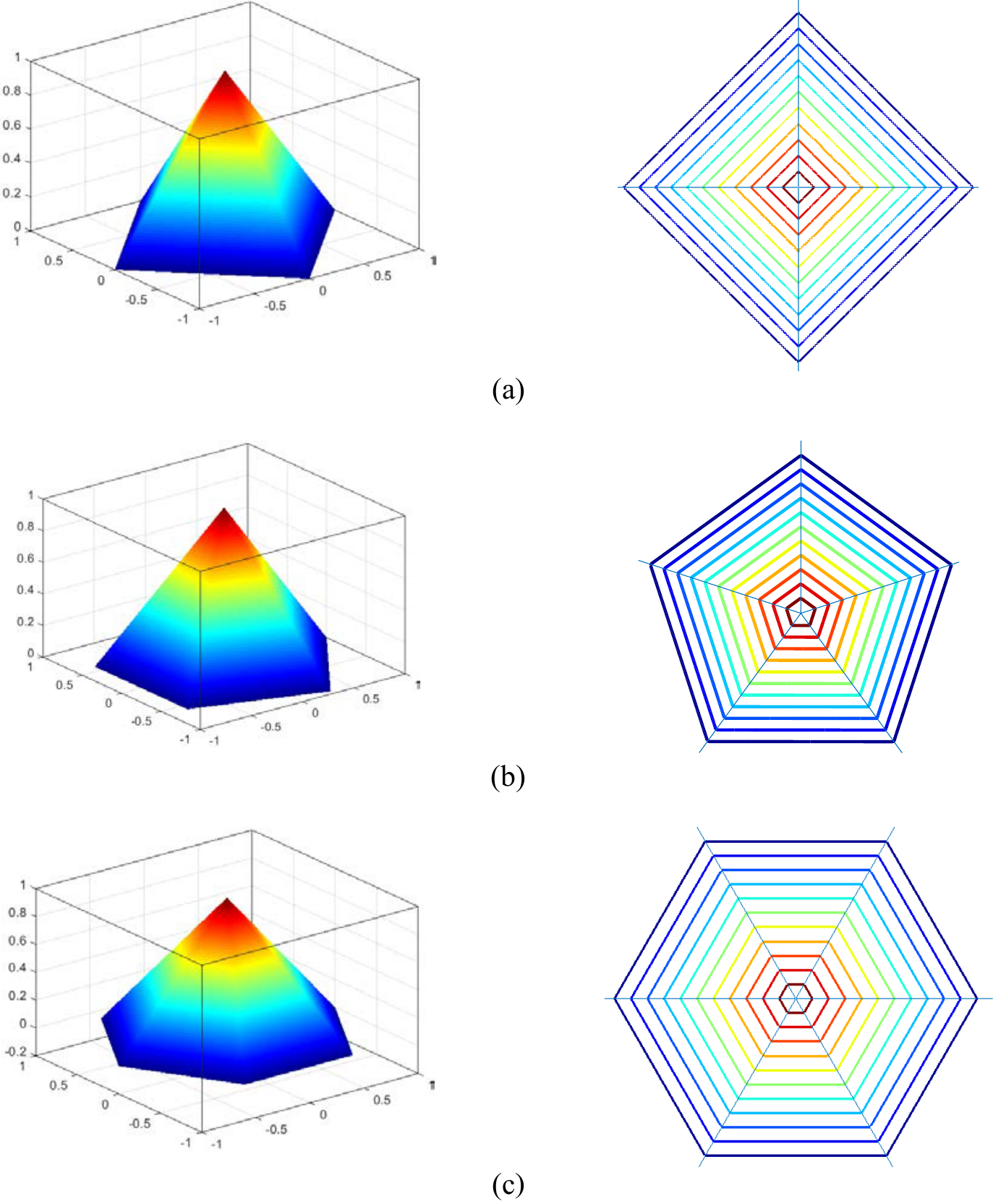}
  \caption{3D view and contour of linearly piecewise bubble shape function of a polygon using the barycentric coordinates attached with an internal node located at the centroid of element: a) quadrilateral; b) pentagon and c) hexagon.}
  \label{Fig: Fig2a}
\end{figure}

In addition, we introduce piecewise-linear bubble shape functions which help to enrich a space of the velocity field for solving incompressible constraints. On each sub-triangle $\Omega_{\triangle}$, we now construct a linear shape function   using barycentric coordinates with vertices including $\bm{x}_c$ and two endpoints of the edges. In other words, $N^e_b$ is a linear function on $\Omega_{\triangle}$. A 3D view of the bubble functions is illustrated in Fig. \ref{Fig: Fig2a}. It is evident that the bubble shape function is zero along the elemental boundary and unity at the centroid. Furthermore, to facilitate computation of underlying quantities in elements with hanging nodes, we adopt the following concept of conforming polygonal element\cite{68Tabarraei}.

\subsection{On polygon with side-nodes}\label{Sec3_3}
\begin{figure}[H]
  \centering
  \includegraphics[]{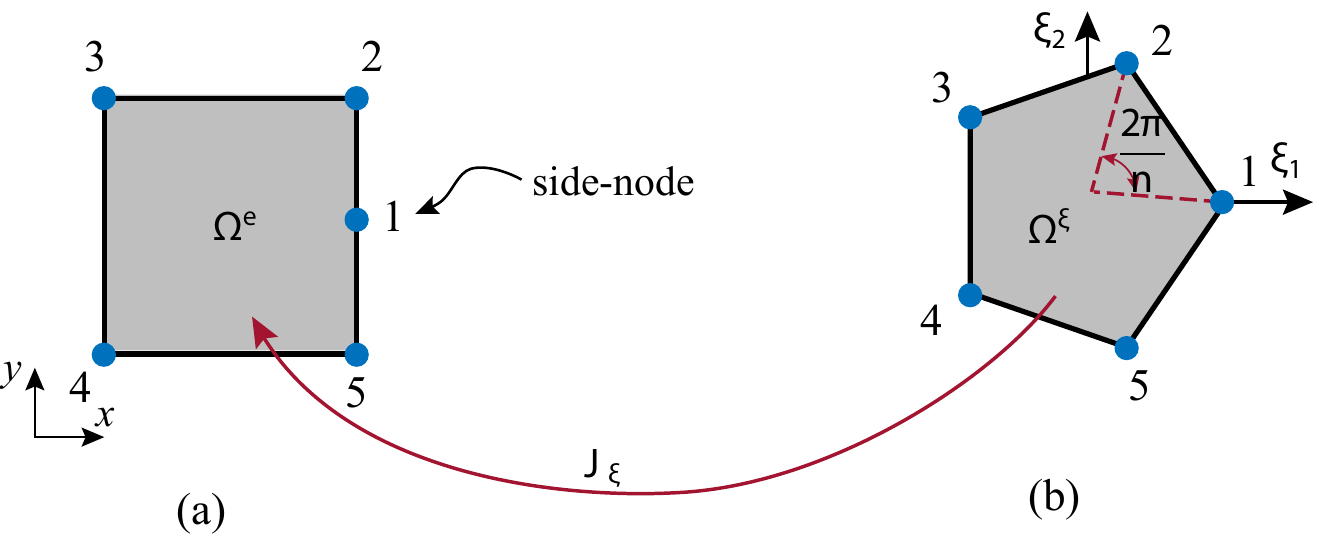}
  \caption{Mapping from a regular quadrilateral element with one hanging node to a pentagonal element:(a) Arbitrary quadrilateral element with one side-node; (b) Reference pentagonal element.}
  \label{Fig: Fig4}
\end{figure}

Consider ``polygonal" element $\Omega^{e}$ displayed in Fig. \ref{Fig: Fig4} (a), which has one side-node along the edge. To construct shape functions for a convex polygon with side nodes, we construct the shape functions on a polygonal reference element $\Omega^{\xi}$ with the reference coordinate $\mathbf{\xi}(\xi_{1},\xi_{2})\in\Omega^{\xi}$ \cite{68Tabarraei}. As shown in the Fig. \ref{Fig: Fig4} (b), the nodes of the polygonal reference element are located at $\xi_{i}=\{\cos(2\pi i/n),\ \sin(2\pi i/n) \} $. The shape functions of polygon with side-nodes are then obtained by mapping from this polygonal reference element to the polygonal physical element. It should be emphasized that there is no distinction in using piecewise-linear interpolation between side-nodes and corner nodes. This means that all the nodes are regarded as corner nodes and the shape functions of all the nodes are obtained in a similar manner.

\subsection{Quadrature scheme}\label{Sec3_4}

It is known that the lower-order quadrilateral element (Q4) does not work well for nearly-incompressible and incompressible materials, and even cohesive-frictional materials. This shortcoming also happens with Q4 having presence of hanging nodes. As an alternative way mentioned in Section \ref{Sec3_2}, piecewise-linear shape functions are defined over the sub-triangles of quadrilateral elements. To reduce incompressibility constraints, we employ an alternative integration technique of a so-called area-averaged edge-based projection technique, which follows the concept of stabilized conforming nodal integration (SCNI)\cite{80Chen} and edge-based smoothed finite element method (ES-FEM) \cite{42Liu}. As a result, the present method suppresses volumetric locking in purely cohesive materials and performs well for cohesive-frictional materials. To use efficiently this alternative integration scheme, a dual mesh is defined over edge-based mesh background $\bar{\mathcal{T}}$ by connecting vertices and centroids of elements of $\mathcal{T}$. Then, $\Omega$ is subdivided into $n_{s}$ edge-based domains $\Omega^{(k)}$ based on edges of elements such that $\Omega=\bigcup\!^{n_s}_{k=1}\Omega^{(k)} $ and $\Omega^{(i)}\cap\Omega^{(j)}=\varnothing$ for $ i\neq j$. The edge-based integration domain $ \Omega^{(k)} $ or smoothing domain associated with the edge \textit{k} is obtained by connecting two end-nodes of the edge \textit{k} to the centroids of adjacent elements, as depicted in Fig. \ref{Fig: Fig5}.

It is now noticeable that in our study assumed strain rates need to be defined over an average projector $\mathcal{P}^{(k)}$ of compatible strain rates on a dual mesh \cite{42Liu}
\begin{align}
  \dot{\boldsymbol\varepsilon} ^ {(k)} = \mathcal{P}^{(k)}(\boldsymbol\nabla_{s} \dot{\mathbf{u}}^{h})=\dfrac{1}{A^{(k)}}\int_{\Omega^{(k)}} \boldsymbol\nabla_{s} \dot{\mathbf{u}}^{h} \mathrm{d}\Omega, \ \text{with} \ \dot{\mathbf{u}}^{h} \in \mathcal{W}^{h}
\end{align}
where $A^{(k)}$ is the area of the edge-based integration domain $\Omega^{(k)}$ and $\boldsymbol\nabla_{s}$ denotes a matrix of differential operators defined as
\begin{equation}
  \boldsymbol\nabla_{s}=
  \begin{bmatrix}
    \partial_{,x} & 0             & \partial_{,y} \\
    0             & \partial_{,y} & \partial_{,x}
  \end{bmatrix}
  ^{\mathrm{T}}
\end{equation}
We write
\begin{equation}
  \dot{\bar{\varepsilon}}^{h}_{m}=\mathcal{P}^{(k)}(\dot{\varepsilon}^{h}_{m})=\mathcal{P}^{(k)}(\boldsymbol\nabla\cdot \dot{\mathbf{u}}^{h})=\dfrac{1}{A^{(k)}}\int_{\Omega^{(k)}}\boldsymbol\nabla \cdot \dot{\mathbf{u}}^{h}\text{d}\Omega, \ \text{with} \ \dot{\mathbf{u}}^{h}\in\mathcal{W}^{h}
  \label{eqn:strainbar}
\end{equation}
In addition, the method must satisfy the incompressibility condition for the purely cohesive material:
\begin{align}
  \forall\Omega^{(k)}\in\bar{\mathcal{T}}, \dot{\bar{\varepsilon}}^{h}_{m}=0
\end{align}
Substituting Eq. \ref{eqn:uh} into Eq. \ref{eqn:strainbar}, the approximation of the constant strain rates on can be expressed as
\begin{align}
  \textcolor{blue}{\dot{\bar{\bm{\varepsilon}}}^h= \sum_{I=1}^{n_{nk}+n_b}\bar{\mathbf{B}}^{(k)}_{I}\dot{\bar{\mathbf{d}}}^{(k)}_{I}=\bar{\mathbf{B}}^{(k)}\dot{\bar{\mathbf{d}}}^{(k)}}
\end{align}
where $ n_{nk} $ is the number of the neighboring nodes of edge $k$, $n_b$ is the number of central nodes or bubble nodes in $\Omega^{(k)}$, $\dot{\bar{\mathbf{d}}}^{(k)}_{I}$ is the nodal degrees of freedom at the $ I^{th} $ node of domain $\Omega^{(k)}$, and $\bar{\mathbf{B}}^{(k)}_{I}$ is the area-averaged edge-based strain-displacement matrix on $\Omega^{(k)}$ computed by
\begin{equation}
  \bar{\mathbf{B}}^{(k)}_{I}=\dfrac{1}{A^{(k)}}\sum_{i=1}^{\textit{nc}}\mathbf{B}^{*}_{i}, \quad \text{\textit{nc} is the number of sub-triangles in the edge \textit{k}}
  \label{eq:B}
\end{equation}
where
\begin{align}\label{eq:B1}
  \mathbf{B}^{*}_{i}=\int_{\Omega^{(k)}_{i}} \boldsymbol\nabla_s N_I(\mathbf{x}) \mathrm{d}\Omega,
\end{align}
Eq. \ref{eq:B} shows that the computations are performed on sub-domain $\Omega^{(k)}_{i}$ of $\Omega^{(k)}$. To achieve numerical integration, the polygonal reference element $\Omega^{\xi}$ introduced in Section \ref{Sec3_3} is divided into sub-triangles and the Gaussian integration is performed over sub-triangle $\Omega^{\xi}_{\triangle}$ of the reference element $\Omega^{\xi}$ corresponding to sub-domain $\Omega^{(k)}_{i}$ of the physical element $\Omega^{e}$, see Fig. \ref{Fig: Fig6}. The piecewise-linear shape functions and their derivatives at all vertices evaluated at the Gaussian point $\boldsymbol\xi_{\triangle m}$ in the sub-triangle $\Omega^{\xi}_{\triangle}$ are computed as follows
\begin{align}
  N^{e}_{I}\big(\boldsymbol\xi_{\triangle m}\big)        & = \sum\limits_{l=1}^{3}N_{\triangle}^{l}\big(\boldsymbol\xi_{\triangle m}\big)\phi^{e}_{I}(\boldsymbol\xi_{\triangle m}\big)\label{Eq_shape1}              \\
  \boldsymbol\nabla N^{e}_{I}\big(\boldsymbol\xi_{\triangle m}\big) & = \sum\limits_{l=1}^{3}\boldsymbol\nabla N_{\triangle}^{l}\big(\boldsymbol\xi_{\triangle m}\big)\phi^{e}_{I}(\boldsymbol\xi_{\triangle m}\big)\label{Eq_deri_shape11}
\end{align}
The numerical integration in \ref{eq:B1} becomes:
\textcolor{blue}{
\begin{align}
  	\mathbf{B}^{*}_{i}=\sum_{m=1}^{n_{g}} \boldsymbol\nabla_s N^{e}_{I}(\mathbf{x}_{\triangle Gm}) |\mathbf{J}_{\xi}| \underbrace{w_{\triangle m} ^ {\xi}}_{|\mathbf{J}_{\eta}| w_{\triangle m}^{\eta}}
\end{align}
}
where $n_{g}$ indicates the number of Gauss points, $\nabla_s N^{e}_{I}(\mathbf{x}_{\triangle Gm})$ denotes a matrix of shape function derivatives at Gauss point $\mathbf{x}_{\triangle Gm}$ in the physical coordinates \textcolor{blue}{and $w_{\triangle m} ^ {\xi}, w_{\triangle m} ^ {\eta}$ represents the corresponding Gaussian quadrature weights defined on polygonal reference element and on three-node triangle natural element, respectively}. From the Jacobian matrix $\mathbf{J}_{\xi}$, the shape function derivatives $\boldsymbol\nabla N ^{e}_{I}(\mathbf{x}_{\triangle Gm})$ in the physical coordinates are determined via the relation: $\boldsymbol\nabla N^{e}_{I}(\mathbf{x}_{\triangle Gm}) = \mathbf{J}^{-1}_{\xi}\boldsymbol\nabla N_I^{e}(\boldsymbol\xi_{\triangle m})$, in which $\boldsymbol\nabla N_{I}^{e}(\boldsymbol\xi_{\triangle m})$ are shape function derivatives which are computed at Gauss points $\boldsymbol{\xi}_{\triangle m}$ in the polygonal reference element $\Omega^{\xi}$.
\subsection{Bubble-enriched finite element limit analysis}
The discrete problem \ref{eqn:LA} is rewritten in the following form
\begin{subequations}
  \begin{align}
    \alpha^{+} & = \min \bar{D}^{h}(\dot{\mathbf{u}}^{h})\nonumber \\
    s.t.       &
    \begin{cases}
      \bar{\lambda}^{h}      & \geq2\sqrt{J_{2}(\bar{\dot{\mathbf{e}}}^{h})}=\|\bar{\dot{ \mathbf{e}}}^{h}_{s}\| \\
      \zeta\bar{\lambda}^{h} & =\dot{\bar{\varepsilon}}^{h}_{m}                                                  \\
      \dot{\mathbf{u}}^{h} \in \mathcal{C}^{h}
    \end{cases}
  \end{align}
  \label{eqn:LAbar}
\end{subequations}
where
\begin{align}
  \bar{\dot{\mathbf{e}}}^{h}_{s} = \Big[2\dot{e}^{h}_{11}  \ \ 2\dot{e}^{h}_{12}\Big]^{T} \ \text{and}\ \bar{D}^{h}(\dot{\mathbf{u}}^{h})=\int_{\Omega}\kappa\bar{\lambda}^{h}\text{d}\Omega=\sum_{k=1}^{n_{s}}\kappa A^{(k)}\bar{\lambda}^{k} \\
  \mathcal{C}^h=\{\dot{\mathbf{u}}^h\in\mathcal{W}^h|W_{int}(\dot{\mathbf{u}})=1 \}
\end{align}
It is evident that the flow rule can be satisfied at every point over $\Omega^{(k)}$, and likewise everywhere in a set $ \bar{\mathcal{T}}$ of the problem domain. On the other hand, \textcolor{blue}{the edge-based averaging strains rates defined on a dual mesh somewhat debilitates the compatibility due to the smoothed strain rate which  is constant over the smoothing domain}. Hence, the present method cannot produce properly a strict upper bound, but the collapse load factor found is reliable and applicable.
\section{Solution procedure of the discrete problem}\label{Sec4}
The aforementioned limit analysis problem can be solved by a general non-linear optimization solver \cite{15Capsoni}. Furthermore, it can be rewritten in a sum of norms \cite{17Christiansen} in second-order cone programming (SOCP), which is solved by the interior point method \cite{43Andersen,44Andersen}. To address this, the limit analysis problem with $ n_{s} $ constraints can be formed as follows
\begin{align}
  \min \sum_{k=1}^{n_{s}}c_{k}t_{k} \qquad\qquad\qquad\nonumber \\
  s.t. \ \ \big\|\mathbf{H}_{k}\mathbf{t}+\mathbf{v}_{k}\big\|\leq\mathbf{y}_{k}^{\mathrm{T}}\mathbf{t}+z_{k} \ \text{for} \ k=1,...,n_{s}
\end{align}
where $t_{k}\in \mathcal{R}^{+}, k=1,...,n_{s} $ or $\mathbf{t}\in\mathbf{R}^{n_{s}}$ denotes optimization variables and the coefficients are $c_{k}\in R, \mathbf{R}_{k} \in R^{m_{dim}\times n_{s}}, \mathbf{v}_{k}\in R^{m_{dim}}, \mathbf{y}_{k}\in R^{n_{s}}$ and $z_{k}\in R$. The SOCP problem is used for $ m_{dim} $ = 2 or $ m_{dim} $ = 3. When $ m_{dim} $ = 1, the SOCP problem reduces to a linear programming problem. Now we represent the limit analysis problem in the form of the second-order quadratic cone as
\begin{align}
  Z=\Big\{\mathbf{t}\in \mathbf{R}^{n_{s}} | t_{1}\geq\sqrt{\sum_{k=2}^{n_{s}}t_{k}^{2}} =\big\|\mathbf{t}_{2\rightarrow n_{s}} \big\|\Big\}
\end{align}
\textcolor{blue}{The limit analysis problem given in \ref{eqn:LAbar} can be reduced to the problem of minimizing a sum of norms \cite{17Christiansen}.} Now the optimization problem can be written in a common way of minimizing a sum of norms as follows
\begin{subequations}
  \begin{align}
    \alpha^{+} & = \min \sum_{k=1}^{n_{s}}\kappa A^{(k)}\bar{\lambda}^{(k)} \nonumber \\
    s.t.       &
    \begin{cases}
      \bar{\lambda}^{(k)}           & \geq2\|\mathbf{\rho}^{k}\|                                          \\
      \zeta\bar{\lambda}^{(k)}      & = \mathcal{P}^{(k)}(\dot{\varepsilon}^{h}_{m}n_{s}),\  k=1,...,n_{s} \\
      \dot{\mathbf{u}}^{h}          & = \dot{\mathbf{u}}_{0} \ \text{on} \ \varGamma_{\dot{u}}            \\
      W_{ext}(\dot{\mathbf{u}}^{h}) & =1
    \end{cases}
  \end{align}
  \label{eqn:LAES}
\end{subequations}
where $\mathbf{\rho}^{k}$ denotes auxiliary variables.
The optimization problem in \ref{eqn:LAES} is classified into a second-order cone programming (SOCP) with the constraint of quadratic cones in Eq. \ref{eqn:LAbar}. As a result, we can use the fast solver of the Mosek optimization package for this problem. The robust feature of Mosek is to provide high computational efficiency and accuracy for solving the large-scale optimization problems in practice.
\section{On an adaptive quadtree mesh refinement}\label{Sec5}
Since uniform meshes are computationally expensive to solve plastic collapse problems, adaptive mesh refinements are very useful for adjusting mesh resolution in order to improve the solution with the desired accuracy. In the limit analysis, the solution accuracy depends on the fast detection of the yield lines or plastic collapse zones over the problem domain. This means that a mesh should be refined along the yield lines while it is gradually coarse away from the yield lines. Regarding computational efficient, the mesh size should be adjusted and refined automatically or adaptively \cite{49Ciria}. Theoretically, an effective error estimator or indicator to guide the mesh refinement process is required yet it is available for elasticity. Nevertheless, in plastic limit analysis, finding such an error indicator is challenging. Therefore, we in this study introduce an alternative indicator based on the $\mathcal{L}^{2}$ - norm measure of quantities in plastic collapse zones based on plastic strain rates.
One of our important contributions in the present method is highly capable of generating the automatic local refinement with a quadtree mesh. Ideally, the quadtree mesh is a recursive spatial decomposition of the quadrilateral element into four new child elements (\textit{children}). It is worth emphasizing that such a mesh procedure is performed easily together with the proposed indicator.
\subsection{$\mathcal{L}^{2}$ - norm-based indicator}
We know that the presence of localized plastic deformations or high strain rates at limit state very significantly affects the accuracy of numerical solutions. Using uniform meshes leads to huge computational cost to gain the limit load factor of the desired accuracy due to refined meshes at unnecessary regions. Therefore, adaptive meshes should be done so that the approximated velocity/strains rates converge the actual ones with the lowest computational cost as possible. To achieve this, we introduce an alternative indicator in adaptive refinement procedure based on the $\mathcal{L}^{2}$ - norm of plastic strain rates, which can be defined over each element as
\begin{align}
  \Theta^{e}=\dfrac{1}{n^{e}_{edge}}\sum_{k=1}^{n^{e}_{edge}}\Theta^{(k)}, \ e=1,2,...,n_{e}
\end{align}
where $n^{e}_{edge}$ indicates the number of elemental sides and $\textcolor{blue}{\Theta^{(k)}}$ is the $\mathcal{L}^{2}$ - norm of plastic strain rates on side \textit{k} of the element $e$ and is calculated by Eq. \ref{eqn:L2} as
\begin{align}
  \Theta^{(k)} = \Big\|\dot{\bar{\boldsymbol\varepsilon}}^{(k)}\Big\|_{\mathcal{L}^{2}(\Omega^{(k)})}, \ k=1,2,...,n_{s}
\end{align}
\subsection{Refinement strategy}
Once plastic strain rates are found, it is used to evaluate the corresponding $\mathcal{L}^{2}$ - norm and to guide the mesh refinements. The global indicator, $ \Theta $, can be described as the contribution of the local refinement indicator, $ \Theta^{e} $, for all the individual elements
\begin{align}
  \Theta=\sum_{e=1}^{n^{e}}\Theta^{e}, \ e=1,2,...,n_{e}
\end{align}
We furthermore mark a set $\mathcal{M}\subseteq \mathcal{T}$ of $\{\Omega^{e}\} \in\mathcal{T}$ such that
\begin{align}
  \sum_{\Omega^{e}\in \mathcal{M}}\Theta^{e}\geq\theta\Theta, \ \text{for some}\ \theta \in(0, 1)
\end{align}
A new mesh $\mathcal{T}^{'}$ is then derived from $\mathcal{T}$ by refining the marked elements $\Omega^{e}\in \mathcal{M}$ by using quadtree meshes.
\subsection{Description of quadtree meshes}
The quadtree meshes were introduced in \cite{01Samet}, where a spatial decomposition structure is used for quad mesh refinements. In the quadtree mesh, each father element is partitioned recursively into four child quadrants. In Fig. \ref{Fig: Fig7}, a quadtree mesh and its representative tree are presented. After each refinement, hanging nodes are created if new elements and their neighbours are not at same levels of refinement. Hanging nodes consist of the vertices of the child element that is placed on the side of the father element (e.g., \textcolor{blue}{diamond-shape nodes} in Fig. \ref{Fig: Fig7} (a)). The presence of these hanging nodes leads to the incompatibilities in finite element approximations. Several methods have been devised to treat these issues such as Lagrange multipliers or penalty method \cite{82Hansbo}, adding temporary elements \cite{83Palle}, constraining hanging nodes to corner nodes \cite{84Ainsworth}, hierarchical enrichment \cite{85Krysl}, B-splines \cite{86Kagan}, natural neighbor basis functions \cite{68Tabarraei}. In \cite{68Tabarraei}, authors constructed natural neighbor (Laplace) basis functions to obtain $C^{0}$ continuous approximations along sides containing hanging nodes. These shape functions were defined over the polygonal reference elements with an affine map. Hence, the number of hanging nodes on each side is arbitrary as shown in several available techniques \cite{82Hansbo,83Palle,84Ainsworth,85Krysl,86Kagan}.
In this study, we utilize $C^{0}$ admissible approximations along sides by using a mapping from the polygonal reference element but we employ the barycentric coordinates to construct piecewise-linear shape functions of quadrilaterals with hanging nodes. Note that the classical Wachspress coordinates are not easy to construct the shape functions of polygon with side-nodes (including hanging nodes) on the quadtree meshes. As described in the Section \ref{Sec3_3}, we can construct shape function for quadrilaterals with side-nodes using a mapping $\mathbf{J}_{\xi}=\partial\mathbf{x}/\partial{\xi}$. Note that hanging nodes and vertices are now regarded as vertices. By this way, conforming shape functions for quadtree meshes with the number of hanging nodes can be well established.
\section{Numerical implementation}\label{Sec6}
In this section, we summarize algorithms for the implementation of the present method. Three main algorithms are given as

\subsection{\textbf{Algorithm 1}: Main program}
\begin{itemize}
  \item[-] Define the problem domain (coordinate, load and boundary conditions): $\Omega, \Gamma_{\dot{\mathbf{u}}}, \Gamma_{\mathbf{t}}$.
  \item[-] Give number of iterations, $ n_{iter} $, and refinement factor $ \theta $.
  \item[-] Given plastic parameters:  $c$, $\varphi$.
\end{itemize}
{
\setstretch{2.5}
Create $\mathbf{\Lambda} = \displaystyle\sum_{e=1}^{n_{e}}\{\Lambda\}^{e}$ \% \emph{element connectivity matrix}\\
Create $\mathbf{\Lambda}_{p}=\mathbf{\Lambda}$ \% \emph{element connectivity matrix on quadtree}\\
\textbf{for} \textit{i} = 1 $\rightarrow n_{iter} $\\
\text{\quad } Add boundary conditions $\Gamma_{\dot{\mathbf{u}}}, \Gamma_{\mathbf{t}}$.\\
\text{\quad } $\mathbf{\Lambda} \rightarrow$ add one node at the centroid of each element $\rightarrow \mathbf{\Lambda}_{c}$\\
\text{\quad } Create edge-based integration domains $\Omega=\bigcup\!^{n_{s}}_{k=1}\Omega^{(k)}$ \\
\text{\quad } \textbf{for} \textit{k} = 1 $\rightarrow n_{s} $ \% $ n_{s} $ \textit{is the number of edge-domains}\\
\text{\quad } \text{\quad } Call \textbf{algorithm 2}: compute strain matrix using $\mathbf{\Lambda}_{c}$\\
\text{\quad } \textbf{end for}\\
\text{\quad } Compute auxiliary variables and define SOCP\\
\text{\quad } Call \textit{MOSEK}: optimization solution\\
\text{\quad } Evaluate element-based strain-norm indicator $\boldsymbol\eta = \displaystyle\sum_{e=1}^{n_{e}}\eta^{e}$\\
\text{\quad } Call \textbf{algorithm 3}: adaptation on quadtree meshes\\
\textbf{end for}
}
\subsection{\textbf{Algorithm 2}: Compute strain matrix}
Input: $\mathbf{\Lambda}_{c}$, information of $\Omega^{(k)}$\\
Give $\bar{\mathbf{B}}^{(k)}=0$\\
\text{\quad\ } \textbf{for} \textit{ic} = 1 $\rightarrow $ number of neighbor cell that \textit{k} edge influence\\
\text{\qquad\ \textbf{+}} Find $ \mathbf{x}_{ic} $ of element $ \Omega^{(k)}_{ic} $ contain the edge domain \textit{k}\\
\text{\qquad\ \textbf{+}} Define reference element $ \Omega^{\xi} $ and find  ${\boldsymbol\xi} $ $\rightarrow $ Find Gauss point $ \boldsymbol\xi_{\triangle} $ and its weight $ w_{\triangle m} ^ {\xi} $ of sub-triangle $\triangle$ corresponding to element $ \Omega^{(k)}_{ic} $\\
\text{\qquad\ \textbf{+}} Define function $ \phi^{e}_{I} $ for all vertex of element $\rightarrow $ Find $ \phi^{\triangle}_{I} $ corresponding to element $ \Omega^{(k)}_{ic} $\\
\text{\qquad\qquad} \textbf{for} \textit{m} = 1 $\textcolor{blue}{ \rightarrow}$ $ n_{g} $ \% where $ n_{g} $ is number of Gauss point in sub-triangle element $\triangle$\\
\text{\qquad\qquad\ \ \textbf{+}} Compute $ N_{b}(\boldsymbol\xi_{\triangle m}),  \boldsymbol\nabla N_{b}(\boldsymbol\xi_{\triangle m})$\\
\text{\qquad\qquad\ \ \textbf{+}} Define $ N_{\triangle}(\boldsymbol\xi_{\triangle m}),  \boldsymbol\nabla N_{\triangle}(\boldsymbol\xi_{\triangle m})$\\
\text{\qquad\qquad\ \ \textbf{+}} Compute linear shape function and their derivatives at all vertices:\\\text{\qquad\qquad\ \ \textbf{ }}
$N_{I}^{e}\big(\boldsymbol\xi_{\triangle m}\big)=\displaystyle\sum\limits_{l=1}^{3}N^{l}_{\triangle}(\boldsymbol\xi_{\triangle m})\phi^{\triangle}_{I}$ and $\boldsymbol\nabla N_{I}^{e}\big(\boldsymbol\xi_{\triangle m}\big)= \displaystyle\sum\limits_{l=1}^{3}\boldsymbol\nabla N^{l}_{\triangle}(\boldsymbol\xi_{\triangle m})\phi^{\triangle}_{I}$ \\
\text{\qquad\qquad\ \ \textbf{+}} Compute linear shape function and their derivatives including bubble node:\\ \text{\qquad\qquad\ \ }  $N_{I}^{e}\big(\boldsymbol\xi_{\triangle m}\big)=\big[N_{I}^{e}\big(\boldsymbol\xi_{\triangle m}\big)\ N_{b}(\boldsymbol\xi_{\triangle m})\big]$ and $\boldsymbol\nabla N_{I}^{e}\big(\boldsymbol\xi_{\triangle m}\big)=\big[\boldsymbol\nabla N_{I}^{e}\big(\boldsymbol\xi_{\triangle m}\big)\ \boldsymbol\nabla N_{b}(\boldsymbol\xi_{\triangle m})\big]$\\
\text{\qquad\qquad\ \ \textbf{+}} Compute Jacobian: $\mathbf{J}_{\xi}=\dfrac{\partial \mathbf{x}_{ic}}{\partial \xi}$\\
\text{\qquad\qquad\ \ \textbf{+}} Compute $\boldsymbol\nabla N_{I}^{e}\big(\mathbf{x}_{\triangle Gm}\big) = \mathbf{J}^{-1}_{\xi}\boldsymbol\nabla N_{I}^{e}\big(\boldsymbol\xi_{\triangle m}\big)$\\
\text{\qquad\qquad\ \ \ } $ B^{*}_{m} = B^{*}_{m} + \boldsymbol\nabla N_{I}^{e} \big(\mathbf{x}_{\triangle Gm}\big) \big|\mathbf{J}_{\xi} \big| w_{\triangle m} ^{\xi} $\\
\text{\qquad\qquad} \textbf{end for} \% end loop Gauss point of sub-triangle $\triangle$\\
\text{\qquad\qquad} $\bar{\mathbf{B}}^{(k)}=\bar{\mathbf{B}}^{(k)}+\mathbf{B}^{*}_{i}$\\
\text{\quad\ } \textbf{end for}\\
\text{\quad\ } $\bar{\mathbf{B}}^{(k)}=\dfrac{1}{{A}^{(k)}}\bar{\mathbf{B}}^{(k)}$\\
\text{\quad\ } Output $\bar{\mathbf{B}}^{(k)}$
\subsection{\textbf{Algorithm 3}: Adaptation on quadtree meshes}
\text{\ \ } Input:$\theta, \Theta, \boldsymbol{\Lambda}, \boldsymbol{\Lambda}_{p}$\\
\text{\qquad} Arrange strain-norm indicator $\Theta$ from largest to smallest\\
\text{\qquad \textbf{+}} Mark elements will be adaption\\
\text{\qquad} $\sum\limits_{e=1}^{n_{e}} \Theta^{e} \geq \theta\Theta$ \\
\text{\qquad} \textbf{while}
\text{\ } $\sum_{e=1}^{n_{e}} \Theta^{e} \geq \theta\Theta$ \\
\text{\qquad\quad} $ \tau^{e} $  = 1 \ \% \emph{marking the elements need to be divided}\\
\text{\qquad} \textbf{end while}\\
\text{\qquad \textbf{+}} Divide parent element into child elements and add new element data\\
\text{\qquad} \textbf{for} \textit{e} = 1 $\rightarrow n_{e}$\\
\text{\qquad\quad\ ~} \textbf{if} $\tau^{e}  $ = 1 \textbf{then}\\
\text{\qquad\qquad\ \ \textbf{+}} Divide the parent element into 4 child elements\\
\text{\qquad\qquad\ \ \textbf{+}} Add 4 new child elements data to $\boldsymbol{\Lambda}_{p}$: $\boldsymbol{\Lambda}_{p} = \boldsymbol{\Lambda}_{p}+\big\{\Lambda_{p} \big\}^{e}_{new}$ and $\boldsymbol{\Lambda} = \boldsymbol{\Lambda}+\big\{\Lambda_{p} \big\}^{e}_{new}$\\
\text{\qquad\qquad\ \ \textbf{+}} Delete parent element\\
\text{\qquad\quad\ ~} \textbf{end if}\\
\text{\qquad} \textbf{end for}\\
\text{\qquad \textbf{+}} Find hanging nodes\\
\text{\qquad} \textbf{for} \textit{e} = 1 $\rightarrow n_{e}$\\
\text{\qquad\quad\ ~} \textbf{for} \textit{iedg} = 1 $\rightarrow n_{edg}$  \% \emph{where $ n_{edg} $ is the number of edges of element}\\
\text{\qquad\qquad\quad \textbf{+}} Find nodes $\mathbf{x}_{new} $  on edge\\
\text{\qquad\qquad\quad } \textbf{if} $\mathbf{x}_{new} \neq \varnothing$ \textbf{then}\\
\text{\qquad\qquad\qquad } $\big\{\Lambda_{p} \big\}^{e}=\big\{\Lambda_{p} \big\}^{e}+\mathbf{x}_{new}\% \emph{ update element data information}\\
  \text{\qquad\qquad\qquad } \rightarrow \mathbf{\Lambda}$\\
\text{\qquad\qquad\quad } \textbf{end if}\\
\text{\qquad\quad\ ~} \textbf{end for} \% \emph{end loop on edges of elements}\\
\text{\qquad} \textbf{end for}\\
\text{\ \ } Output: $\boldsymbol{\Lambda}, \boldsymbol{\Lambda}_{p}$
\section{Numerical validations}\label{Sec7}
In this section, we examine the performance of the proposed approach through three benchmark problems. The program is complied on \textit{Macbook Air (Intel Core I7, 2.0GHz CPU, 8G RAM)}. The conic interior-point optimizer of the academic MOSEK package is employed.
\subsection{Strip footing bearing capacity}
The first example considered in our analysis is a smooth strip footing of width $B$ on a weightless cohesive--frictional soil ($c \geq 0$, $\varphi \geq 0$) as given in Fig. \ref{Fig: Conv_LAL_StripFooting_Fullmodel}a. We investigate the collapse plastic load for the structure subjected to loading $Q$. The analytical solution of the limit load factor for the first case ($\varphi = 0$) is $ \alpha^{*} = 2 + \pi $ and for the second case is given as $ \alpha^{*} = [e^{\pi\tan\varphi} \tan^{2}(\pi/4 + \varphi/2)-1]\cot\varphi $ \cite{99Prandtl}. Applying the symmetry boundary condition, a half of the foundation with length $L$ and height $H$ is considered.

For purely cohesive soil case ($\varphi = 0$), the problem is modelled with $L = 2.5B$ and $H = B$. A uniform mesh of $40$ quadrilateral (Q4) elements is indicated in Fig.\ref{Fig: Conv_LAL_StripFooting_Fullmodel}b. Some adaptive meshes are depicted in Fig.\ref{Fig: Conv_LAL_StripFooting_AdaptiveMeshStep_phi0}. Fig. \ref{Fig: Conv_LAL_Stripfooting_phi0} exhibits the convergence of the limit load factor using uniform and adaptive meshes. Table \ref{tab: table1} shows the convergence of solution using adaptive meshes. As expected, the accuracy of solution is improved a lot after using adaptive meshes. Regarding computational efficiency, to gain a limit load factor value $ \alpha^{+} = 5.149$ for bL--Q4, optimization Mosek time takes around 2.7s for 16542 optimization variables. For comparison, Table \ref{tab: table2} lists the present result and several other ones. Again, we show that the present method is strongly competitive to other researches \cite{16Sloan,23Makrodimopoulos}. Taking 16542 variables, the present method using adaptive bL--Q4 meshes yields ultra-accurate solution with a small error of 0.14\% in comparison with the error of 1.33\% reported in \cite{16Sloan}. Note that the approach in \cite{16Sloan} based on the discontinuous velocity field can leads to the poor performance for unstructured meshes \cite{23Makrodimopoulos}. Makrodimopoulos and Martin \cite{23Makrodimopoulos} introduced a simplex strain element formulation for unstructured meshes. Here, it is numerically proved that our approach works well for unstructured meshes. It is seen that the present limit load value matched well with a strict upper bound solution proposed by Makrodimopoulos and Martin \cite{23Makrodimopoulos} using 18719 6-node triangular meshes (approximately 149752 variables). Also, the present solutions are more accurate than that of mixed FE formulation\cite{15Capsoni} and are acceptable with that of lower bound formulation \cite{22Makrodimopoulos}. Fig. \ref{Fig: Strip_dissipation} plots plastic dissipation distribution of a purely cohesive soil problem using an adaptive mesh. Last but not least, the convergence of relative error is shown in Fig. \ref{Fig: ConvError_LAL_Stripfooting_phi0}. As we can seen from this figure, the present method has a good convergence in comparison with uniform mesh (bL--Q4).

For cohesive--frictional material case ($\varphi = 35^{\circ}$), the problem is modeled with $L = 13B$ and $H = 10B$. Several adaptive meshes are displayed in Fig. \ref{Fig: Conv_LAL_StripFooting_AdaptiveMeshStep_phi35}. The approximate upper bound value against adaptive meshes is listed in Table \ref{tab: table1}. As expected, the collapse load factor $ \alpha^{+} = 46.292$ using bL--Q4 for the last adaptive mesh has a quite small error of $0.36\%$ while optimization Mosek time takes approximately $8$s with $43967$ variables. The efficiency of adaptive refinement is that more than $90$ percent of the elements are located in the plastic localization zone. For comparison, Table \ref{tab: table2} shows that the present method is very good competitor to the linear strain element \cite{23Makrodimopoulos} using $18719$ 6-node triangular meshes (approximately $149752$ variables).

Additionally, a series of limit load factors are evaluated by both the present and analytical approaches for various internal frictional angles. For the present approach, domain size and initial meshes are the same for all angles and the limit load factors are taken at the eighth step of the successive adaptive mesh refinement loops. The final adaptive meshes are different for each internal frictional angle which lead to the discrepancies in the number of optimization variables. These variables tend to increase correspondingly with the expansion of the internal frictional angles and the final values are $150442$, $204437$, $256457$, $369947$, $449972$, $649247$, $921087$, $1295417$, $1456077$, $1502162$ for the angles of $0^\circ, 5^\circ, 10^\circ, 15^\circ, 20^\circ, 25^\circ, 30^\circ, 35^\circ, 40^\circ, 45^\circ$, respectively. These evaluations are then put side by side in Table~{\ref{tab: footing_present_prandt}} as well as illustrated in Fig.~{\ref{fig:analytical_prandtl}} for comparison purpose. As it pointed out in this table, the relative error values are smaller than $0.1\%$ in all cases except for the last one which is mainly attributed to the computational domain modelled is not large enough to capture the behaviour of plastic dissipation at large frictional angles (see Fig.~{\ref{Fig: strip_dissipation_various_angles}}). These analyses once again demonstrate the high accuracy and flexibility of the present method.

\subsection{Block with two symmetric holes}
The second benchmark is a rectangular domain with two small holes of same radius which has a geometry and applied load as shown in Fig. \ref{Fig: Conv_LAL_2holes_Fullmodel}. This problem also investigated by Zouain \textit{et al.}\cite{18Zouain}, Makrodimopoulos \textit{et al.}\cite{22Makrodimopoulos}, Munoz \textit{et al.}\cite{50Munoz}, H. Nguyen-Xuan \textit{et al.}\cite{98Hung}. The upper bound solutions of this problem was reported in \cite{23Makrodimopoulos}, namely $ \alpha^{+}$ = 1.825 for $\varphi$ = 0$^{\circ}$ and $ \alpha^{+}$ = 1.063 for  $\varphi$ = 30$^{\circ}$ using 177426 stress variables and in \cite{98Hung}, namely $ \alpha^{+}$ = 1.817 for $\varphi$ = 0$^{\circ}$  using 34948 stress variables and $ \alpha^{+}$ = 1.061 for  $\varphi$ = 30$^{\circ}$ using 29768 stress variables are used as reference upper bound values. Also, the lower bound solutions given in \cite{22Makrodimopoulos}, namely $ \alpha^{-}$ = 1.8089 for $\varphi$ = 0$^{\circ}$ and $ \alpha^{-}$ = 1.0562 for $\varphi$ = 30$^{\circ}$ and in \cite{50Munoz} , namely $ \alpha^{-}$ = 1.8119 for $\varphi$ = 0$^{\circ}$ and $ \alpha^{-}$ = 1.0581 for $\varphi$ = 30$^{\circ}$.  All computations are then performed over adaptive mesh refinements.  Fig. \ref{Fig: Conv_LAL_2holes_AdaptiveMeshStep_phi0} and Fig. \ref{Fig: Conv_LAL_2holes_AdaptiveMeshStep_phi30} plot several steps of an adaptive mesh process with respect to the internal friction angle $\varphi$ = 0$^{\circ}$ and $\varphi$ = 30$^{\circ}$, respectively. It is seen that the present method is well captured in accurately capture the plastic zones with adaptive meshes.
The present solution versus adaptive meshes is indicated in Table \ref{tab:table3}. Fig. \ref{Fig: Conv_LAL_2holes_phi0} and Fig. \ref{Fig: Conv_LAL_2holes_phi30} exhibit the limit load factors using uniform and adaptive meshes. Present results agree well a strict upper bound value using the linear strain elements \cite{23Makrodimopoulos}. It is observed that the adaptive bL-Q4 element produces the most accurate solutions for this problem. The global errors are less than 1\% for a slightly coarse mesh of $ N_{var} $ = 6028 for $\varphi$ = 0$^{\circ}$ and $ N_{var} $ = 5713 for $\varphi$ = 30$^{\circ}$. These do not exceed 0.2\% of the global errors for slightly fine meshes with variables around $ N_{var} $ = 56818 for $\varphi$ = 0$^{\circ}$ and $ N_{var} $ = 67458 for $\varphi$ = 30$^{\circ}$. Additionally, a comparison for bL-Q4 listed in Table \ref{tab:table4} demonstrates good performance of the present method. Also, Fig. \ref{Fig: 2holes_dissipation} present accurately the representation of plastic collapse regions which is confirmed by other studies \cite{23Makrodimopoulos,50Munoz}.

\subsection{Slope stability}
The next example is a homogeneous slope of cohesive-frictional soil with inclination 70$^{\circ}$ and height \textit{H}. The problem model and boundary conditions are given in Fig. \ref{Fig: Conv_LAL_Slope_Fullmodel}. The load in this problem is only the soil weight, $ \gamma $. The limit load factor or the stability factor is defined as $\alpha^{s} = \gamma H/c$. The analytical solution (upper bound) for the homogeneous slope stability was investigated by Chen\cite{97Chen}. Some authors have also studied numerical solutions by the upper bound models (see the works of Krabbenhoft \textit{et al.}\cite{19Krabben}, Makrodimopoulos \textit{et al.}\cite{23Makrodimopoulos}, Lyamin \textit{et al.}\cite{21Lyamin}) or the lower bound models (see the work\textbf{s} of Makrodimopoulos \textit{et al.}\cite{22Makrodimopoulos}) for the slope stability in past decades.
\\In this example, we consider $c = 1, H = 1$ and two internal friction angles ($\varphi = 20^{\circ}$ and $\varphi =35^{\circ}$) in our analyses. Here we demonstrate the high performance of the present approach for unstructured meshes. Several adaptive mesh steps for internal friction angles $\varphi = 20^{\circ}$ and $\varphi =35^{\circ}$ are shown in Fig \ref{Fig: Conv_LAL_Slope_AdaptiveMeshStep_phi20} and Fig. \ref{Fig: Conv_LAL_Slope_AdaptiveMeshStep_phi35}, respectively.
The results of analyses are presented in Table \ref{tab:table5} and Fig. \ref{Fig: Conv_LAL_Slope_phi20} for $ \varphi = 20^{\circ}$ and Fig. \ref{Fig: Conv_LAL_Slope_phi35} for $ \varphi = 35^{\circ}$. The present solutions agree well with available ones in the literature as listed in Table \ref{tab:table6}. In detail, obtained solutions are between the upper bound value \cite{23Makrodimopoulos} and lower bound value \cite{22Makrodimopoulos}. Fig.\ref{Fig: Slope_dissipation} displays plastic dissipation distribution. Such plastic collapse zones in structure is detected well by our approach.

\subsection{\textcolor{blue}{Failure in porous media}}
\textcolor{blue}{Final example is a square specimen measures $L = 10$ in length and $H = 10$ in width which contains $16$ equal holes measure $r_c = 0.5$ in radius. The distances between holes' centers in both direction equal to $l = 2$. The applied load and boundary conditions are shown in Fig \ref{Fig: Fullmodel_final}. This example was investigated by Nguyen-Xuan and Liu \cite{29Nguyen} with aims to analyze the failure mechanism in materials with cavities and showed the applicability of the limit analysis approach to homogenization of strength properties in multiscale modeling of porous materials. Fig \ref{Fig: PorousPlateAdaptiveMeshStep_phi0} and Fig \ref{Fig: PorousPlateAdaptiveMeshStep_phi20} depict several meshes for detecting failure mechanisms using adaptive procedure for two cases of internal friction angles $\varphi = 0^{\circ}$ and $\varphi = 20^{\circ}$, respectively. The plastic dissipation distribution of the present approach is provided in Figs \ref{Fig:PorousDissipation} and \ref{Fig:PorousDissipation20}. The present solution is in good agreement with the result reported in Ref \cite{29Nguyen}. Furthermore, our approach is capable of producing properly failure mechanisms under several frictional angles for this problem.}

\section{Conclusion}\label{Sec8}
We presented an efficient and fast numerical scheme over adaptive quadtree mesh generation for limit analysis of structures. The inherent difficulty of the presence of hanging nodes during quadtree mesh generation is simply solved by its definition over arbitrary polygons. The approximate velocity field used piecewise-linear shape functions defined on the primal mesh, while the plastic dissipation was computed over the dual mesh based on  the edge-based integration background. The area-averaged edge-based projection was adopted to redistribute the plastic strain rates on the edge-based integration domains. As expected, the flow condition is satisfied at everywhere in the whole structure. The limit analysis model converted into the form of the second-order cone programming (SOCP) was significantly solved by interior-point solvers for the large-scale optimization problems. A quadtree-based adaptive refinement approach controlled by the $\mathcal{L}^{2}$ -norm-based indicator of plastic strain rates showed the high effectiveness in plastic collapse analysis of structures. The numerical results demonstrated excellent agreement with the reference solutions and compared very well with several other numerical ones. The present method is very simple yet efficient to implement into the existing limit analysis packages and enables us to further extend to other plastic models. Finally, another interesting feature of this study will be extensible to a more general form of yield criteria.

\section*{Acknowledgements}

The author would like to thank Mr. Son Nguyen-Hoang for his comments and suggestions to this work. The financial support provided by the Vietnam National Foundation for Science and Technology Development (NAFOSTED) under grant number 107.02-2016.19, and RISE-project BESTOFRAC (734370) is gratefully acknowledged.

\section*{References}

\pagebreak

\section*{Figures}


\begin{figure}[H]
  \centering
  \includegraphics[]{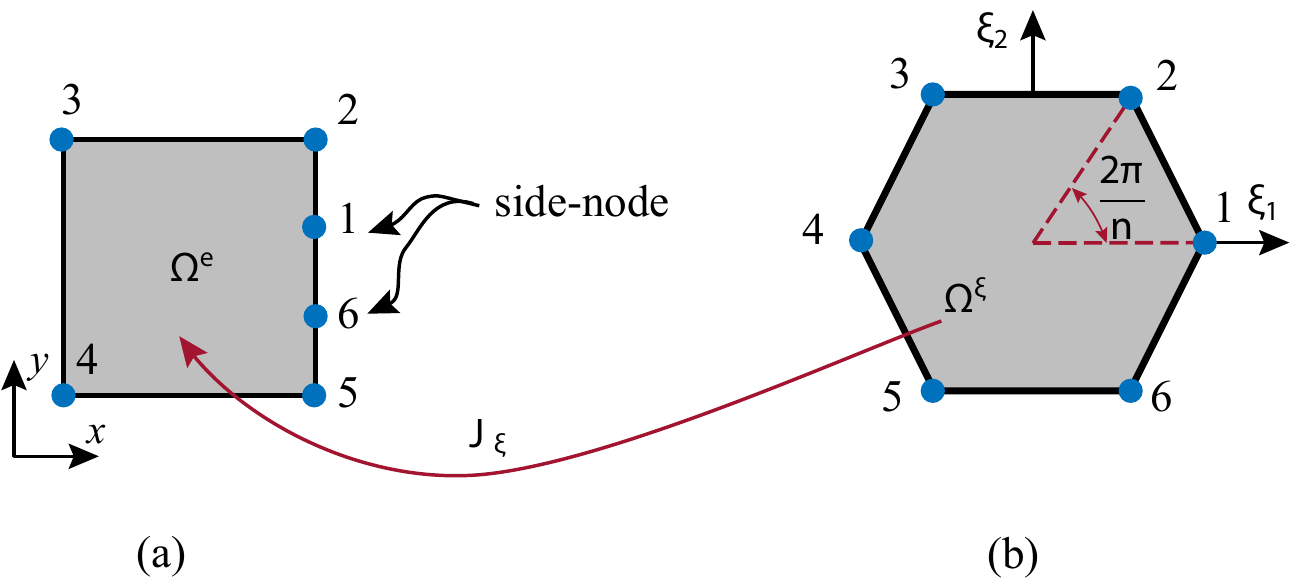}
  \caption{Mapping from a regular quadrilateral element with two
    hanging nodes to a hexagonal element with :(a) Arbitrary quadrilateral element with two side-nodes; (b) Reference hexagonal element.}
  \label{Fig: Fig4a}
\end{figure}
\begin{figure}[H]
  \centering
  \includegraphics[]{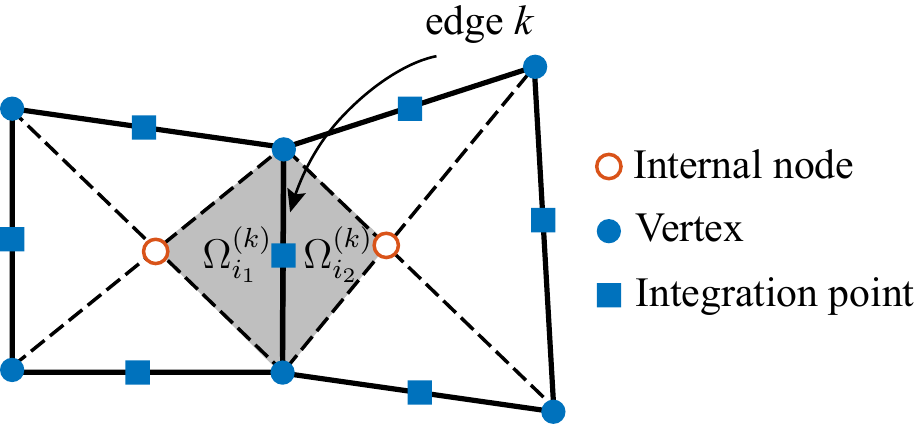}
  \caption{Edge-based integration domain associated with side \textit{k}.}
  \label{Fig: Fig5}
\end{figure}
\begin{figure}[H]
  \centering
  \includegraphics{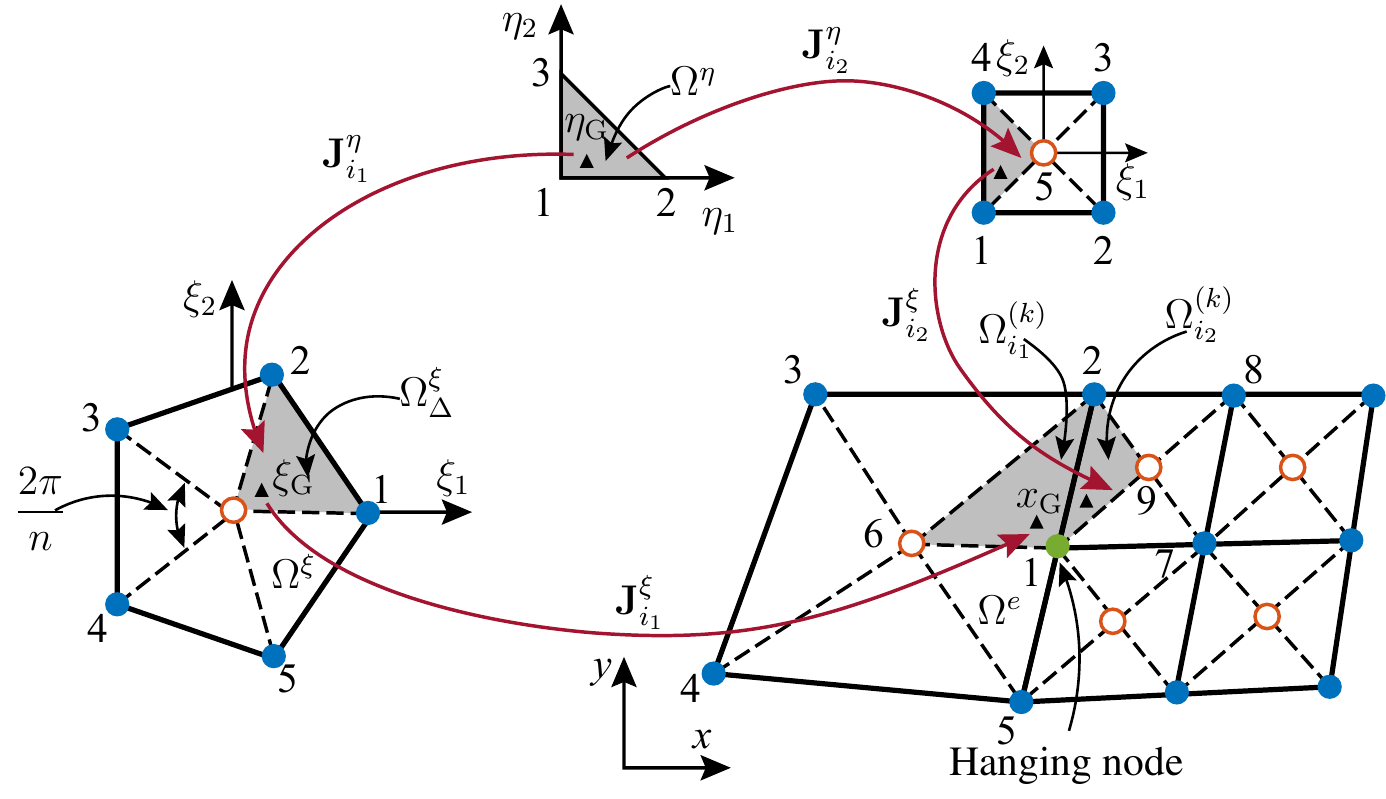}
  \caption{Numerical integration performed on the reference element $\Omega^{\xi}$ and mapping of quadrature points from a parent element $\Omega^{\eta}$.}
  \label{Fig: Fig6}
\end{figure}
\begin{figure}[H]
  \centering
  \includegraphics[]{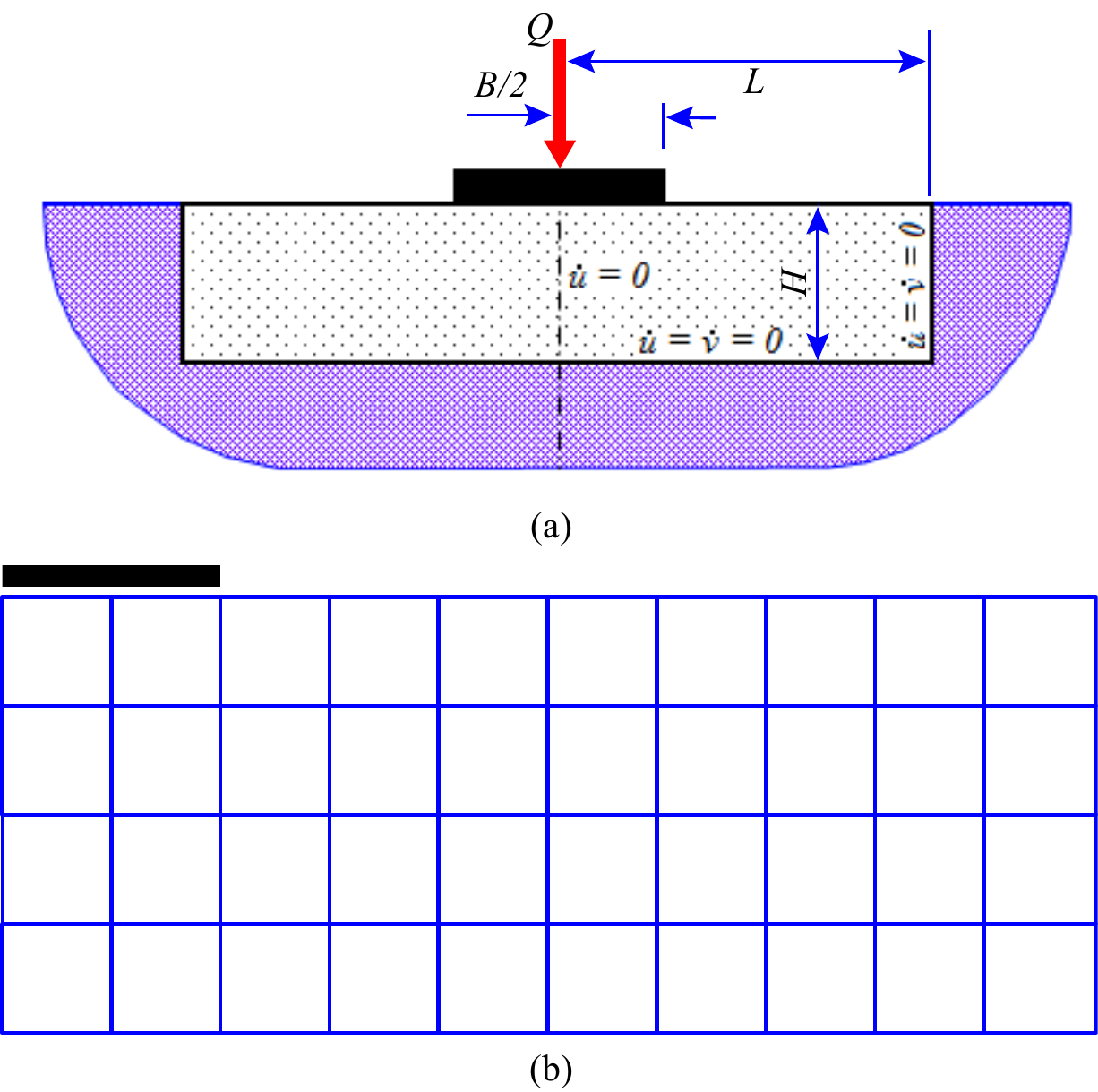}
  \caption{A strip footing problem model on purely cohesive soil: (a) Full model with geometry and boundary condition; (b) Typically uniform Q4 mesh in case of symmetry.}
  \label{Fig: Conv_LAL_StripFooting_Fullmodel}
\end{figure}
\begin{figure}[H]
  \centering
  \begin{subfigure}[b]{0.45\textwidth}
    {
      \centering
      \includegraphics{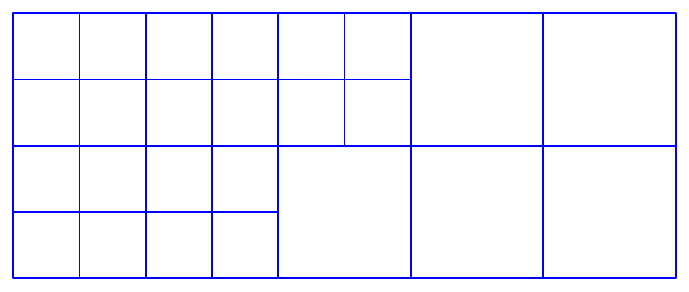}
      \caption{}
    }
  \end{subfigure}%
  \quad
  \begin{subfigure}[b]{0.45\textwidth}
    {
      \centering
      \includegraphics{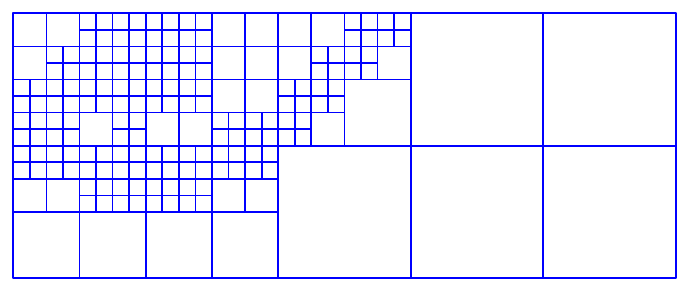}
      \caption{}
    }
  \end{subfigure}%
  \\ \bigskip
  \begin{subfigure}[b]{0.45\textwidth}
    {
      \centering
      \includegraphics{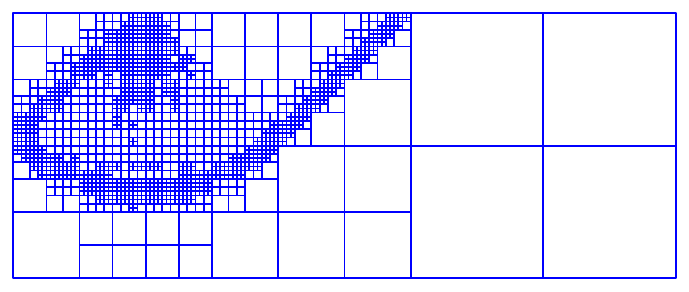}
      \caption{}
    }
  \end{subfigure}%
  \quad
  \begin{subfigure}[b]{0.45\textwidth}
    {
      \centering
      \includegraphics{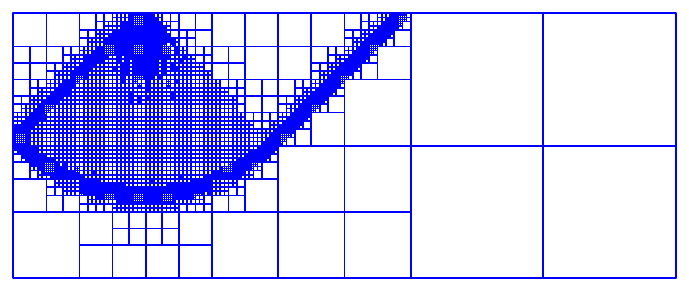}
      \caption{}
    }
  \end{subfigure}%
  \caption{Several adaptive mesh steps for the smooth strip footing$ (\varphi = 0^{\circ} $).}
  \label{Fig: Conv_LAL_StripFooting_AdaptiveMeshStep_phi0}
\end{figure}

\begin{figure}[H]
  \centering
  \includegraphics{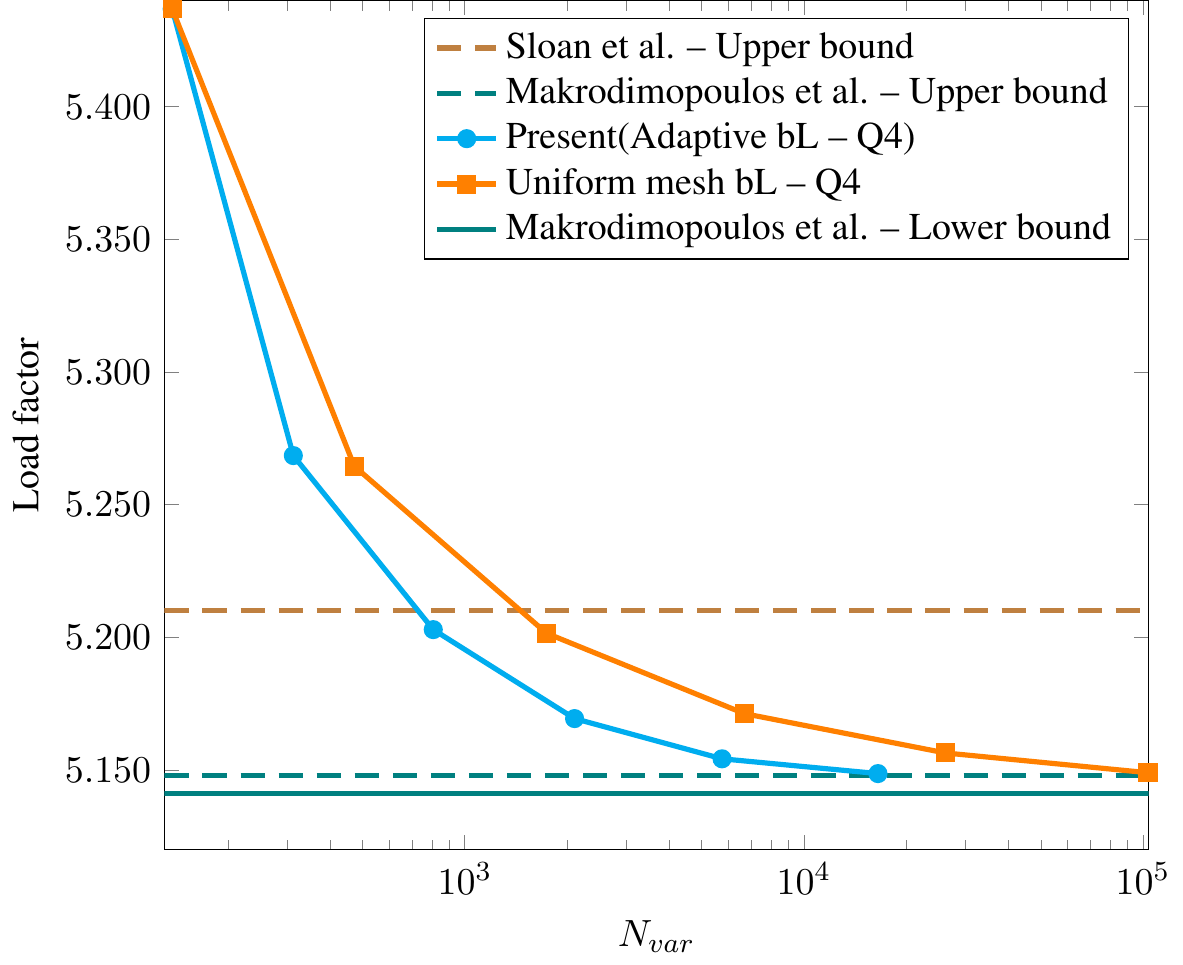}
  \caption{The strip footing of cohesive frictional soil$ (\varphi = 0^{\circ} $): The convergence of limit load factor with respect to optimization variables.}
  \label{Fig: Conv_LAL_Stripfooting_phi0}
\end{figure}
\begin{figure}[H]
  \centering
  \includegraphics{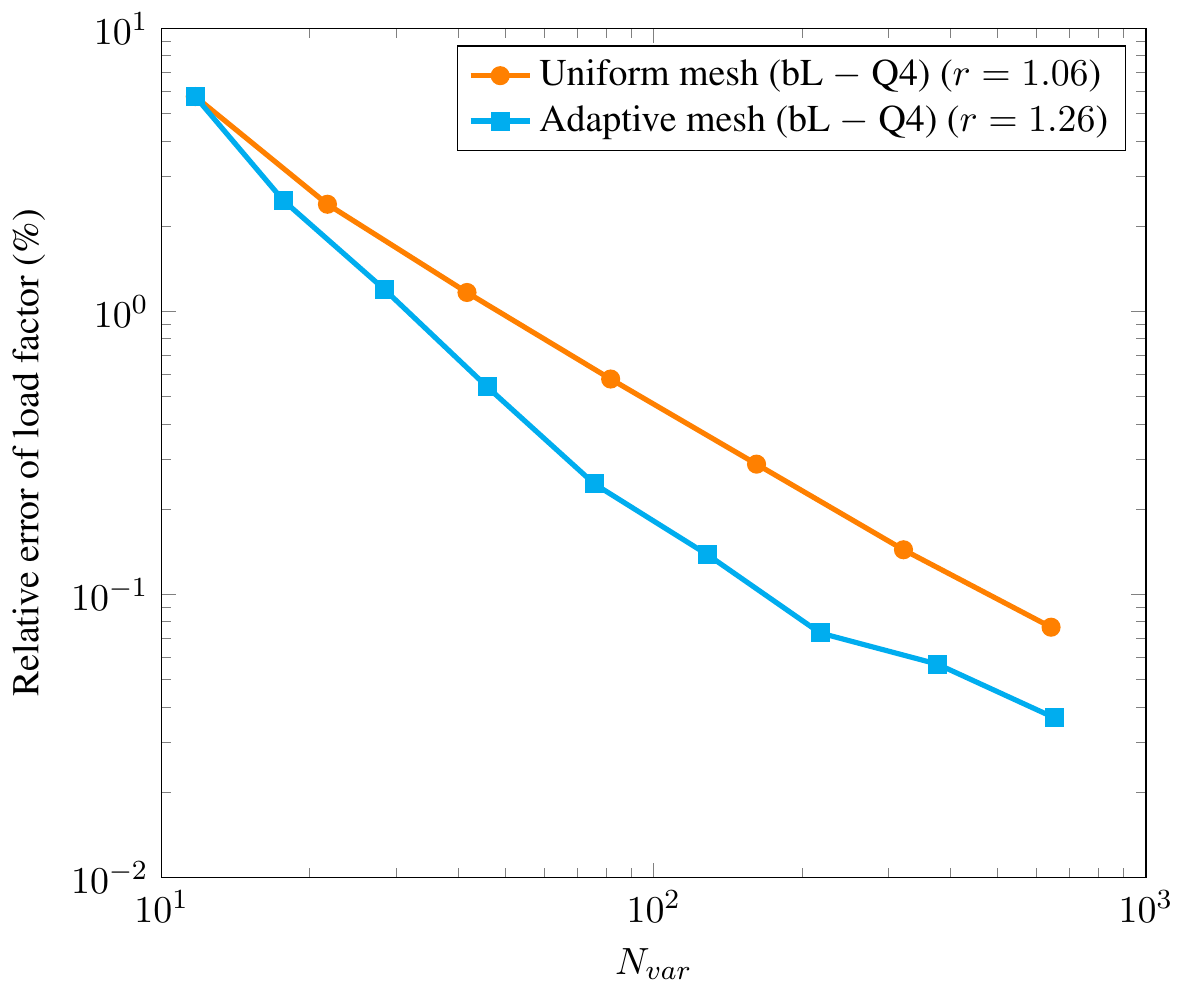}
  \caption{\textcolor{blue}{The strip footing of cohesive frictional soil$ (\varphi = 0^{\circ} $): The relative error of limit load factor with respect to optimization variables.}}
  \label{Fig: ConvError_LAL_Stripfooting_phi0}
\end{figure}
\begin{figure}[H]
  \centering
  \begin{subfigure}[b]{0.45\textwidth}
    {
      \centering
      \includegraphics{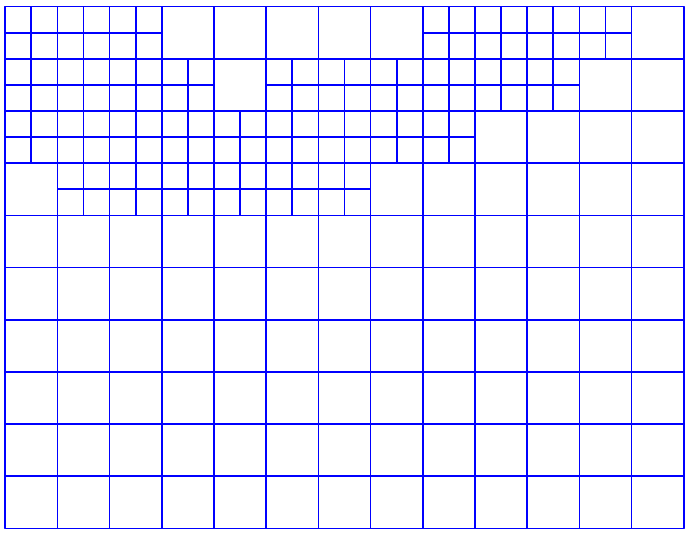}
      \caption{}
    }
  \end{subfigure}%
  \quad
  \begin{subfigure}[b]{0.45\textwidth}
    {
      \centering
      \includegraphics{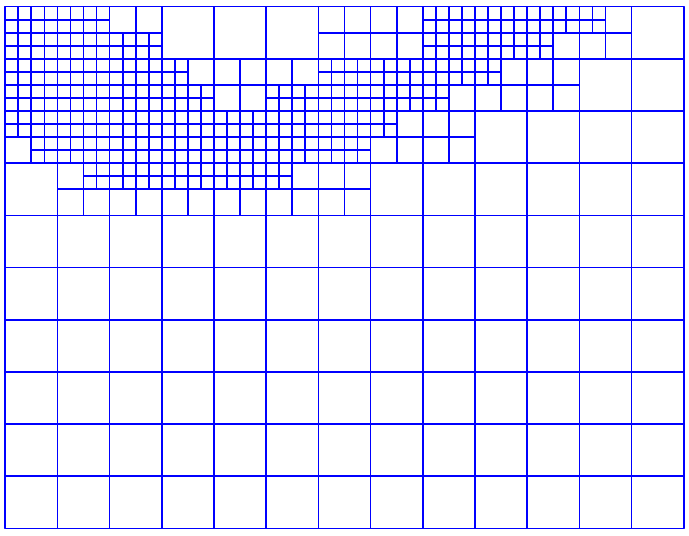}
      \caption{}
    }
  \end{subfigure}%
  \\ \bigskip
  \begin{subfigure}[b]{0.45\textwidth}
    {
      \centering
      \includegraphics{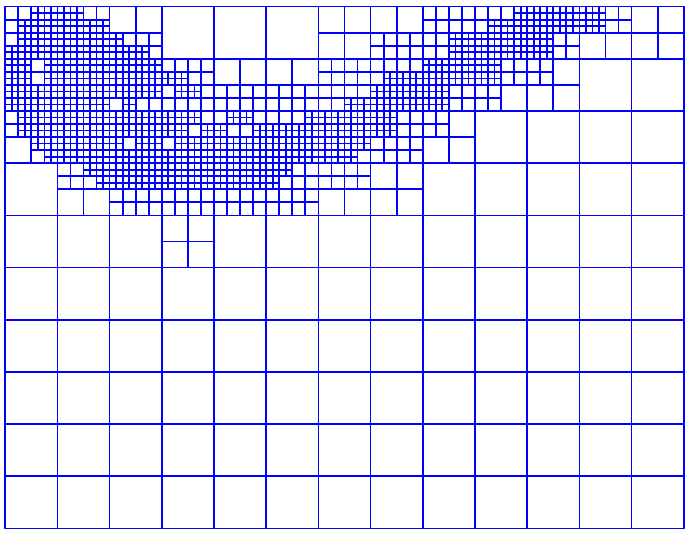}
      \caption{}
    }
  \end{subfigure}%
  \quad
  \begin{subfigure}[b]{0.45\textwidth}
    {
      \centering
      \includegraphics{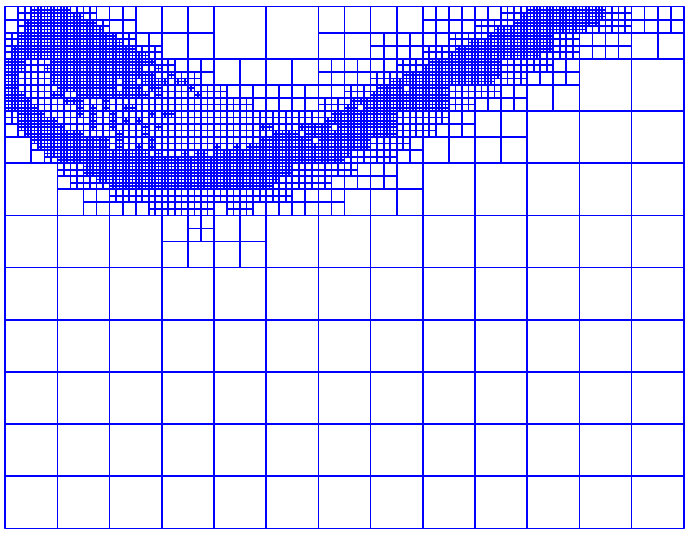}
      \caption{}
    }
  \end{subfigure}%
  \caption{Several adaptive mesh steps for the smooth strip footing$ (\varphi = 35^{\circ} $).}
  \label{Fig: Conv_LAL_StripFooting_AdaptiveMeshStep_phi35}
\end{figure}
\begin{figure}[H]
  \centering
  \includegraphics{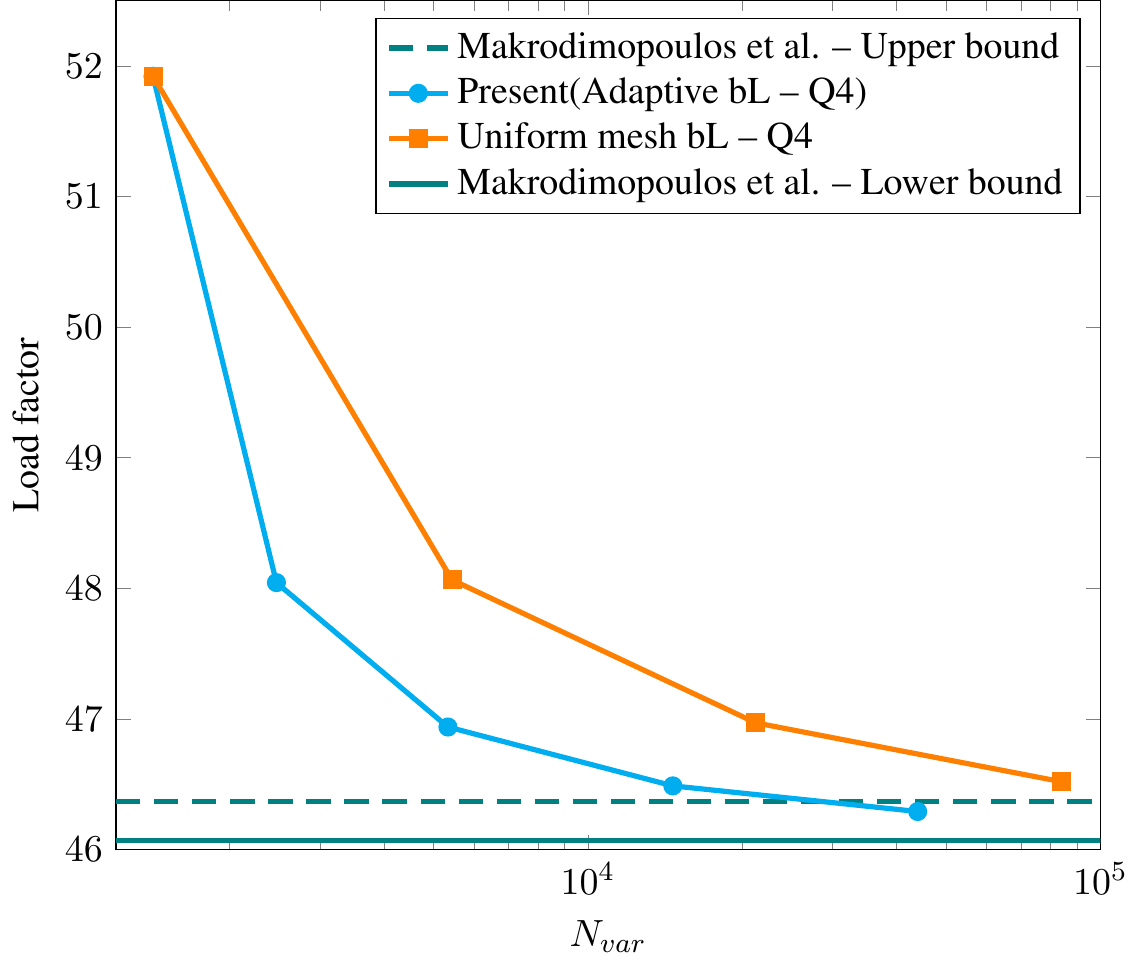}
  \caption{The strip footing of cohesive frictional soil$ (\varphi = 35^{\circ} $): The convergence of limit load factor with respect to optimization variables.}
  \label{Fig: Conv_LAL_Stripfooting_phi35}
\end{figure}
\begin{figure}[H]
  \centering
  \begin{subfigure}[b]{0.45\textwidth}
    {
      \centering
      \includegraphics[width=\textwidth]{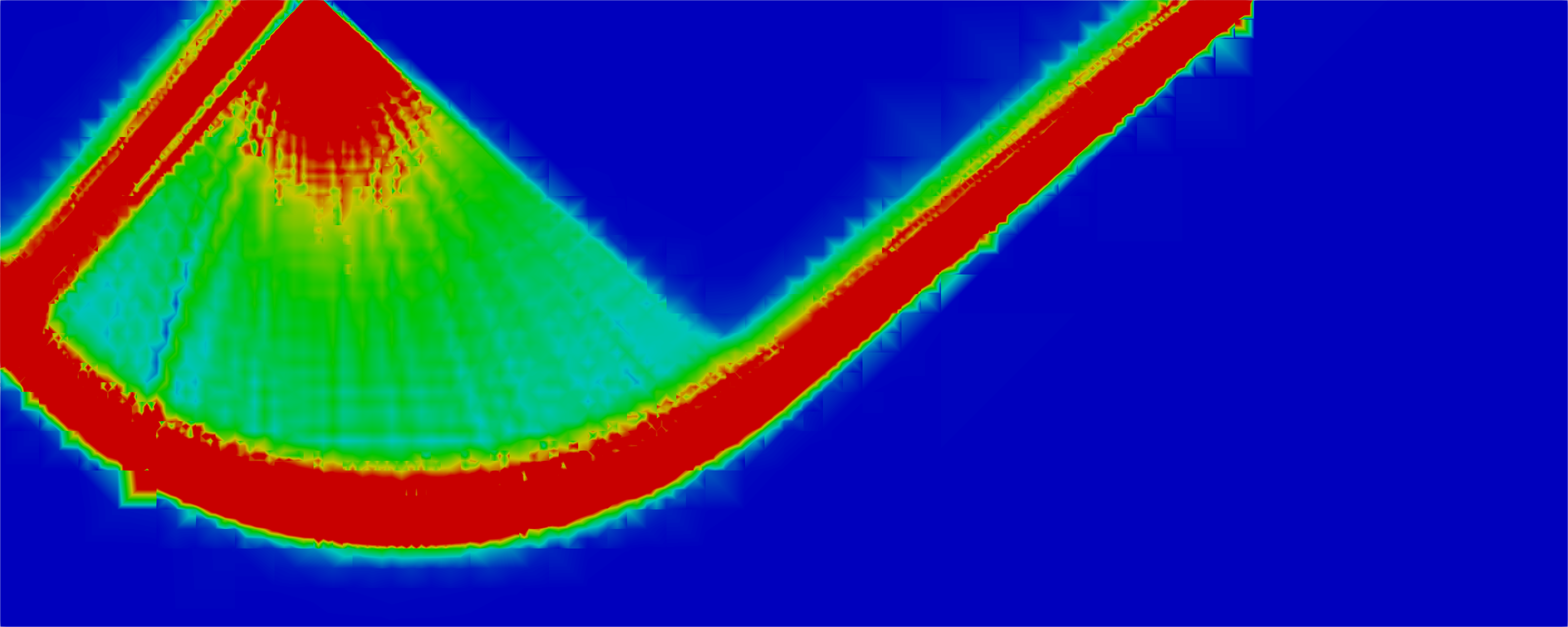}
      \caption{$\varphi = 0^\circ$}
    }
  \end{subfigure}%
  \qquad
  \begin{subfigure}[b]{0.45\textwidth}
    {
      \centering
      \includegraphics[width=\textwidth]{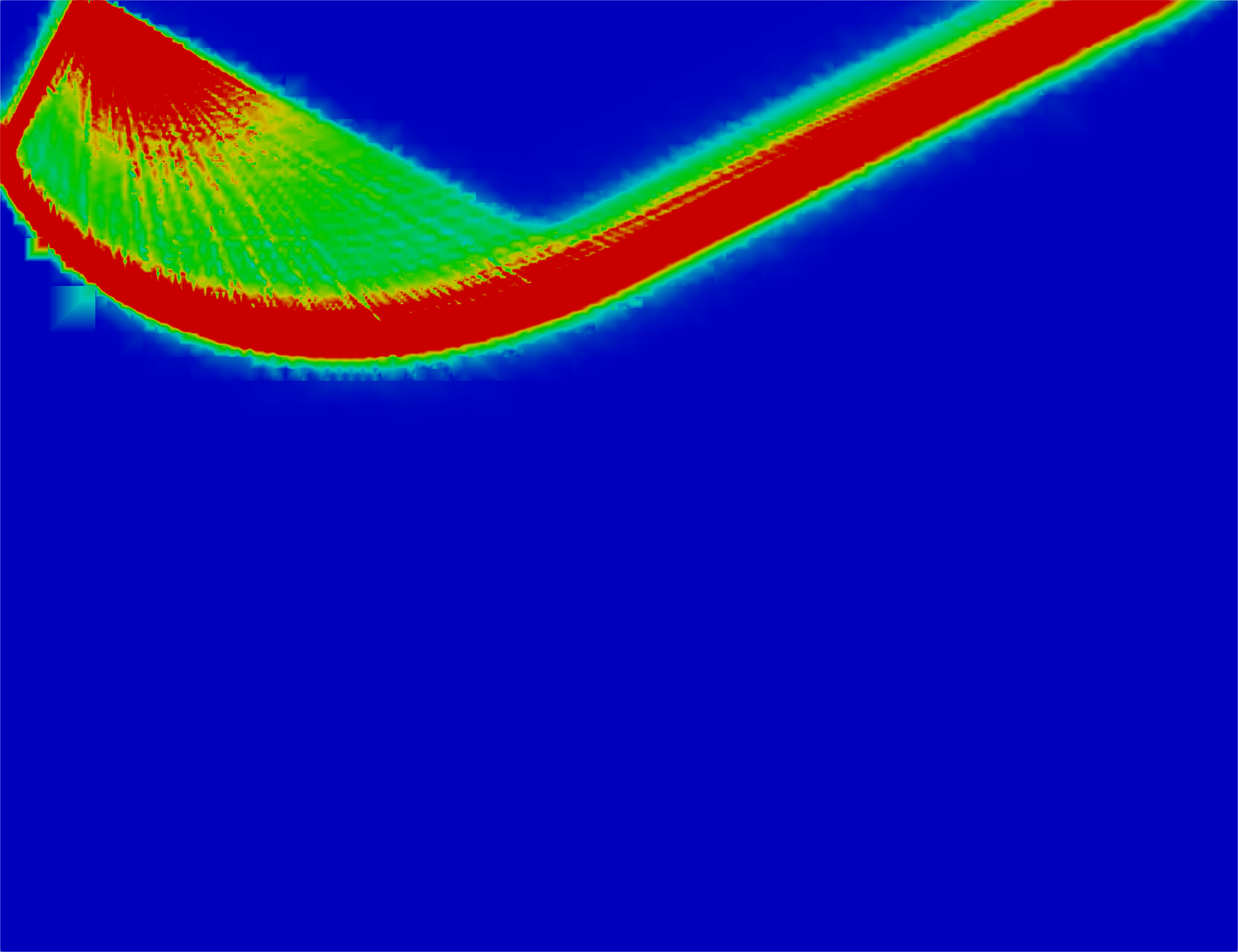}
      \caption{$\varphi = 35^\circ$}
    }
  \end{subfigure}%
  \caption{Plastic dissipation.}
  \label{Fig: Strip_dissipation}
\end{figure}

\begin{figure}[H]
  \centering
  \includegraphics{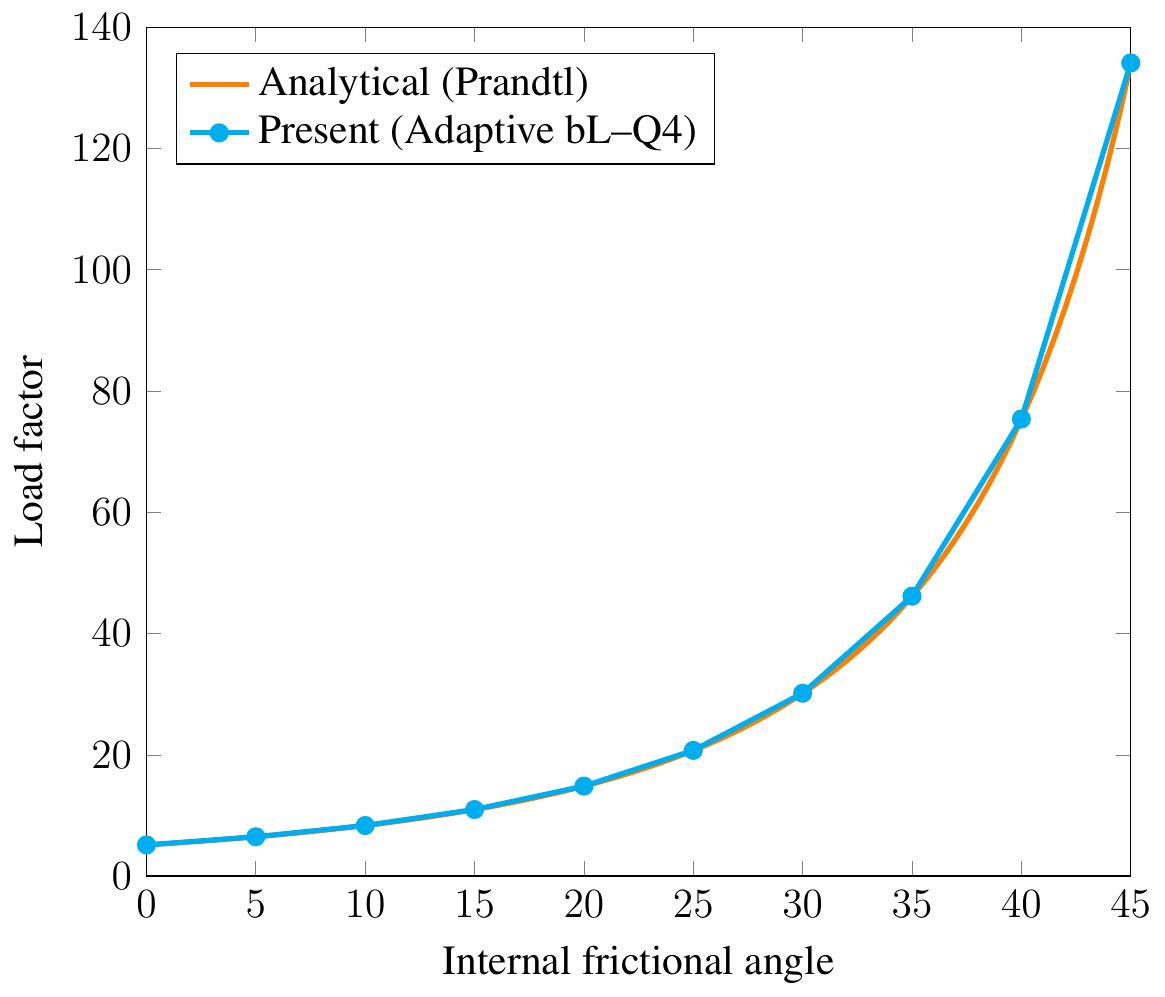}
  \caption{A smooth strip footing: Comparison with the analytical solution using adaptive mesh for a variety of internal frictional angles.}
  \label{fig:analytical_prandtl}
\end{figure}

\begin{figure}[H]
  \centering
  \begin{subfigure}[b]{0.3\textwidth}
    {
      \centering
      \includegraphics[width=\textwidth]{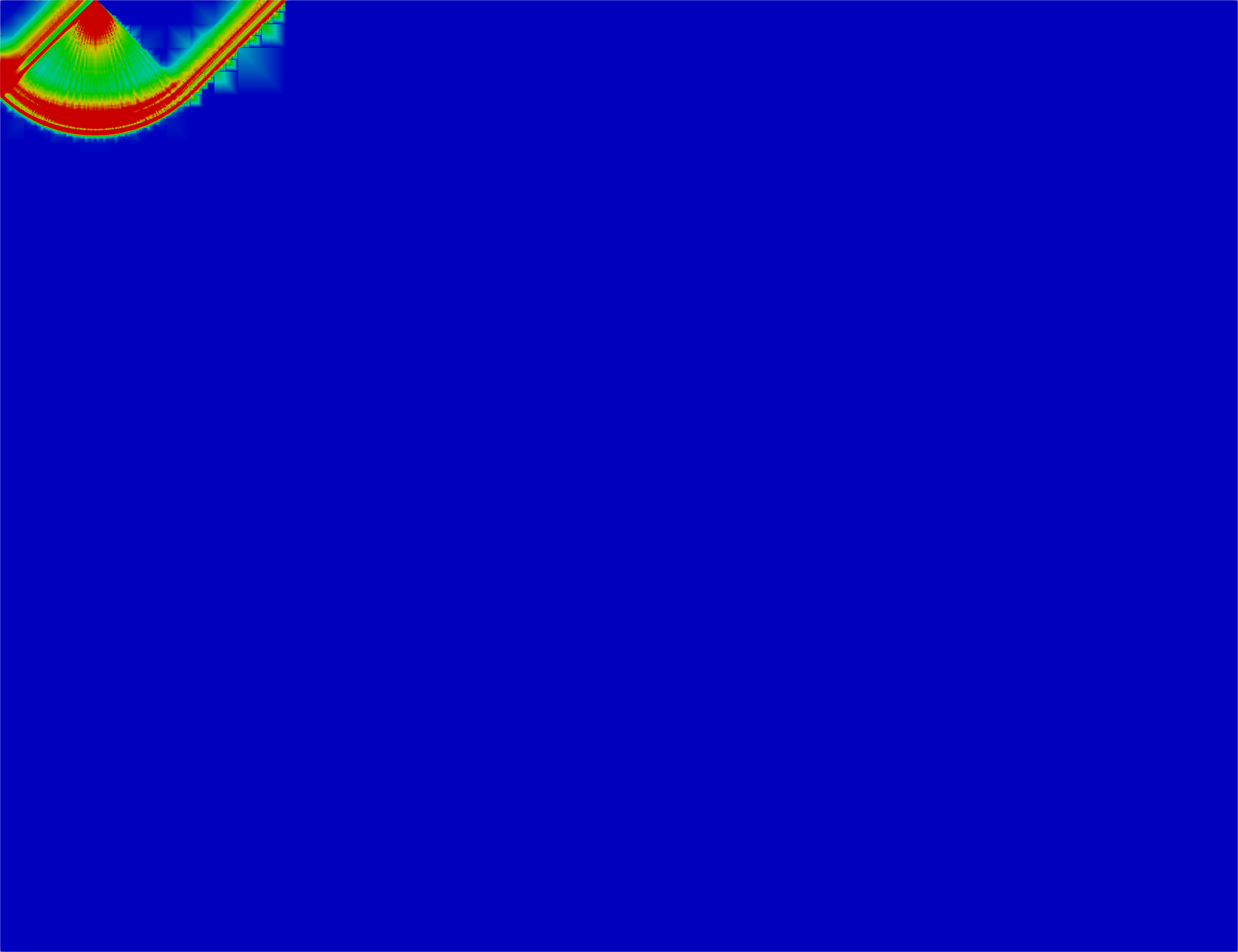}
      \caption{$\varphi = 0^\circ$}
    }
  \end{subfigure}%
  \quad
  \begin{subfigure}[b]{0.3\textwidth}
    {
      \centering
      \includegraphics[width=\textwidth]{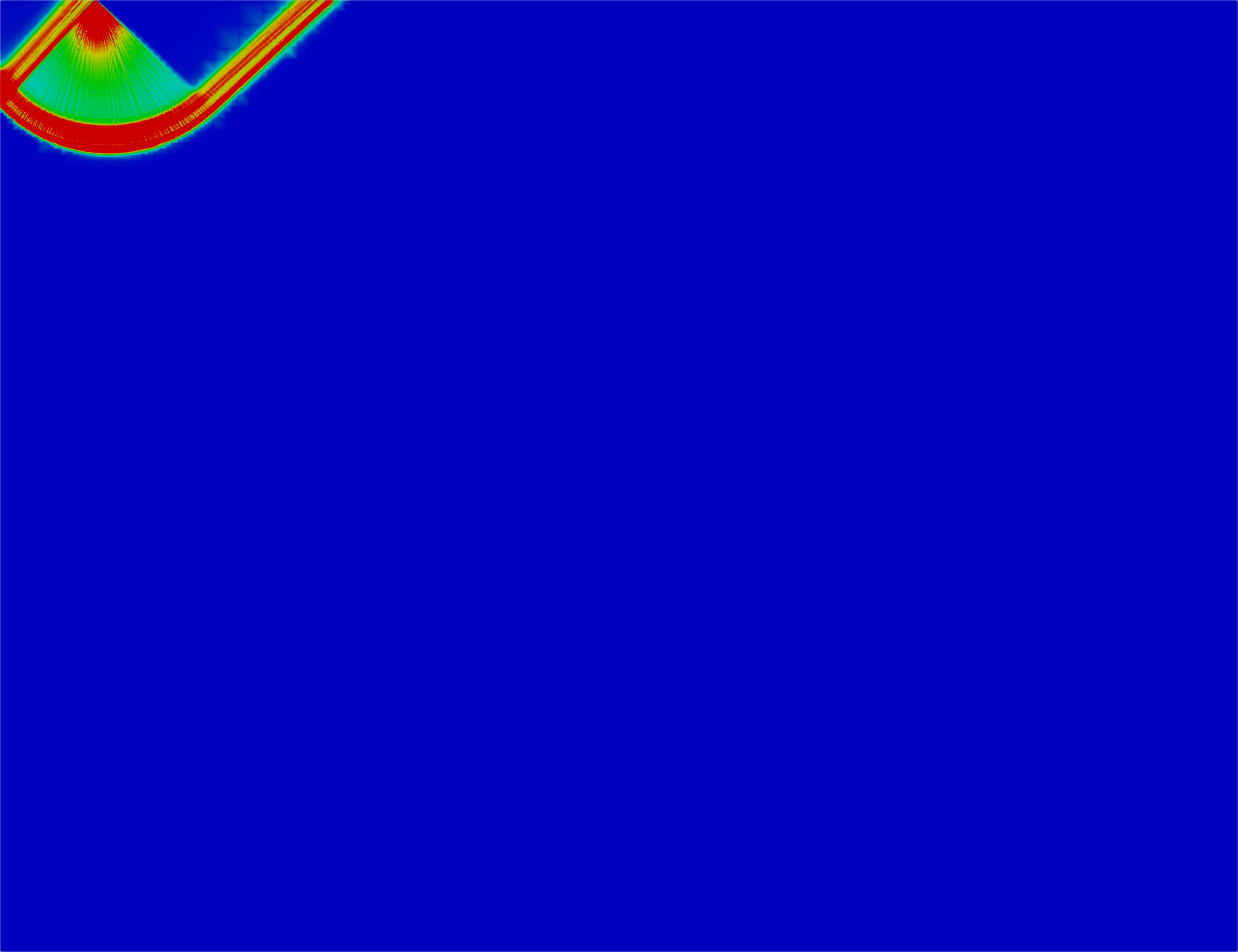}
      \caption{$\varphi = 5^\circ$}
    }
  \end{subfigure}%
  \quad
  \begin{subfigure}[b]{0.3\textwidth}
    {
      \centering
      \includegraphics[width=\textwidth]{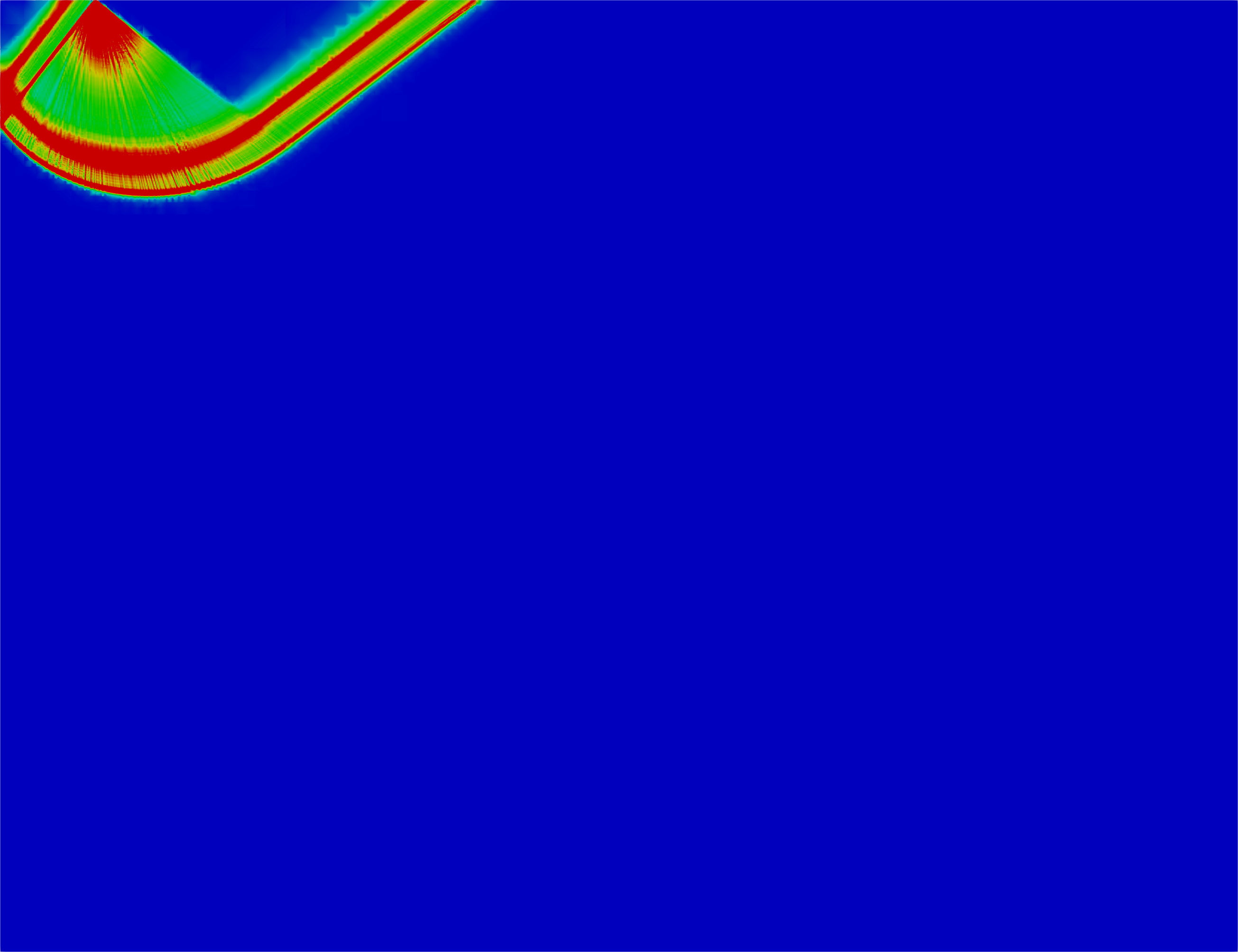}
      \caption{$\varphi = 15^\circ$}
    }
  \end{subfigure}%
  \\ \bigskip
  \begin{subfigure}[b]{0.3\textwidth}
    {
      \centering
      \includegraphics[width=\textwidth]{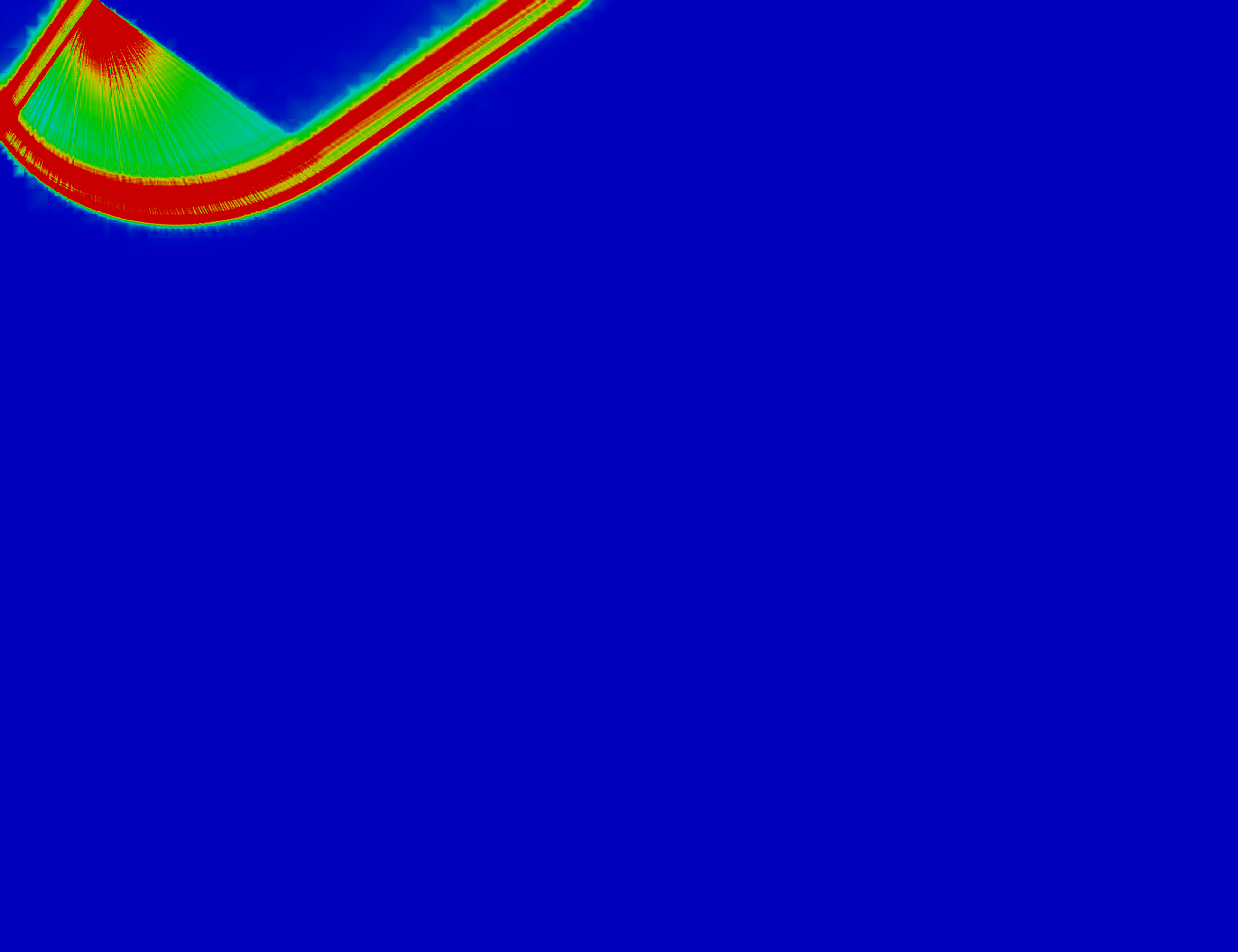}
      \caption{$\varphi = 20^\circ$}
    }
  \end{subfigure}%
  \quad
  \begin{subfigure}[b]{0.3\textwidth}
    {
      \centering
      \includegraphics[width=\textwidth]{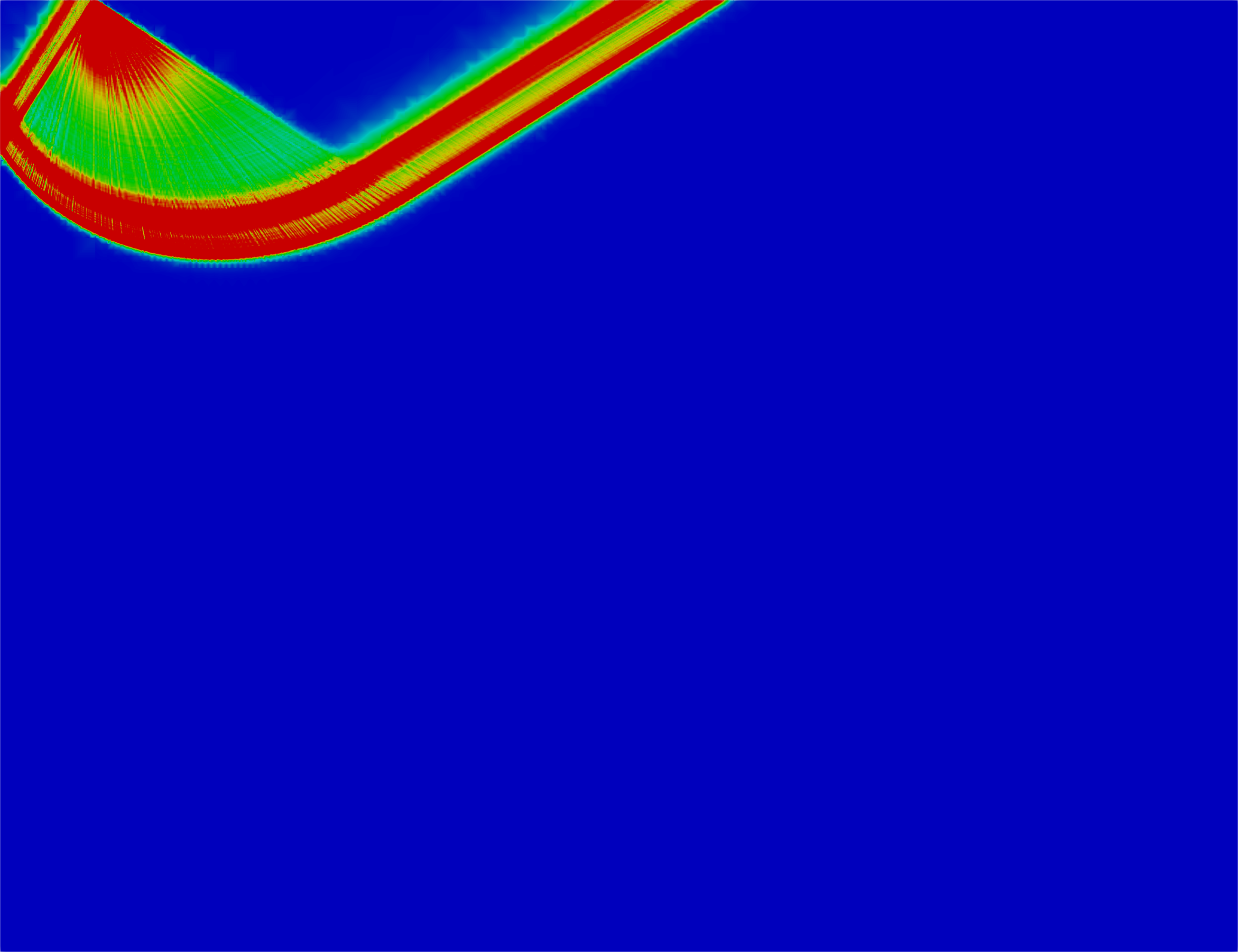}
      \caption{$\varphi = 25^\circ$}
    }
  \end{subfigure}%
  \quad
  \begin{subfigure}[b]{0.3\textwidth}
    {
      \centering
      \includegraphics[width=\textwidth]{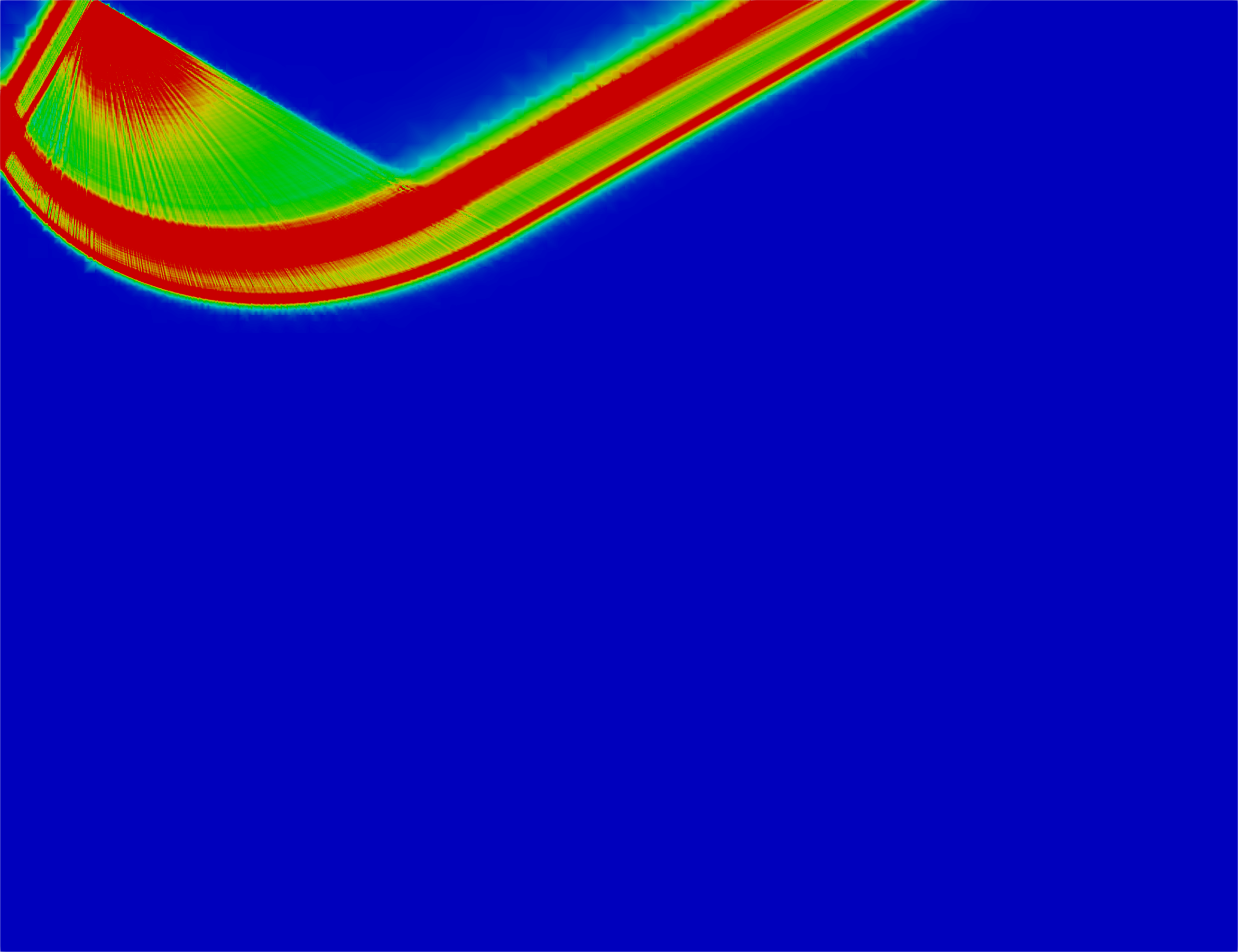}
      \caption{$\varphi = 30^\circ$}
    }
  \end{subfigure}%
  \\ \bigskip
  \begin{subfigure}[b]{0.3\textwidth}
    {
      \centering
      \includegraphics[width=\textwidth]{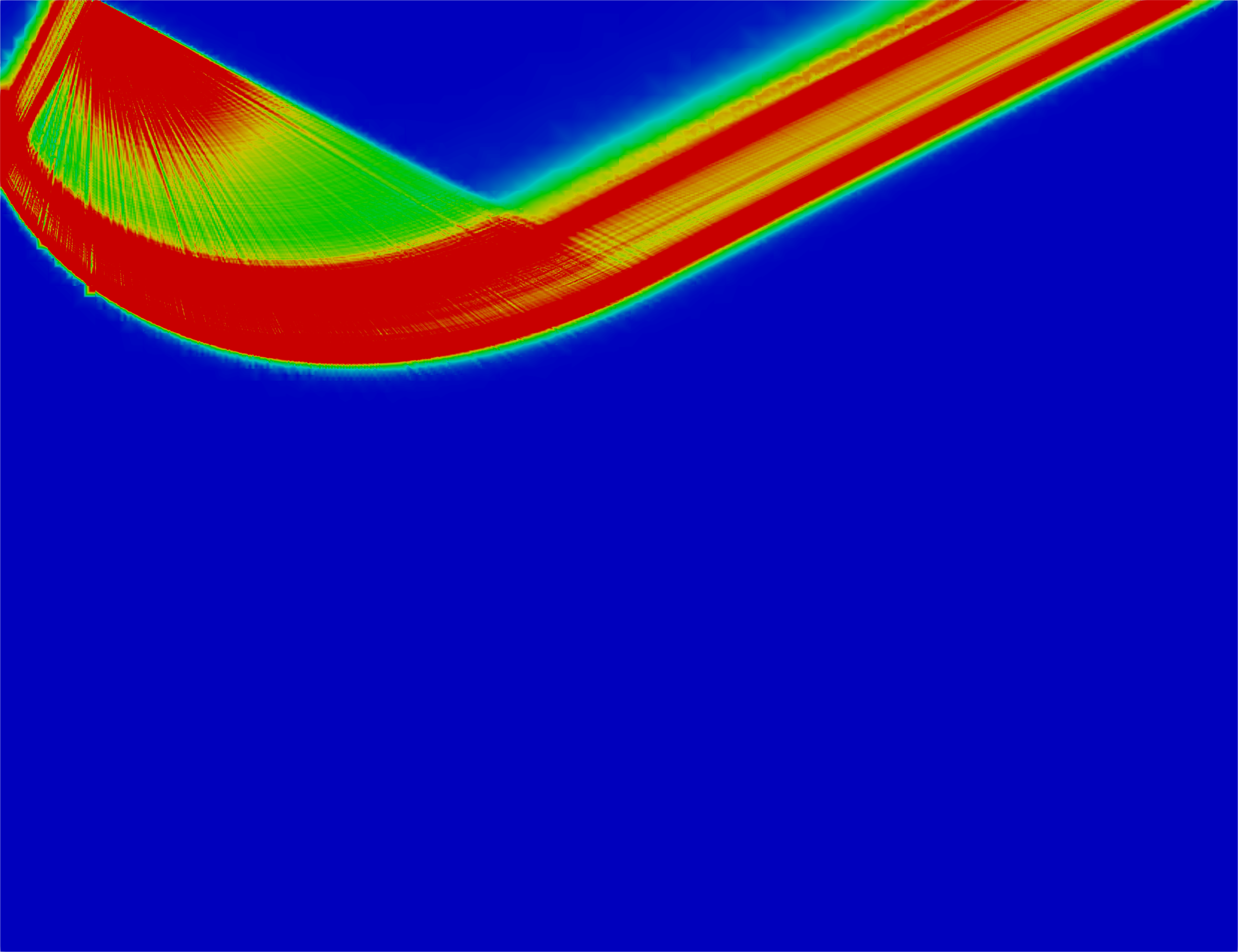}
      \caption{$\varphi = 35^\circ$}
    }
  \end{subfigure}%
  \quad
  \begin{subfigure}[b]{0.3\textwidth}
    {
      \centering
      \includegraphics[width=\textwidth]{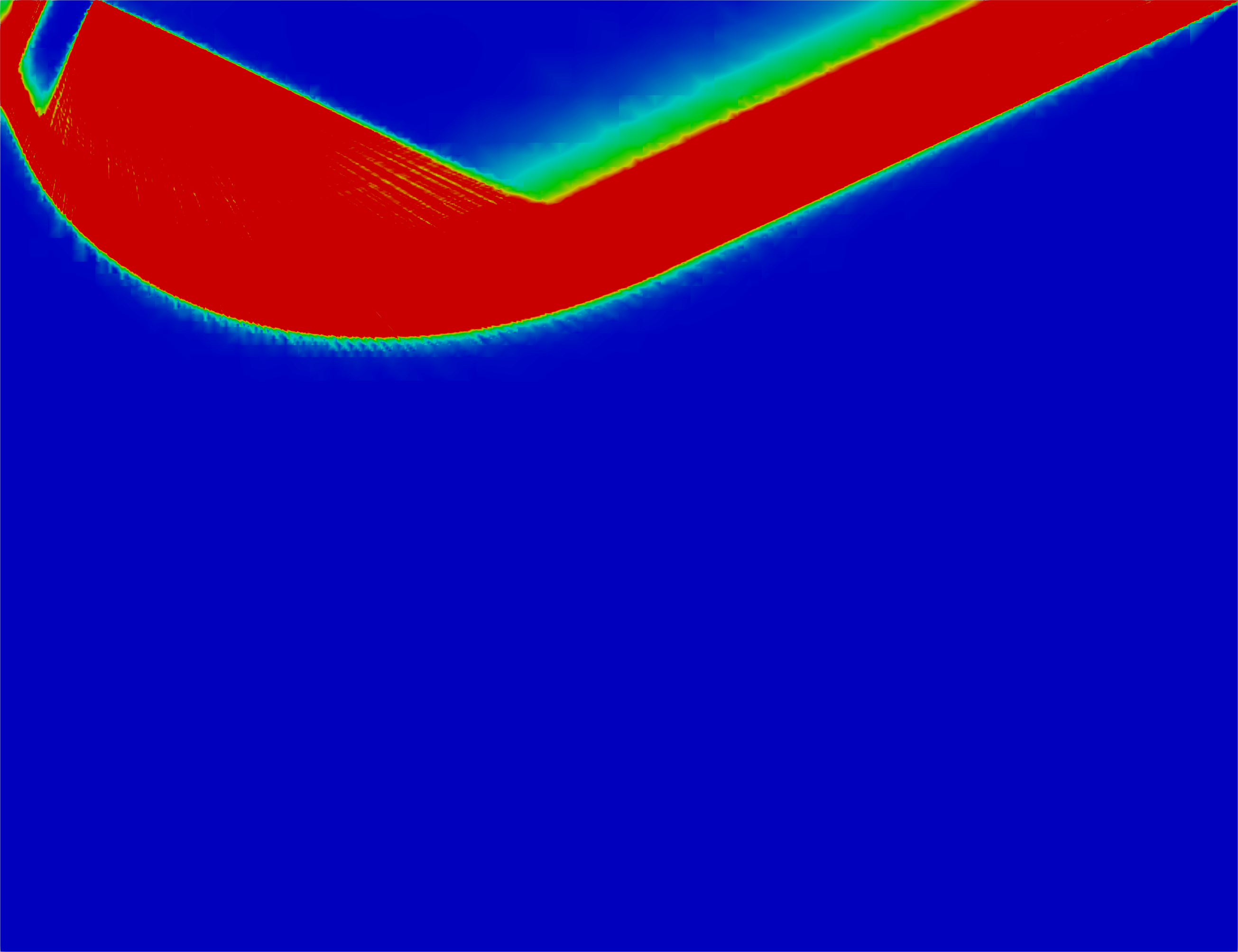}
      \caption{$\varphi = 40^\circ$}
    }
  \end{subfigure}%
  \quad
  \begin{subfigure}[b]{0.3\textwidth}
    {
      \centering
      \includegraphics[width=\textwidth]{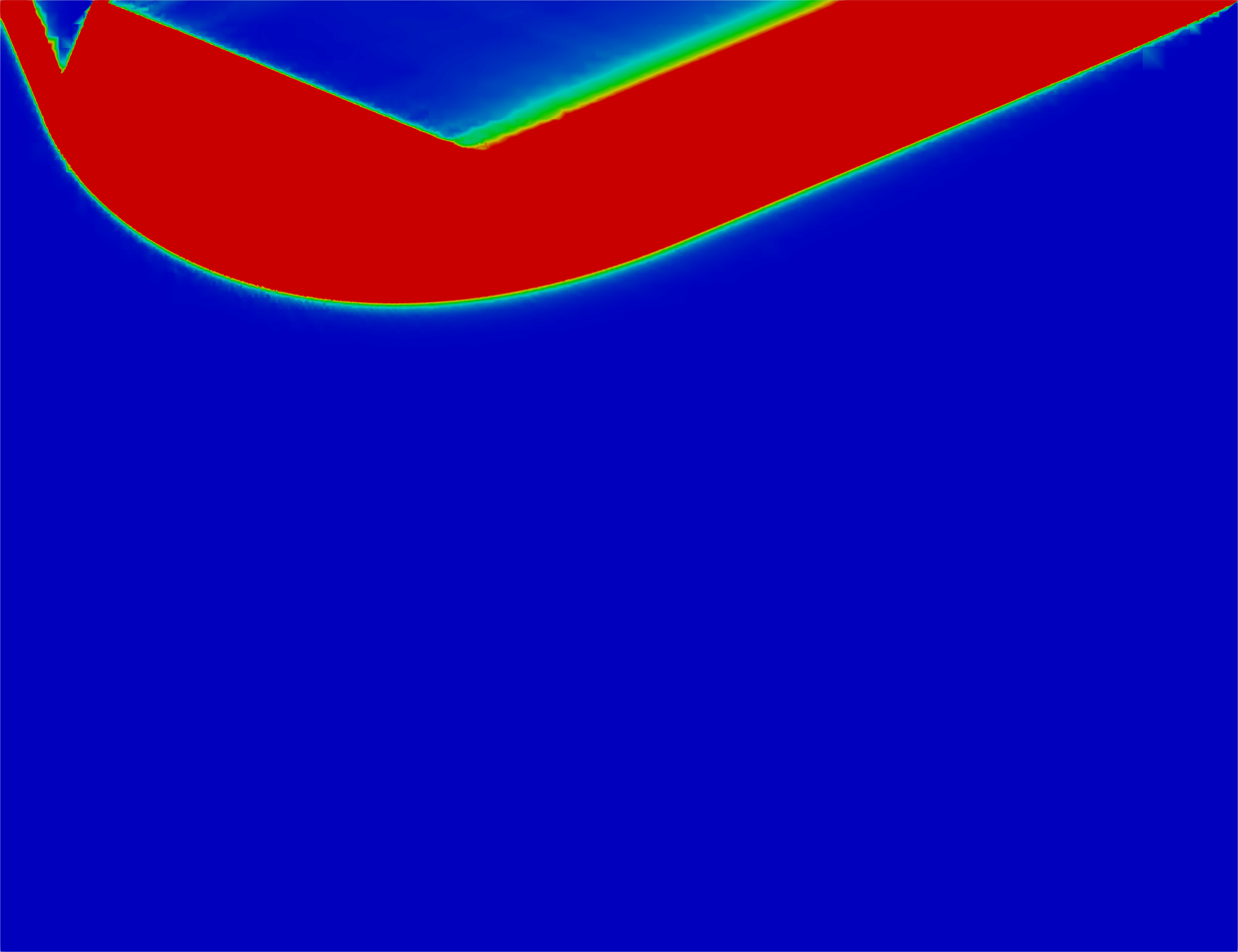}
      \caption{$\varphi = 45^\circ$}
    }
  \end{subfigure}%
  \caption{A smooth strip footing: Plastic dissipation for a variety of internal frictional angles.}
  \label{Fig: strip_dissipation_various_angles}
\end{figure}
\begin{figure}[H]
  \centering
  \includegraphics[viewport=10  380 650 570,clip=true,scale=0.8]{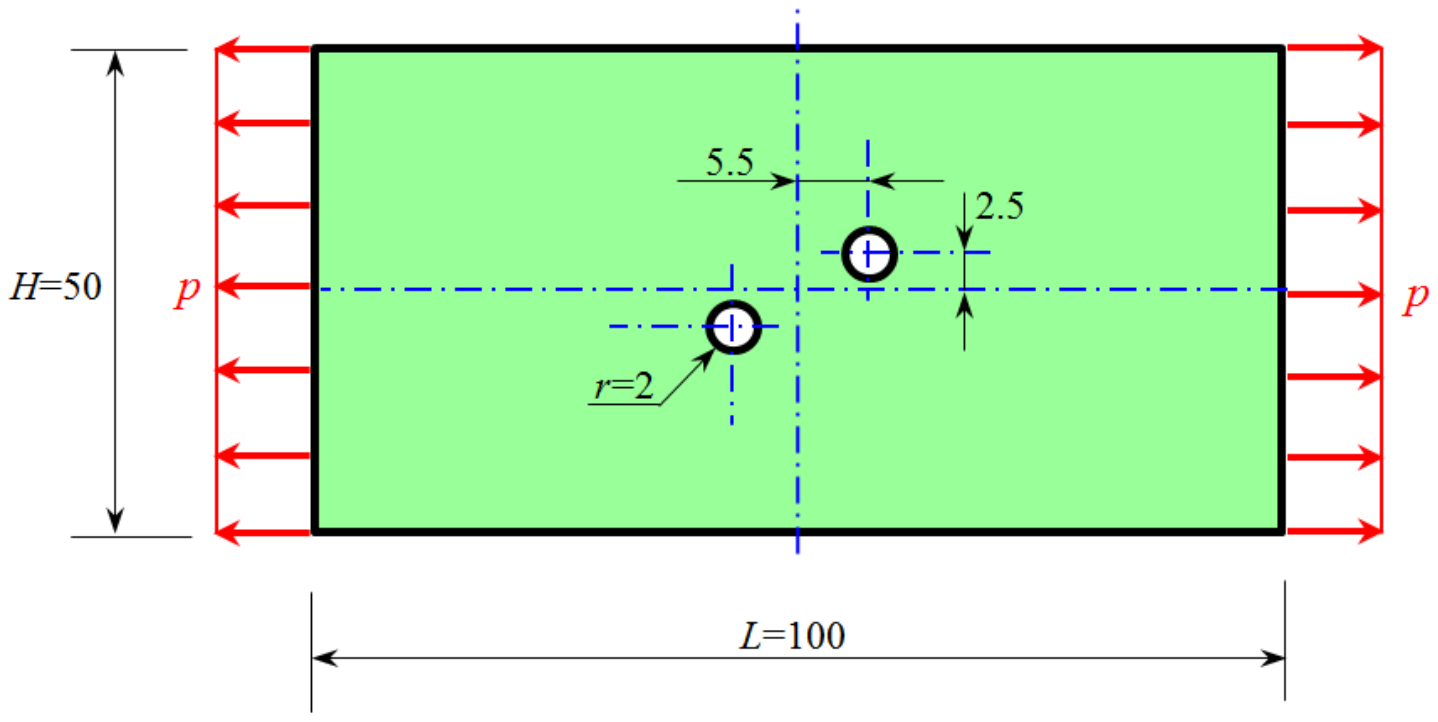}
  \caption{A full model of block with two symmetric holes.}
  \label{Fig: Conv_LAL_2holes_Fullmodel}
\end{figure}
\begin{figure}[H]
  \centering
  \begin{subfigure}[b]{0.45\textwidth}
    {
      \centering
      \includegraphics{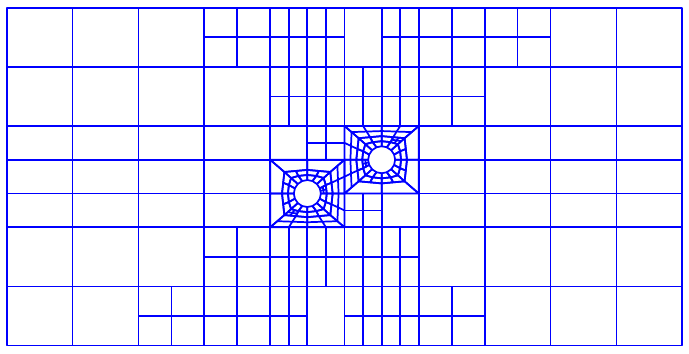}
      \caption{}
    }
  \end{subfigure}%
  \quad
  \begin{subfigure}[b]{0.45\textwidth}
    {
      \centering
      \includegraphics{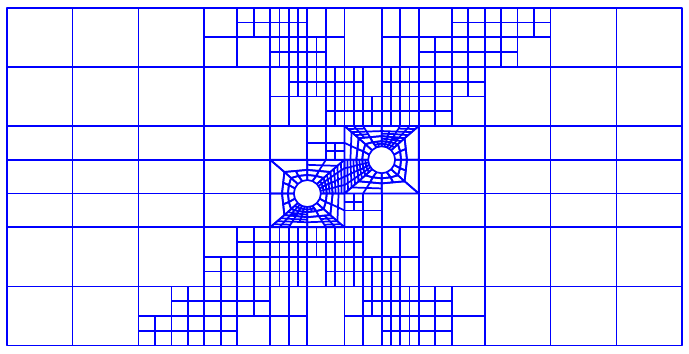}
      \caption{}
    }
  \end{subfigure}%
  \\ \bigskip
  \begin{subfigure}[b]{0.45\textwidth}
    {
      \centering
      \includegraphics{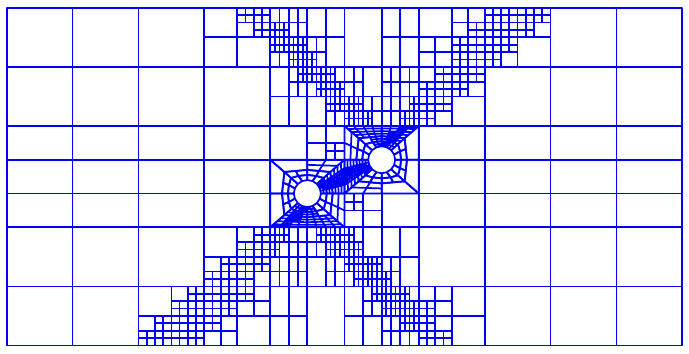}
      \caption{}
    }
  \end{subfigure}%
  \quad
  \begin{subfigure}[b]{0.45\textwidth}
    {
      \centering
      \includegraphics{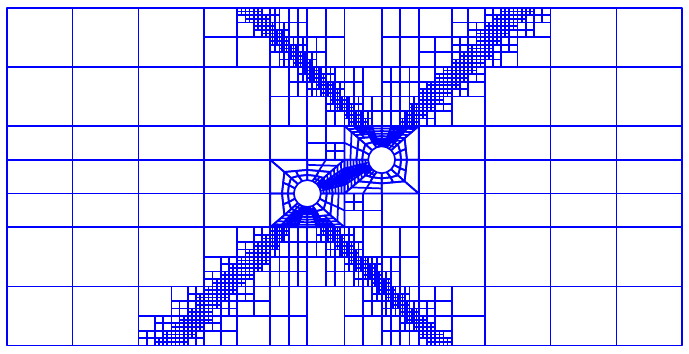}
      \caption{}
    }
  \end{subfigure}%
  \caption{Several adaptive mesh steps for the block with two symmetric holes$ (\varphi = 0^{\circ} $).}
  \label{Fig: Conv_LAL_2holes_AdaptiveMeshStep_phi0}
\end{figure}
\begin{figure}[H]
  \centering
  \begin{subfigure}[b]{0.45\textwidth}
    {
      \centering
      \includegraphics{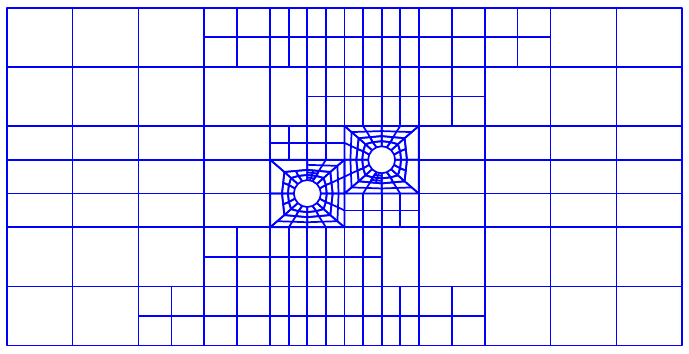}
      \caption{}
    }
  \end{subfigure}%
  \quad
  \begin{subfigure}[b]{0.45\textwidth}
    {
      \centering
      \includegraphics{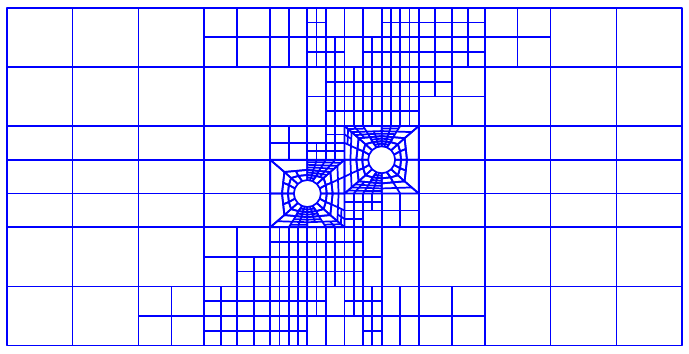}
      \caption{}
    }
  \end{subfigure}%
  \\ \bigskip
  \begin{subfigure}[b]{0.45\textwidth}
    {
      \centering
      \includegraphics{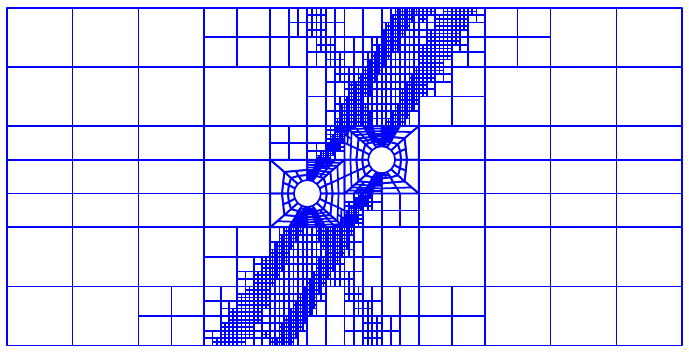}
      \caption{}
    }
  \end{subfigure}%
  \quad
  \begin{subfigure}[b]{0.45\textwidth}
    {
      \centering
      \includegraphics{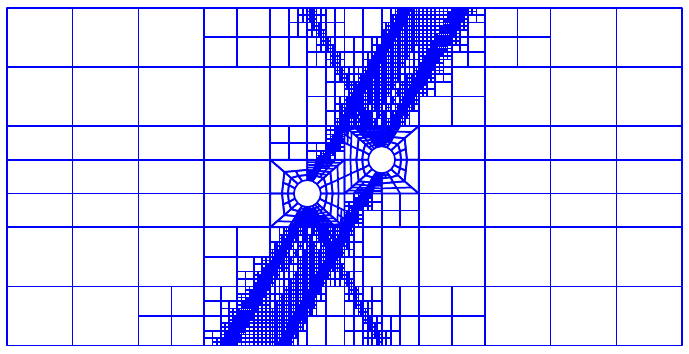}
      \caption{}
    }
  \end{subfigure}%
  \caption{Several adaptive mesh steps for the block with two symmetric holes$ (\varphi = 30^{\circ} $).}
  \label{Fig: Conv_LAL_2holes_AdaptiveMeshStep_phi30}
\end{figure}
\begin{figure}[H]
  \centering
  \includegraphics{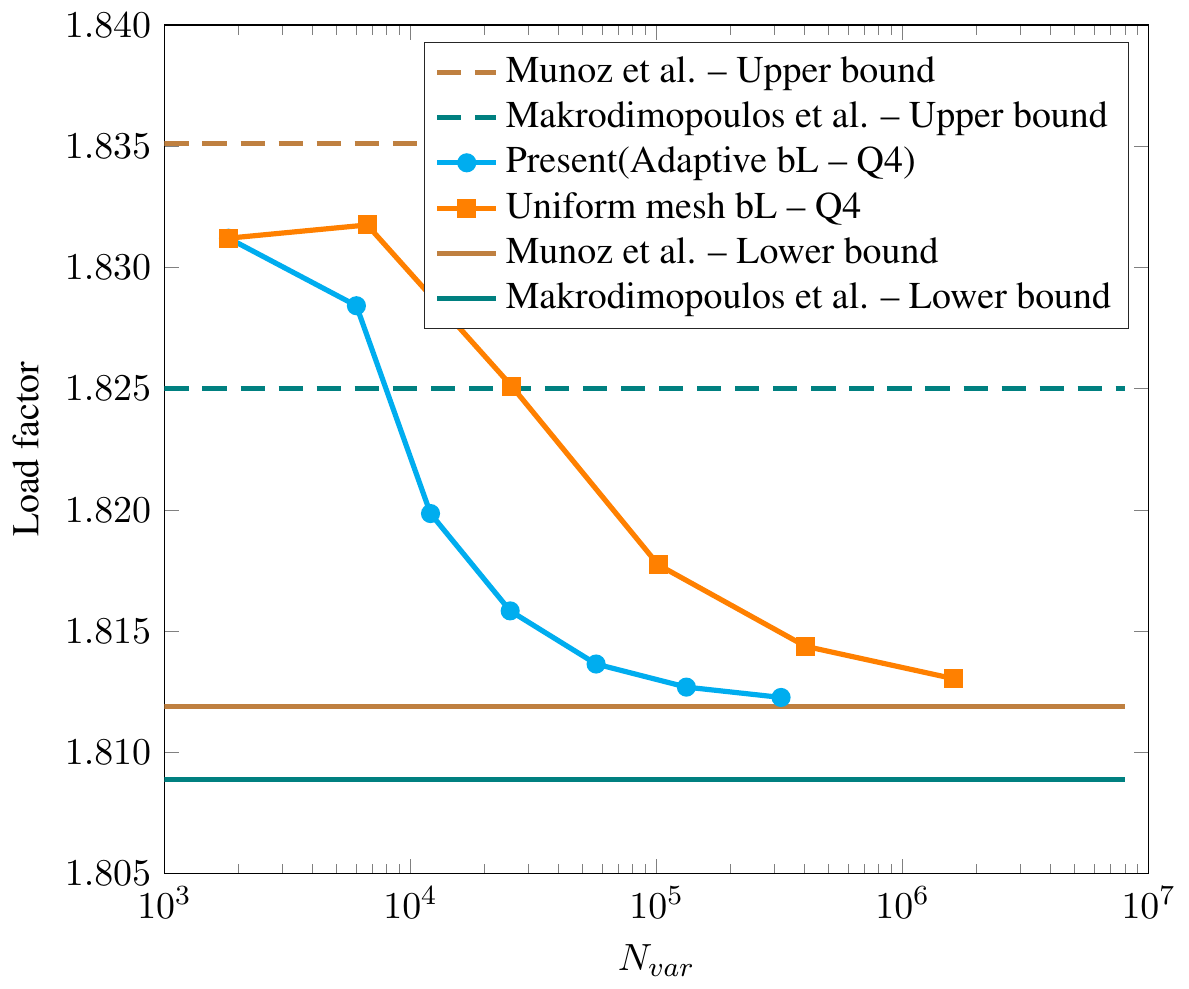}
  \caption{The block with two symmetric holes of cohesive frictional soil$ (\varphi = 0^{\circ} $): The convergence of limit load factor with respect to optimization variables.}
  \label{Fig: Conv_LAL_2holes_phi0}
\end{figure}
\begin{figure}[H]
  \centering
  \includegraphics{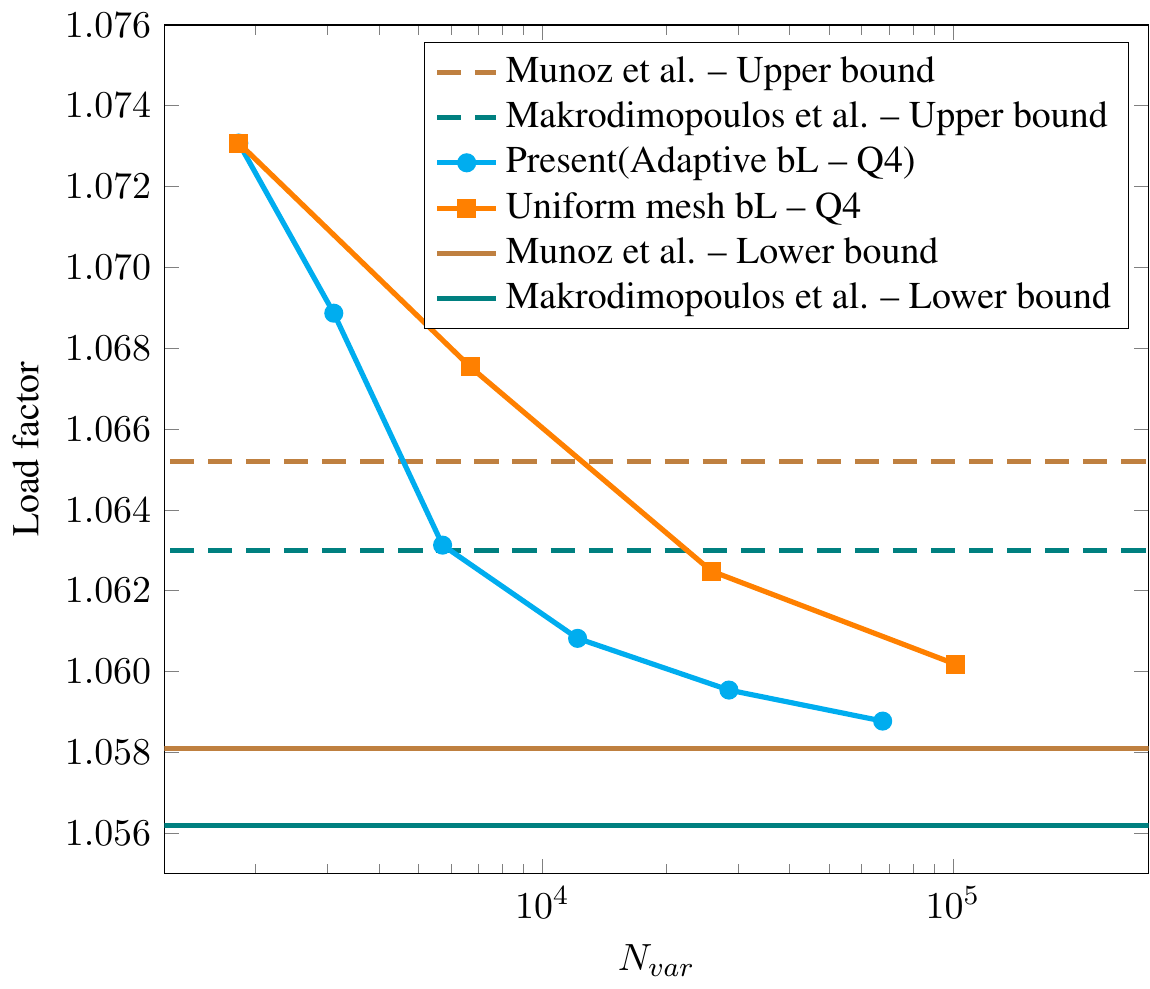}
  \caption{The block with two symmetric holes of cohesive frictional soil$ (\varphi = 30^{\circ} $): The convergence of limit load factor with respect to optimization variables.}
  \label{Fig: Conv_LAL_2holes_phi30}
\end{figure}
\begin{figure}[H]
  \centering
  \begin{subfigure}[b]{0.45\textwidth}
    {
      \centering
      \includegraphics[width=\textwidth]{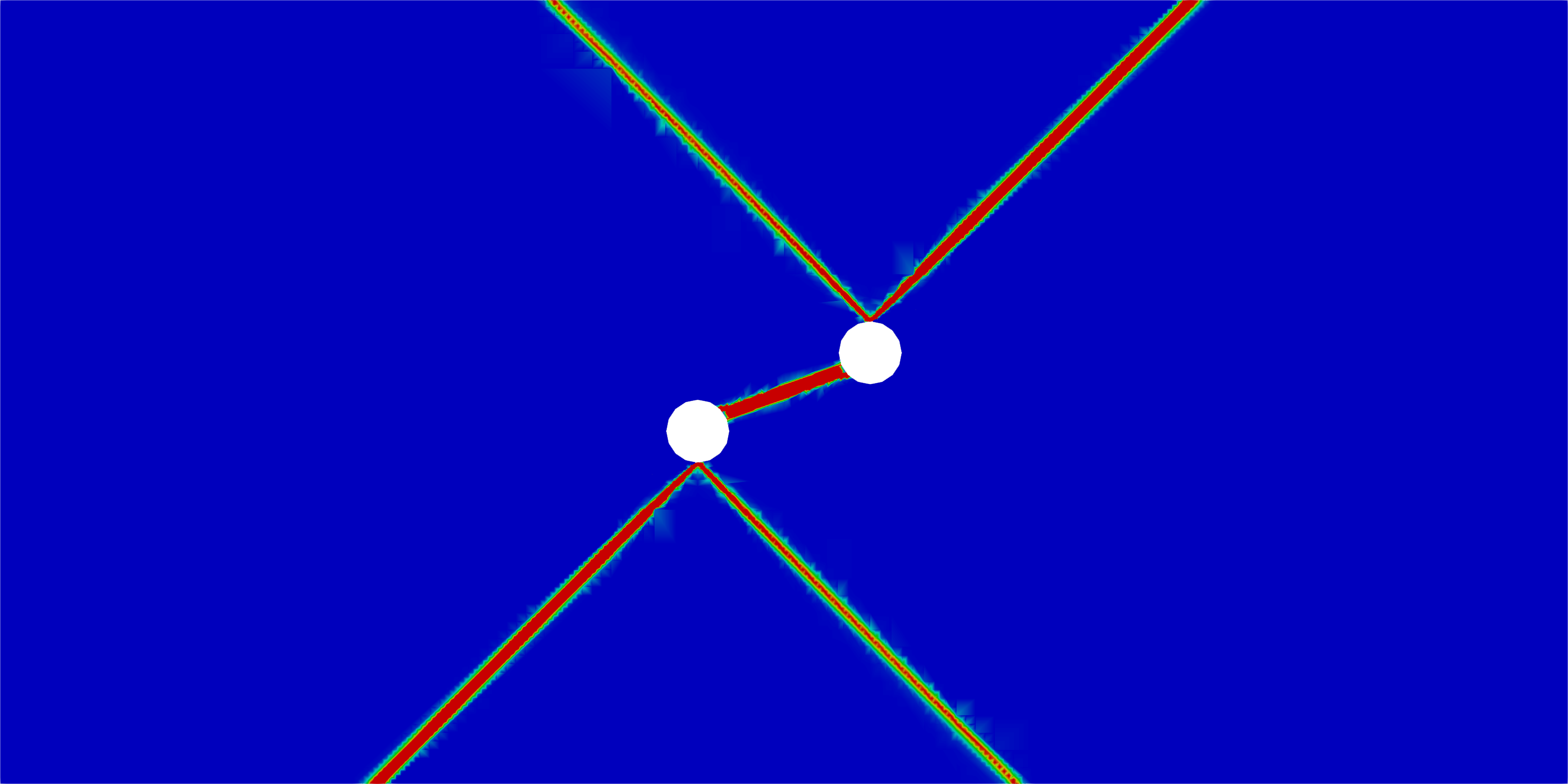}
      \caption{$\varphi = 0^\circ$}
    }
  \end{subfigure}%
  \qquad
  \begin{subfigure}[b]{0.45\textwidth}
    {
      \centering
      \includegraphics[width=\textwidth]{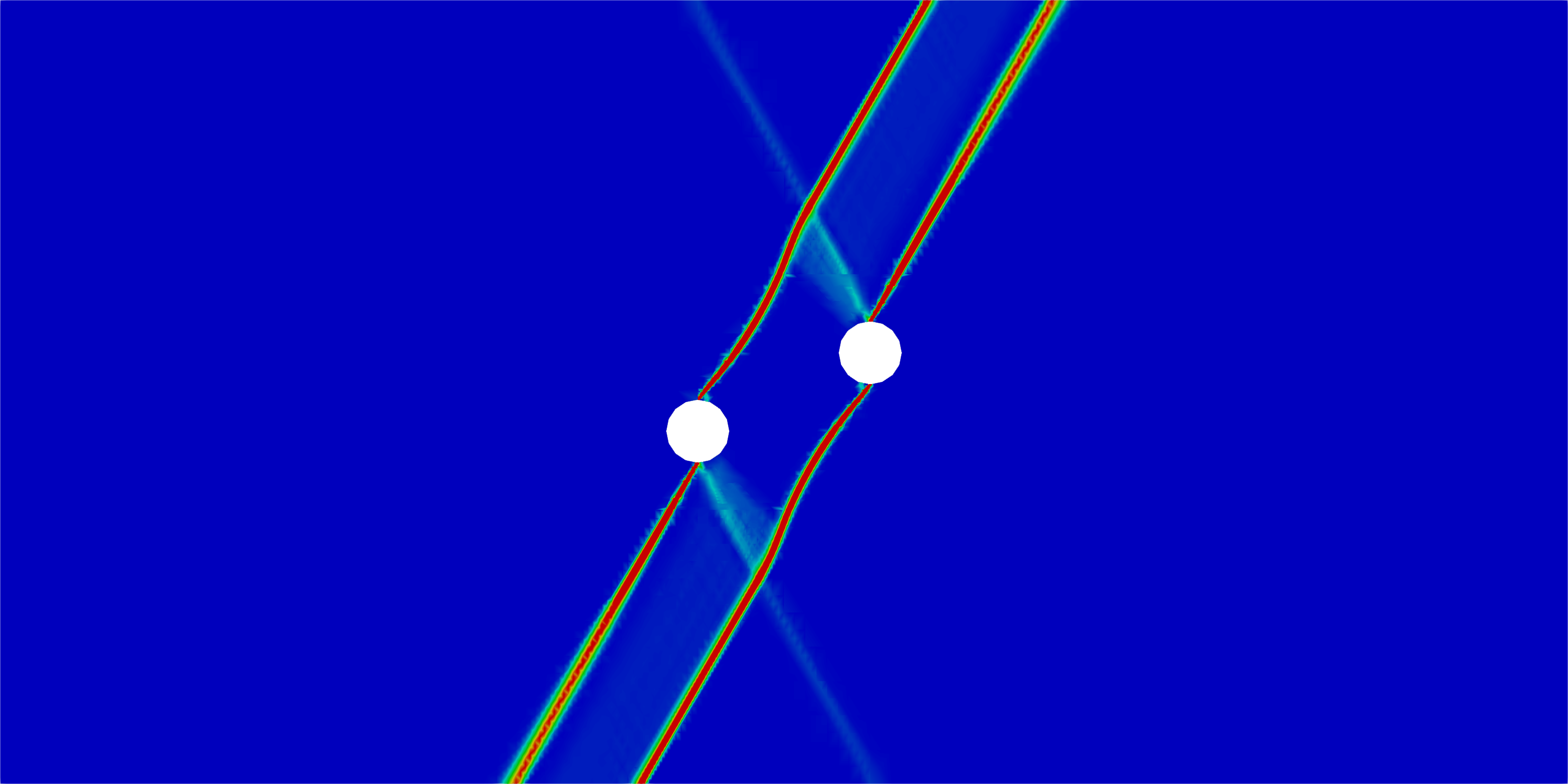}
      \caption{$\varphi = 30^\circ$}
    }
  \end{subfigure}%
  \caption{Plastic dissipation.}
  \label{Fig: 2holes_dissipation}
\end{figure}


\begin{figure}[H]
  \centering
  \includegraphics[width=0.5\textwidth]{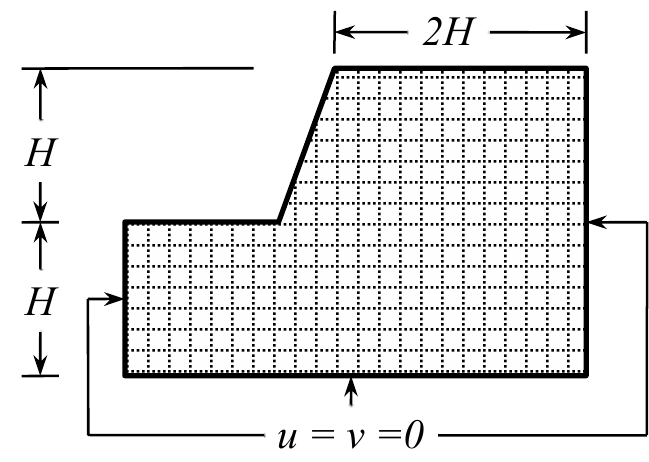}
  \caption{$70^{\circ}$ Slope stability geometry and boundary condition.}
  \label{Fig: Conv_LAL_Slope_Fullmodel}
\end{figure}
\begin{figure}[H]
  \centering
  \begin{subfigure}[b]{0.45\textwidth}
    {
      \centering
      \includegraphics{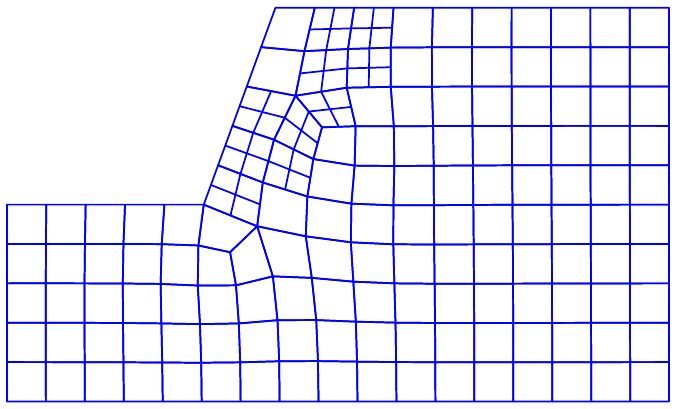}
      \caption{}
    }
  \end{subfigure}%
  \quad
  \begin{subfigure}[b]{0.45\textwidth}
    {
      \centering
      \includegraphics{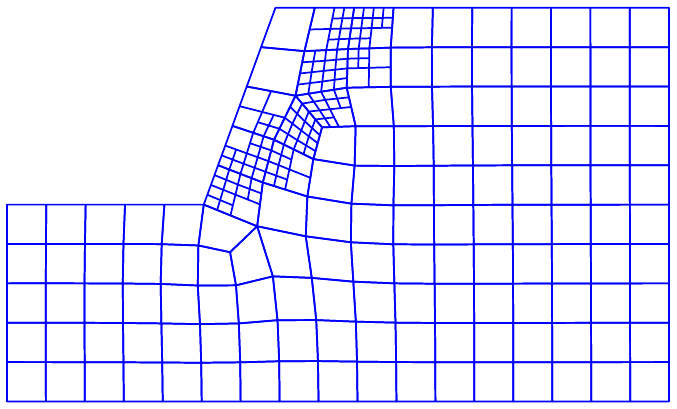}
      \caption{}
    }
  \end{subfigure}%
  \\ \bigskip
  \begin{subfigure}[b]{0.45\textwidth}
    {
      \centering
      \includegraphics{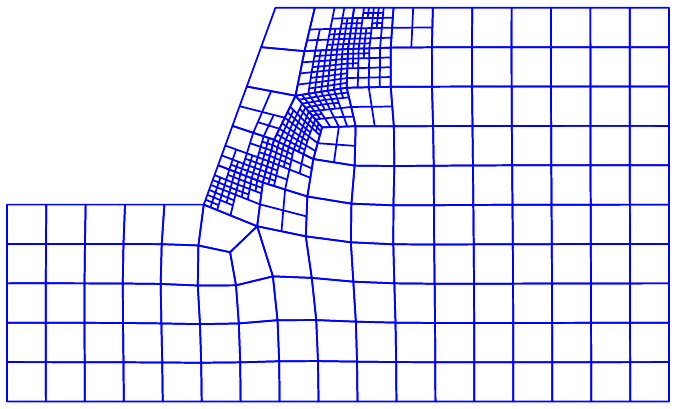}
      \caption{}
    }
  \end{subfigure}%
  \quad
  \begin{subfigure}[b]{0.45\textwidth}
    {
      \centering
      \includegraphics{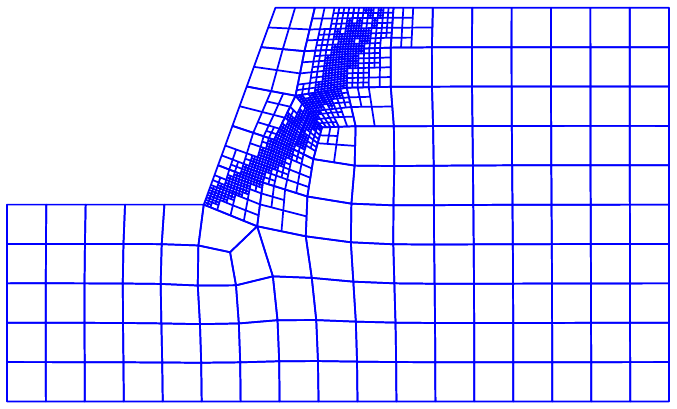}
      \caption{}
    }
  \end{subfigure}%
  \caption{Several adaptive mesh steps for the slope stability$ (\varphi = 20^{\circ} $).}
  \label{Fig: Conv_LAL_Slope_AdaptiveMeshStep_phi20}
\end{figure}

\begin{figure}[H]
  \centering
  \begin{subfigure}[b]{0.45\textwidth}
    {
      \centering
      \includegraphics{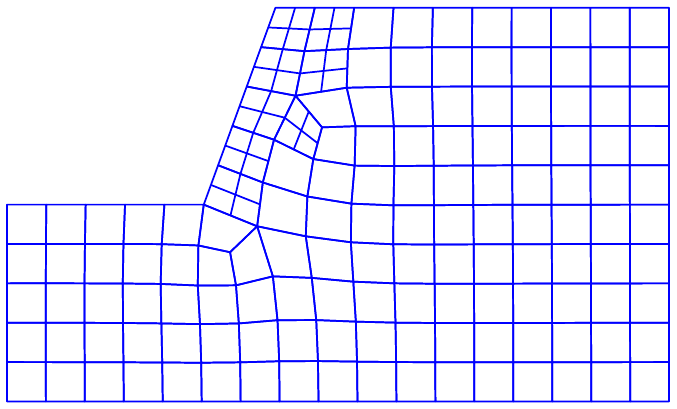}
      \caption{}
    }
  \end{subfigure}%
  \quad
  \begin{subfigure}[b]{0.45\textwidth}
    {
      \centering
      \includegraphics{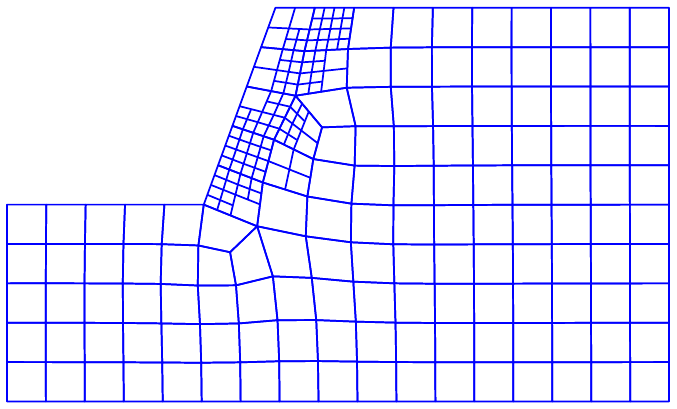}
      \caption{}
    }
  \end{subfigure}%
  \\ \bigskip
  \begin{subfigure}[b]{0.45\textwidth}
    {
      \centering
      \includegraphics{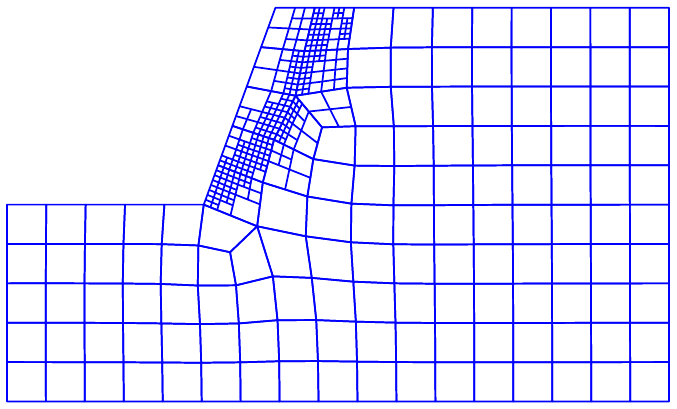}
      \caption{}
    }
  \end{subfigure}%
  \quad
  \begin{subfigure}[b]{0.45\textwidth}
    {
      \centering
      \includegraphics{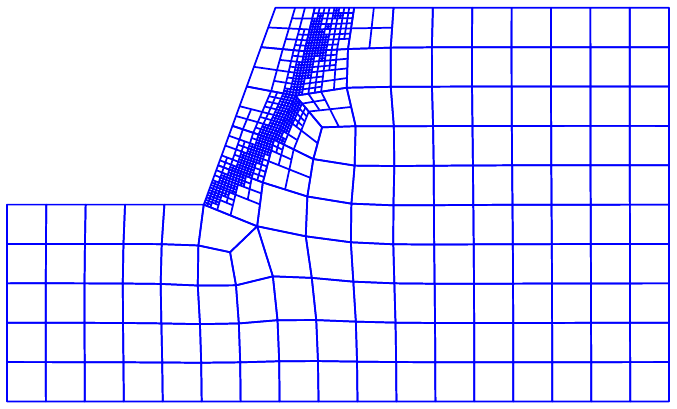}
      \caption{}
    }
  \end{subfigure}%
  \caption{Several adaptive mesh steps for the slope stability$ (\varphi = 35^{\circ} $).}
  \label{Fig: Conv_LAL_Slope_AdaptiveMeshStep_phi35}
\end{figure}

\begin{figure}[H]
  \centering
  \includegraphics{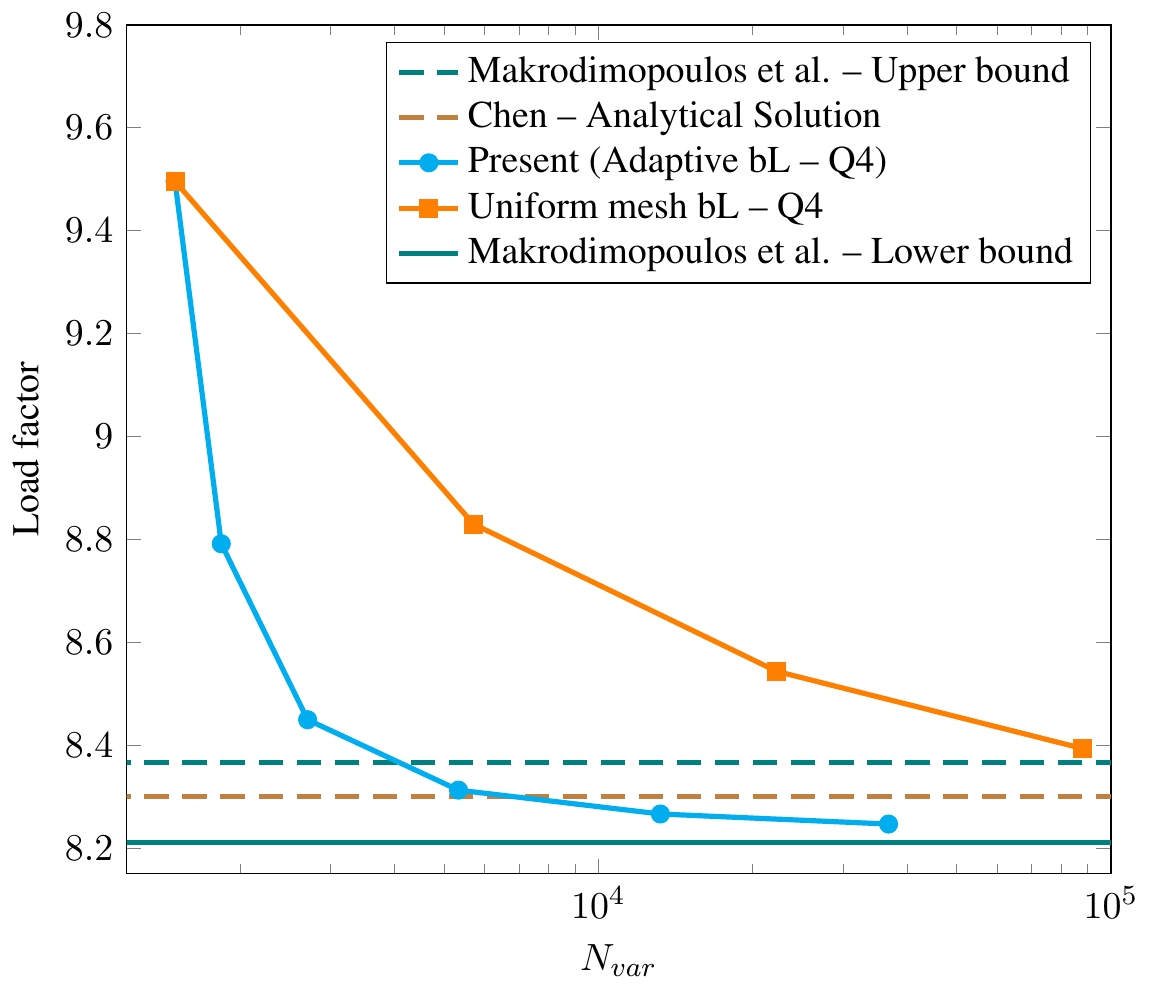}
  \caption{The Slope stability of cohesive frictional soil$ (\varphi = 20^{\circ} $): The convergence of limit load factor with respect to optimization variables.}
  \label{Fig: Conv_LAL_Slope_phi20}
\end{figure}
\begin{figure}[H]
  \centering
  \includegraphics{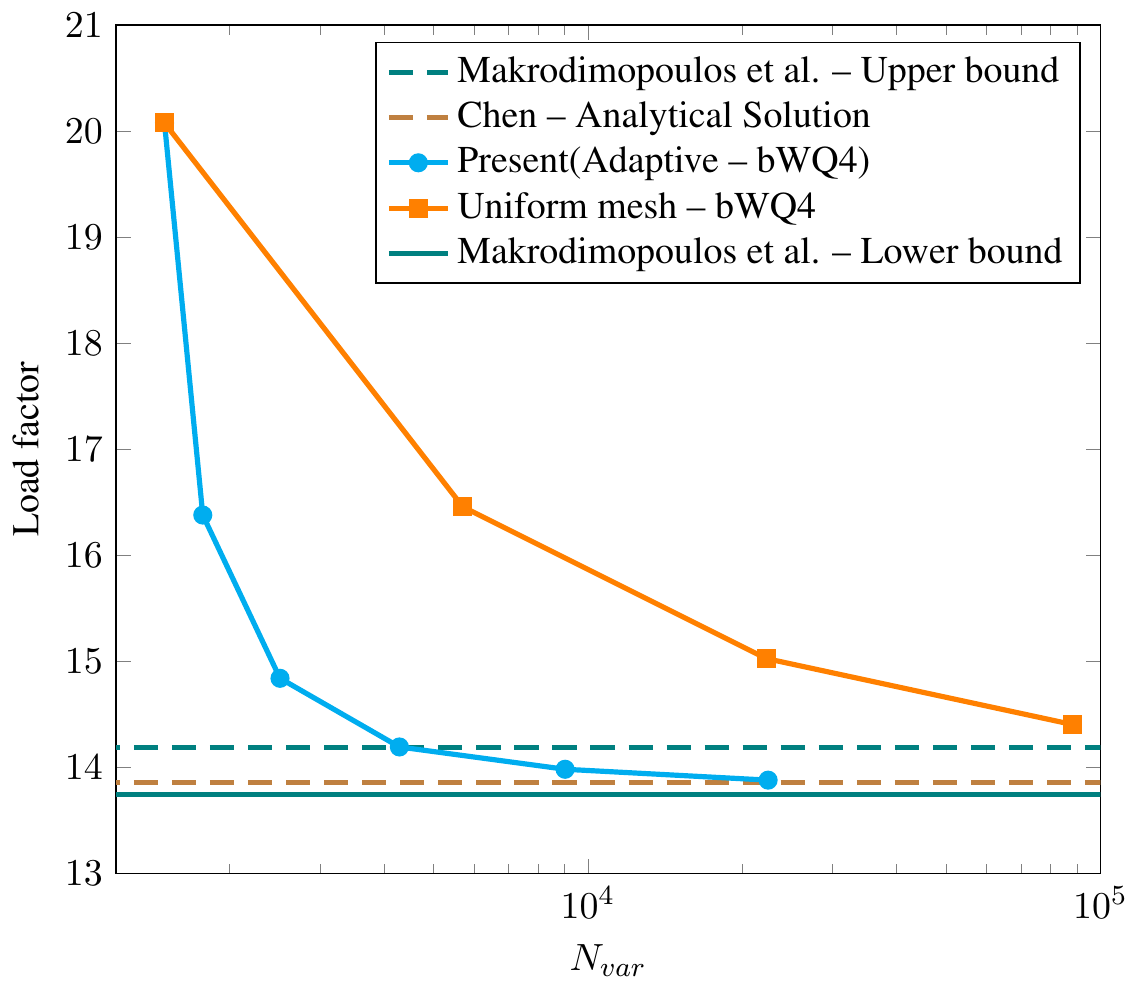}
  \caption{The Slope stability of cohesive frictional soil$ (\varphi = 35^{\circ} $): The convergence of limit load factor with respect to optimization variables.}
  \label{Fig: Conv_LAL_Slope_phi35}
\end{figure}
\begin{figure}[H]
  \centering
  \begin{subfigure}[b]{0.45\textwidth}
    {
      \centering
      \includegraphics[width=\textwidth]{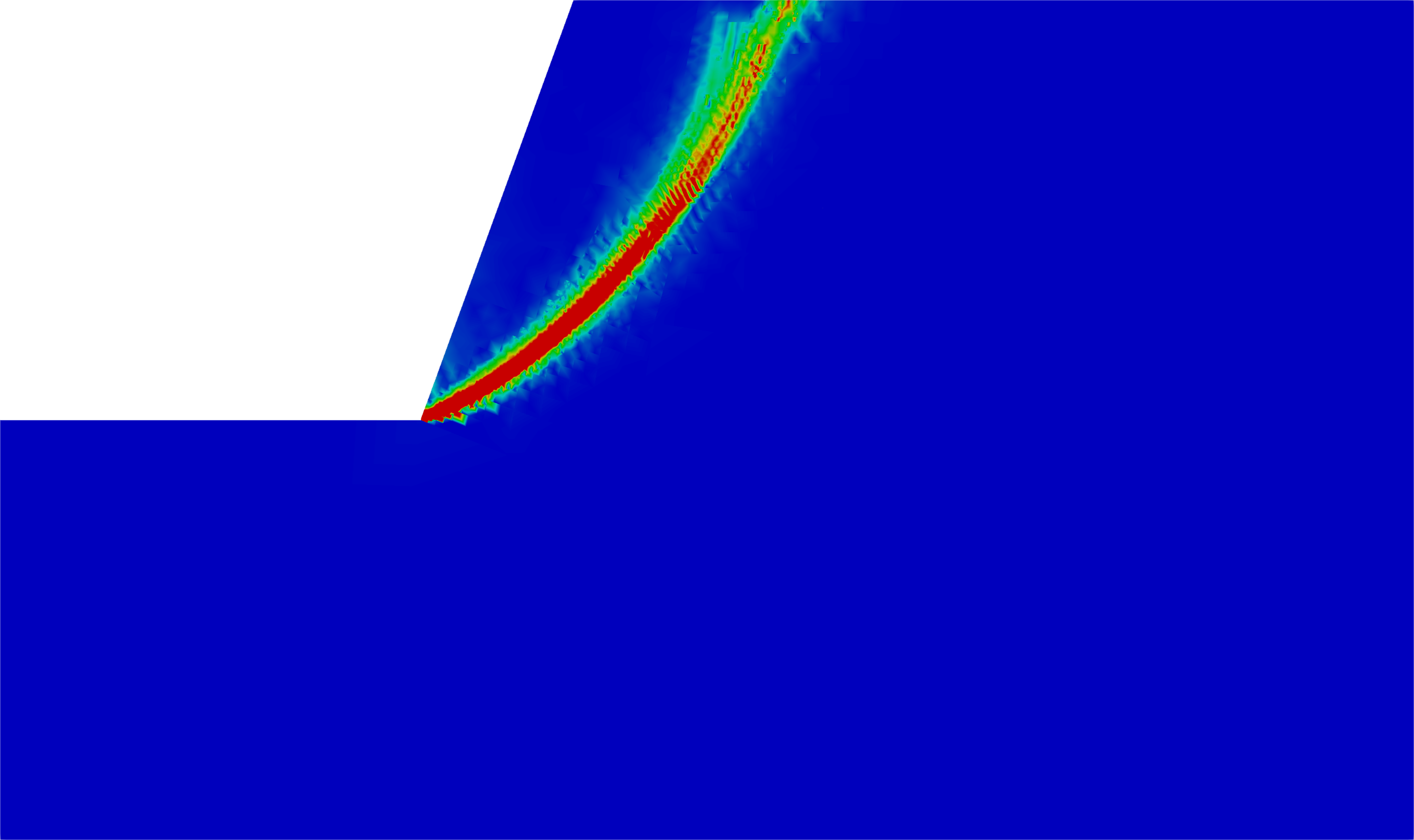}
      \caption{$\varphi = 0^\circ$}
    }
  \end{subfigure}%
  \qquad
  \begin{subfigure}[b]{0.45\textwidth}
    {
      \centering
      \includegraphics[width=\textwidth]{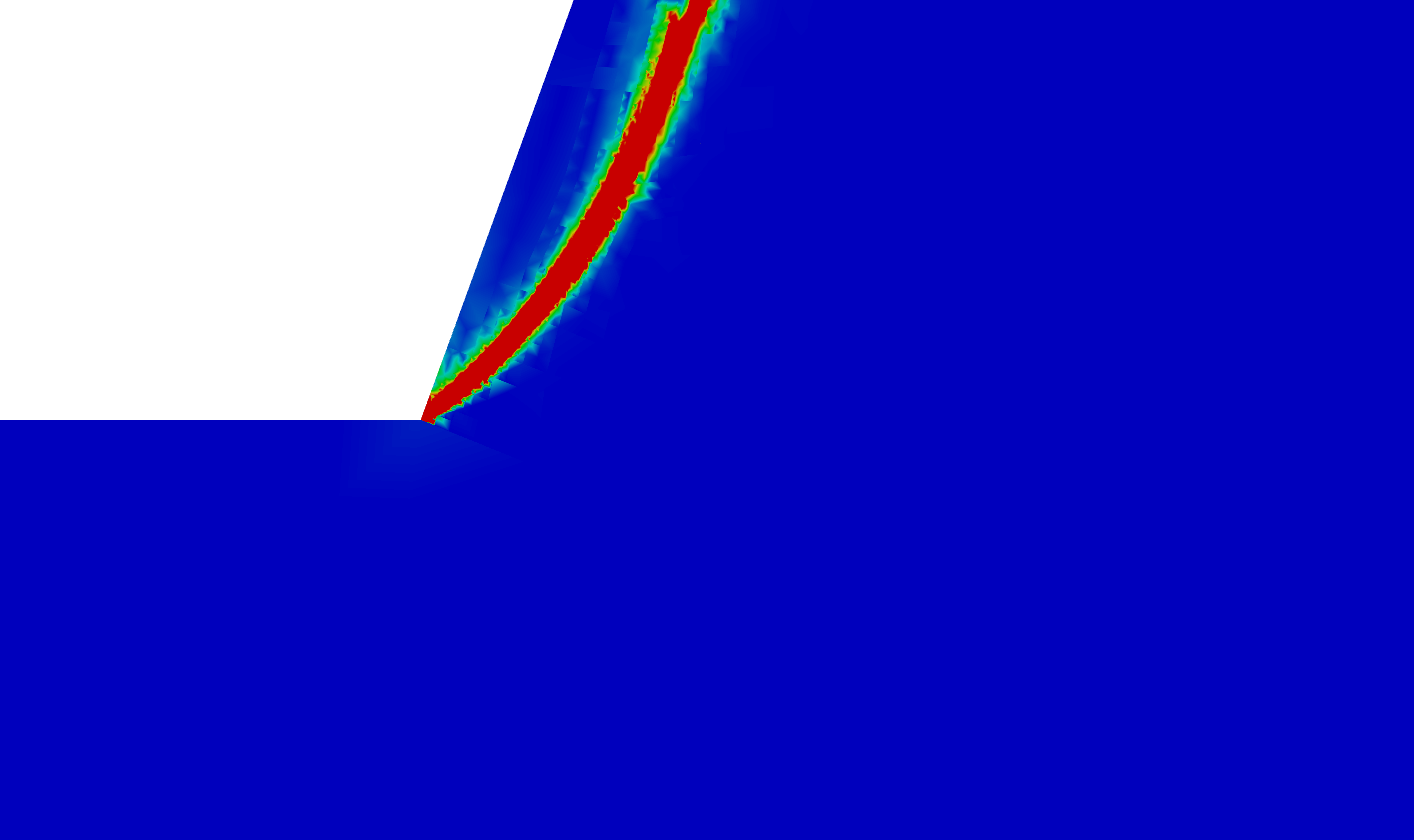}
      \caption{$\varphi = 35^\circ$}
    }
  \end{subfigure}%
  \caption{Plastic dissipation.}
  \label{Fig: Slope_dissipation}
\end{figure}

\begin{figure}[H]
	\centering
	\includegraphics[viewport=90  320 650 530,clip=true,width=1.5\textwidth]{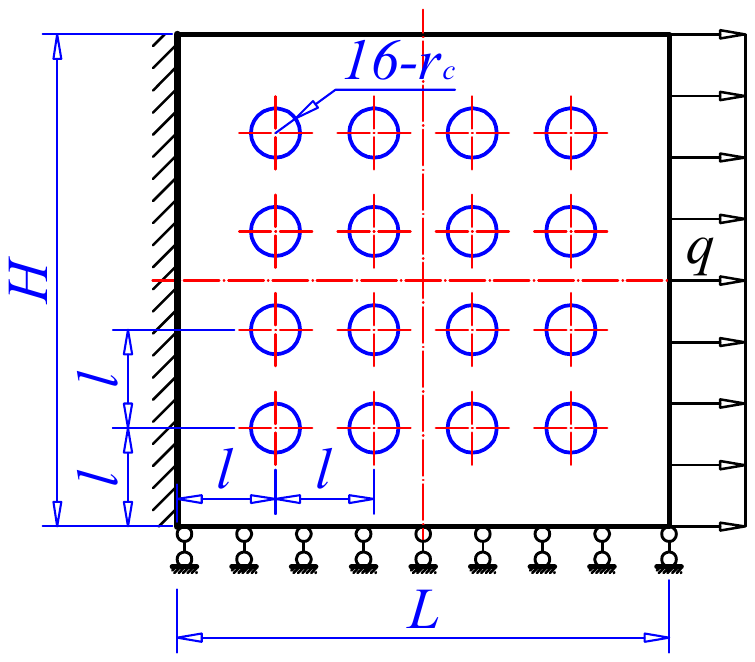}
	\caption{\textcolor{blue}{A model of porous material.}}
	\label{Fig: Fullmodel_final}
\end{figure}

\begin{figure}[H]
	\centering
	\begin{subfigure}[b]{0.4\textwidth}
		{
			\centering
			\includegraphics{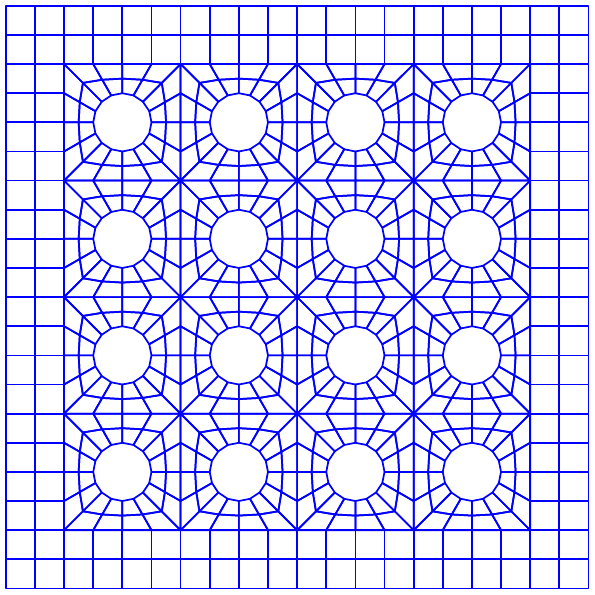}
			\caption{}
		}
	\end{subfigure}%
	\quad
	\begin{subfigure}[b]{0.4\textwidth}
		{
			\centering
			\includegraphics{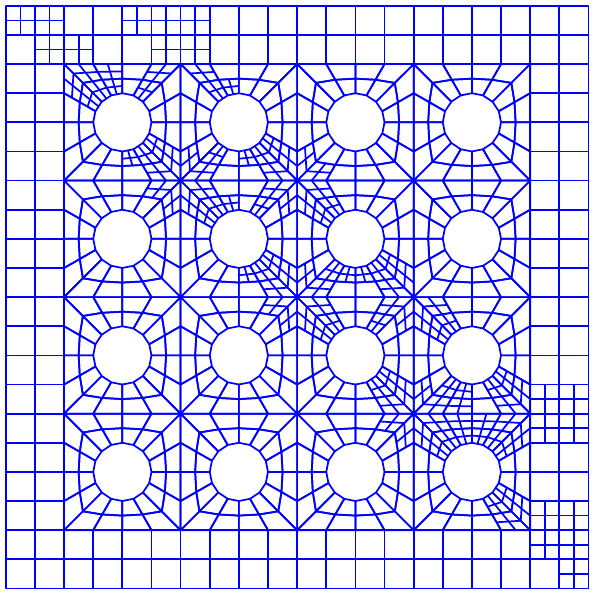}
			\caption{}
		}
	\end{subfigure}%
	\\ \bigskip
	\begin{subfigure}[b]{0.4\textwidth}
		{
			\centering
			\includegraphics{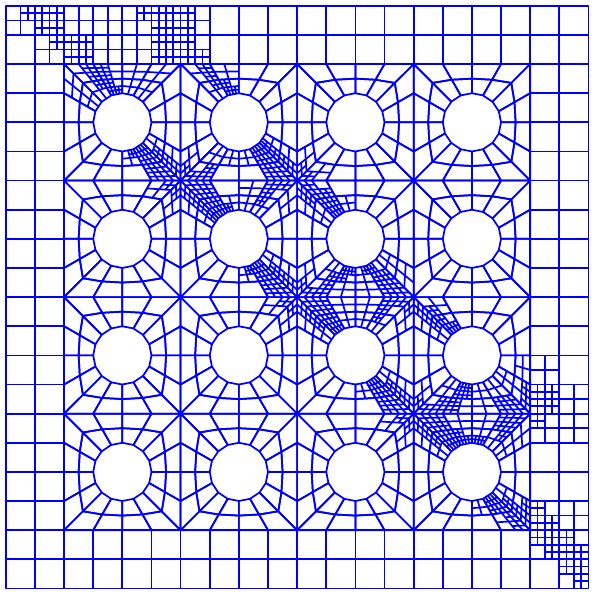}
			\caption{}
		}
	\end{subfigure}%
	\quad
	\begin{subfigure}[b]{0.4\textwidth}
		{
			\centering
			\includegraphics{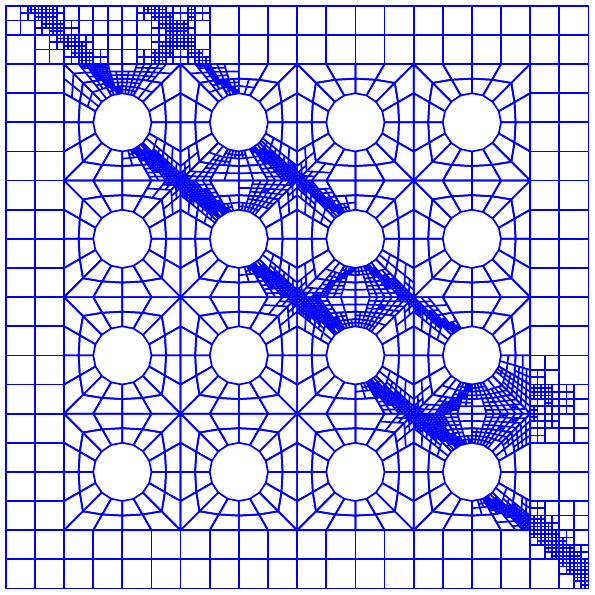}
			\caption{}
		}
	\end{subfigure}%
	\caption{\textcolor{blue}{Several adaptive mesh steps for the porous media$ (\varphi = 0^{\circ} )$.}}
	\label{Fig: PorousPlateAdaptiveMeshStep_phi0}
\end{figure}

\begin{figure}[H]
	\centering
	\includegraphics[width=0.34\textwidth]{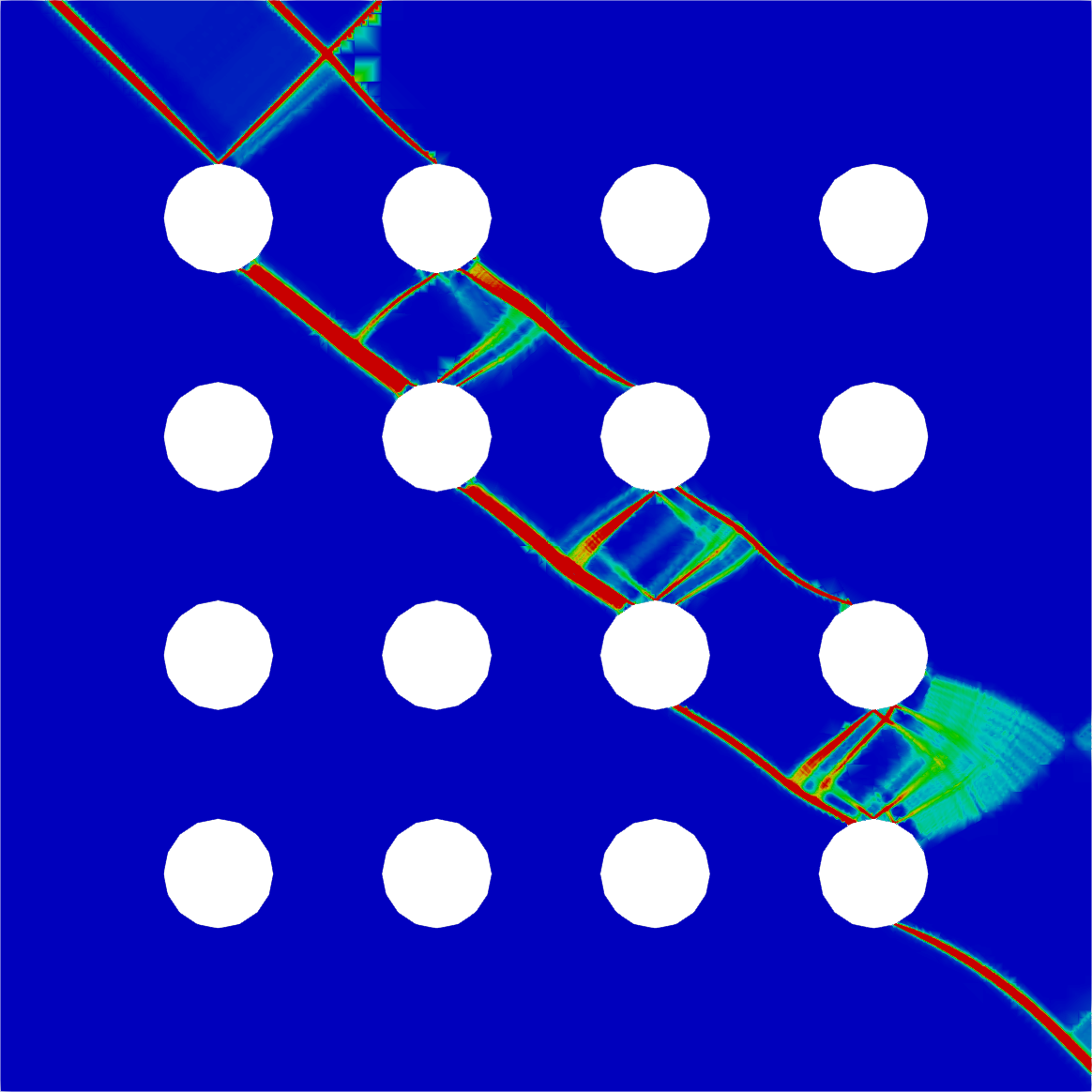}
	\caption{\textcolor{blue}{Plastic dissipation of the porous media$ (\varphi = 0^{\circ} )$.}}
	\label{Fig:PorousDissipation}
\end{figure}

\begin{figure}[H]
	\centering
	\begin{subfigure}[b]{0.4\textwidth}
		{
			\centering
			\includegraphics{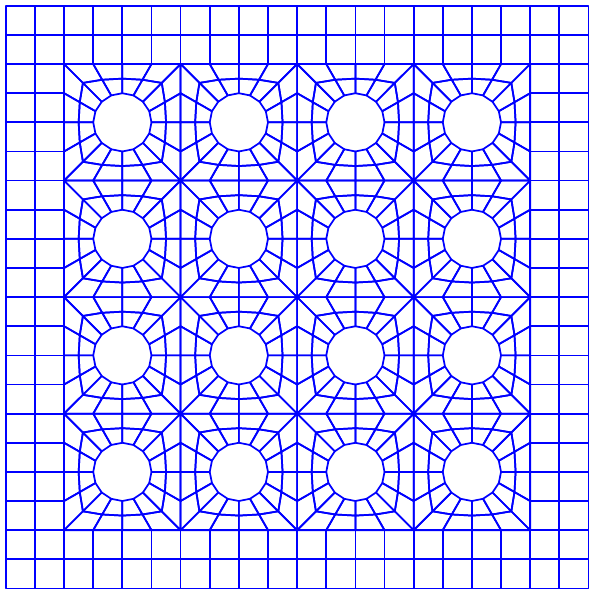}
			\caption{}
		}
	\end{subfigure}%
	\quad
	\begin{subfigure}[b]{0.4\textwidth}
		{
			\centering
			\includegraphics{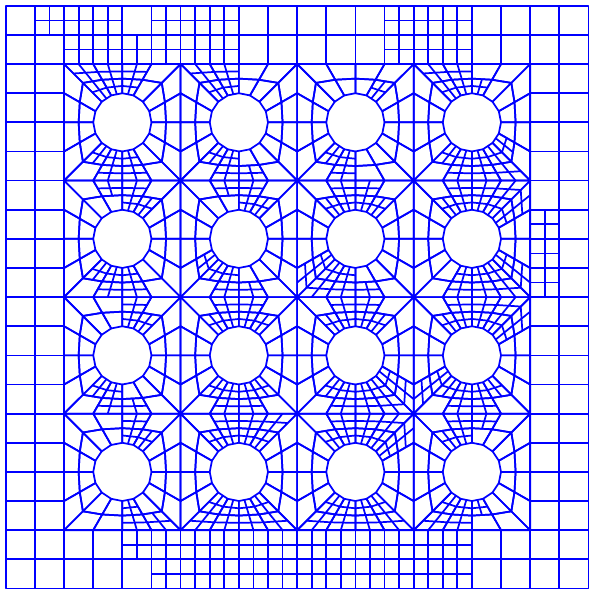}
			\caption{}
		}
	\end{subfigure}%
	\\ \bigskip
	\begin{subfigure}[b]{0.4\textwidth}
		{
			\centering
			\includegraphics{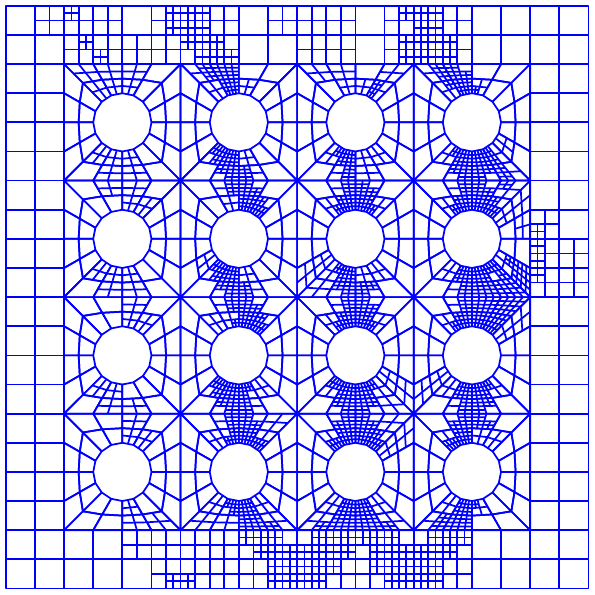}
			\caption{}
		}
	\end{subfigure}%
	\quad
	\begin{subfigure}[b]{0.4\textwidth}
		{
			\centering
			\includegraphics{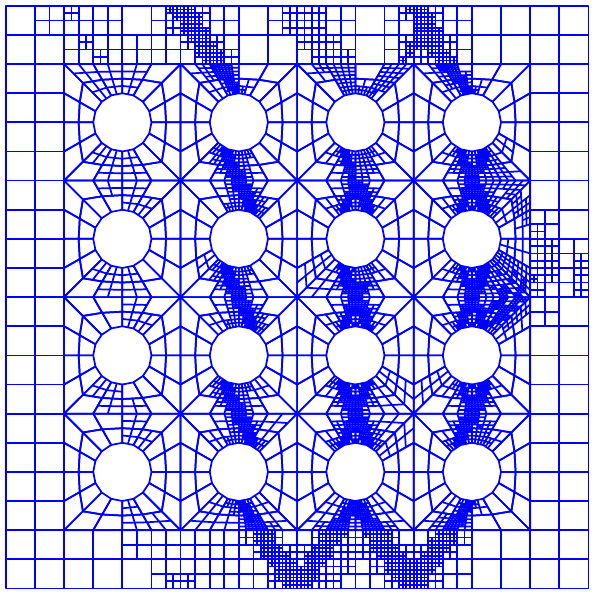}
			\caption{}
		}
	\end{subfigure}%
	\caption{\textcolor{blue}{Several adaptive mesh steps for the porous media$ (\varphi = 20^{\circ} )$.}}
	\label{Fig: PorousPlateAdaptiveMeshStep_phi20}
\end{figure}

\begin{figure}[H]
	\centering
	\includegraphics[width=0.34\textwidth]{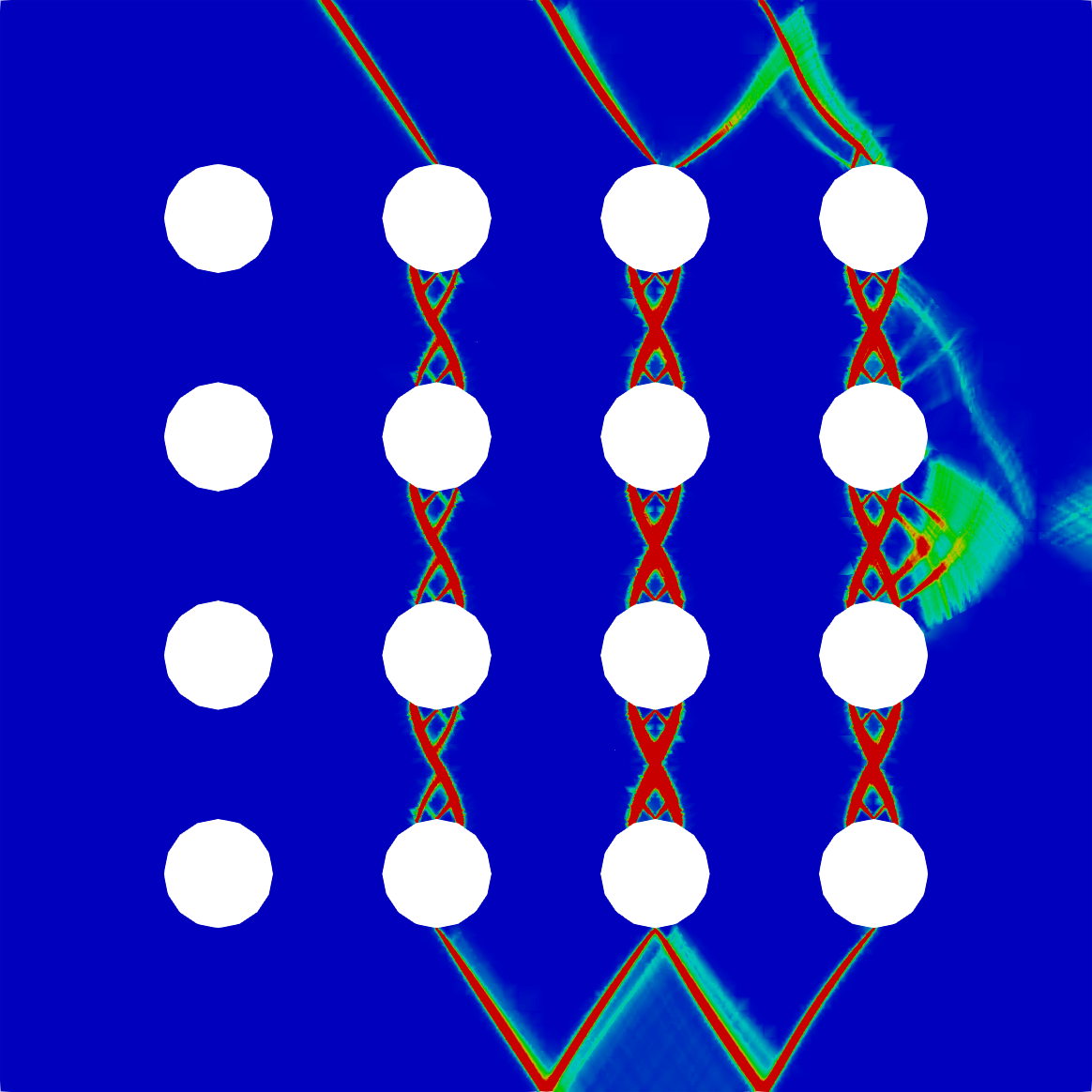}
	\caption{\textcolor{blue}{Plastic dissipation of the porous media$ (\varphi = 20^{\circ} )$.}}
	\label{Fig:PorousDissipation20}
\end{figure}

\pagebreak
\section*{Tables}
\begin{table}[H]
  \centering
  \caption{A smooth strip footing: Comparison with the exact solution using adaptive mesh.}
  \begin{tabular}{ccp{7.855em}ccccc}
    \toprule
    \multirow{8}[16]{*}{$ \varphi = 0^{\circ} $} & \multicolumn{1}{c}{\multirow{4}[8]{*}{\parbox{3.2cm}{Present approach (Adaptive bL-Q4)}}}  & $ N_{var} $                        & 137    & 807    & 2107   & 5737   & 16542  \\
    \cmidrule{3-8}                               &                                                                                            & \multicolumn{1}{l}{Mosek time (s)} & 0.05   & 0.06   & 0.16   & 0.45   & 2.70   \\
    \cmidrule{3-8}                               &                                                                                            & \multicolumn{1}{l}{$ \alpha^{+} $} & 5.437  & 5.203  & 5.169  & 5.154  & 5.149  \\
    \cmidrule{3-8}                               &                                                                                            & Relative error (\%)                & 5.75   & 1.19   & 0.53   & 0.24   & 0.14   \\
    \cmidrule{2-8}                               & \multicolumn{1}{c}{\multirow{4}[8]{*}{\parbox{2.5cm}{Uniform mesh (bL-Q4)}}}               & $ N_{var} $                        & 137    & 1742   & 6682   & 26162  & 103522 \\
    \cmidrule{3-8}                               &                                                                                            & \multicolumn{1}{l}{Mosek time (s)} & 0.05   & 0.11   & 0.58   & 2.38   & 18.67  \\
    \cmidrule{3-8}                               &                                                                                            & \multicolumn{1}{l}{$ \alpha^{+} $} & 5.437  & 5.202  & 5.171  & 5.156  & 5.149  \\
    \cmidrule{3-8}                               &                                                                                            & Relative error (\%)                & 5.75   & 1.17   & 0.57   & 0.28   & 0.14   \\
    \midrule
    \multirow{8}[16]{*}{$ \varphi = 35^{\circ}$} & \multicolumn{1}{c}{\multirow{4}[8]{*}{\parbox{3.15cm}{Present approach (Adaptive bL-Q4)}}} & $ N_{var} $                        & 1417   & 2467   & 5332   & 14647  & 44017  \\
    \cmidrule{3-8}                               &                                                                                            & \multicolumn{1}{l}{Mosek time (s)} & 0.11   & 0.16   & 0.39   & 2
                                                 & 9.5                                                                                                                                                                          \\
    \cmidrule{3-8}                               &                                                                                            & \multicolumn{1}{l}{$ \alpha^{+} $} & 51.922 & 48.044 & 46.939 & 46.488 & 46.292 \\
    \cmidrule{3-8}                               &                                                                                            & Relative error (\%)                & 12.57  & 4.16   & 1.76   & 0.79   & 0.36   \\
    \cmidrule{2-8}                               & \multicolumn{1}{c}{\multirow{4}[8]{*}{\parbox{2.5cm}{Uniform mesh (bL-Q4)}}}               & $ N_{var} $                        & 1417   & 5432   & 21262  & 84122  &        \\
    \cmidrule{3-8}                               &                                                                                            & \multicolumn{1}{l}{Mosek time (s)} & 0.19   & 0.39   & 2.05   & 11.03  &        \\
    \cmidrule{3-8}                               &                                                                                            & \multicolumn{1}{l}{$ \alpha^{+} $} & 51.922 & 48.065 & 46.971 & 46.519 &        \\
    \cmidrule{3-8}                               &                                                                                            & Relative error (\%)                & 12.57  & 4.21   & 1.84   & 0.86   &        \\
    \bottomrule
  \end{tabular}%
  \label{tab: table1}
\end{table}%

\begin{table}[H]
  \centering
  \caption{A smooth strip footing: Comparison with other methods.}
  \begin{tabular}{llcccc}
    \toprule
    \multicolumn{1}{c}{\multirow{2}[4]{*}{Approach}}   & \multicolumn{1}{c}{\multirow{2}[4]{*}{Authors}}    & \multicolumn{2}{c}{$ \varphi = 0^{\circ} $} & \multicolumn{2}{c}{$ \varphi = 35^{\circ}$}                                \\
    \cmidrule{3-6}                                     &                                                    & $ \alpha^{+} $                              & (error (\%))                                & $ \alpha^{+} $ & (error \%)) \\
    \midrule
    \multicolumn{1}{c}{\multirow{3}[6]{*}{Kinematic }} & Sloan \& Kleeman \cite{16Sloan}                    & 5.21                                        & (-1.33)                                     & --             & --          \\
    \cmidrule{2-6}                                     & Makrodimopoulos \& Martin \cite{23Makrodimopoulos} & 5.148                                       & (-0.12)                                     & 46.37          & (-0.52)     \\
    \cmidrule{2-6}                                     & Present (Adaptive bL-Q4)                           & 5.148                                       & (-0.14)                                     & 46.29          & (-0.36)     \\
    \midrule
    Mixed                                              & Capsoni and Corradi \cite{15Capsoni}               & 5.24                                        & (-1.91)                                     & --             & --          \\
    \midrule
    Static                                             & Makrodimopoulos \& Martin \cite{22Makrodimopoulos} & 5.141                                       & (-0.02)                                     & 46.07          & (-0.12)     \\
    \midrule
    Analytical                                         & Prandtl                                            & 2 + $ \pi $                                 & 0                                           & 46.124         & 0           \\
    \bottomrule
  \end{tabular}%
  \label{tab: table2}%
\end{table}%
\begin{table}[H]
  \centering
  \caption{A smooth strip footing: limit load factor for various internal frictional angles.}
  \begin{tabular}{cccc}
    \hline
    $\varphi ^ \circ$ & $\alpha^{+}$ Present Method & $\alpha^{+}$ Prandt \cite{99Prandtl} & Error ($\%$) \\
    \hline
    $0$               & $5.145$                     & $5.142$                              & $0.060$      \\
    $5$               & $6.493$                     & $6.489$                              & $0.066$      \\
    $10$              & $8.350$                     & $8.345$                              & $0.065$      \\
    $15$              & $10.986$                    & $10.977$                             & $0.084$      \\
    $20$              & $14.845$                    & $14.835$                             & $0.072$      \\
    $25$              & $20.735$                    & $20.721$                             & $0.072$      \\
    $30$              & $30.166$                    & $30.140$                             & $0.086$      \\
    $35$              & $46.160$                    & $46.124$                             & $0.079$      \\
    $40$              & $75.384$                    & $75.313$                             & $0.095$      \\
    $45$              & $134.098$                   & $133.874$                            & $0.168$      \\
    \hline
  \end{tabular}%
  \label{tab: footing_present_prandt}%
\end{table}

\begin{table}[H]
  \centering
  \caption{Block with two symmetrical circular holes: The convergence of the present solutions using adaptive meshes.}
  \begin{tabular}{llcp{9.645em}cccc}
    \toprule
    \multicolumn{2}{l}{\multirow{6}[12]{*}{$ \varphi = 0^{\circ} $}}  & \multicolumn{1}{c}{\multirow{3}[6]{*}{\parbox{3.15cm}{Present approach (Adaptive bL-Q4)}}} & $N_{var}$                        & 1821  & 6028  & 25418 & 56818  \\
    \cmidrule{4-8}    \multicolumn{2}{l}{}                            &                                                                                            & \multicolumn{1}{l}{$\alpha^{+}$} & 1.831 & 1.828 & 1.816 & 1.814  \\
    \cmidrule{4-8}    \multicolumn{2}{l}{}                            &                                                                                            & Relative error (\%)              & 1.05  & 0.89  & 0.23  & 0.12   \\
    \cmidrule{3-8}    \multicolumn{2}{l}{}                            & \multicolumn{1}{c}{\multirow{3}[6]{*}{\parbox{2.5cm}{Uniform mesh (bL-Q4)}}}               & $N_{var}$                        & 1821  & 6668  & 25818 & 101558 \\
    \cmidrule{4-8}    \multicolumn{2}{l}{}                            &                                                                                            & \multicolumn{1}{l}{$\alpha^{+}$} & 1.831 & 1.832 & 1.825 & 1.818  \\
    \cmidrule{4-8}    \multicolumn{2}{l}{}                            &                                                                                            & Relative error (\%)              & 1.05  & 1.11  & 0.72  & 0.34   \\
    \midrule
    \multicolumn{2}{l}{\multirow{6}[12]{*}{$ \varphi = 30^{\circ} $}} & \multicolumn{1}{c}{\multirow{3}[6]{*}{\parbox{3.15cm}{Present approach (Adaptive bL-Q4)}}} & $N_{var}$                        & 1821  & 5713  & 28478 & 67458  \\
    \cmidrule{4-8}    \multicolumn{2}{l}{}                            &                                                                                            & \multicolumn{1}{l}{$\alpha^{+}$} & 1.073 & 1.063 & 1.06  & 1.059  \\
    \cmidrule{4-8}    \multicolumn{2}{l}{}                            &                                                                                            & Relative error (\%)              & 1.41  & 0.46  & 0.18  & 0.09   \\
    \cmidrule{3-8}    \multicolumn{2}{l}{}                            & \multicolumn{1}{c}{\multirow{3}[6]{*}{\parbox{2.5cm}{Uniform mesh (bL-Q4)}}}               & $N_{var}$                        & 1821  & 6668  & 25818 & 101558 \\
    \cmidrule{4-8}    \multicolumn{2}{l}{}                            &                                                                                            & \multicolumn{1}{l}{$\alpha^{+}$} & 1.073 & 1.068 & 1.062 & 1.060  \\
    \cmidrule{4-8}    \multicolumn{2}{l}{}                            &                                                                                            & Relative error (\%)              & 1.41  & 0.94  & 0.37  & 0.18   \\
    \bottomrule
  \end{tabular}%
  \label{tab:table3}
\end{table}%

\begin{table}[H]
  \centering
  \caption{Block with two symmetric holes: Comparison with literature solutions.}
  \begin{tabular}{llcc}
    \toprule
    \multicolumn{1}{c}{Approach}   & \multicolumn{1}{c}{Authors}                                           & $ \varphi = 0^{\circ} $ & $ \varphi = 30^{\circ} $ \\
    \midrule
    \multirow{3}[10]{*}{Kinematic} & \multicolumn{1}{l}{Makrodimopoulos \&Martin \cite{23Makrodimopoulos}} & 1.825                   & 1.063                    \\
    \cmidrule{2-4}                 & \multicolumn{1}{l}{Munoz et al. \cite{50Munoz}}                       & 1.8351                  & 1.0652                   \\
    \cmidrule{2-4}                 & \multicolumn{1}{l}{Present (Adaptive bL-Q4)}                          & 1.814                   & 1.059                    \\
    \midrule
    \multirow{2}[4]{*}{Static}     & \multicolumn{1}{l}{Makrodimopoulos \&Martin \cite{22Makrodimopoulos}} & 1.8089                  & 1.0562                   \\
    \cmidrule{2-4}                 & \multicolumn{1}{l}{Munoz et al. \cite{50Munoz}}                       & 1.8119                  & 1.0581                   \\
    \midrule
    Reference                      & Zouain et al \cite{24Zouain}                                          & 1.8131                  & --                       \\
    \bottomrule
  \end{tabular}%
  \label{tab:table4}%
\end{table}%
\begin{table}[H]
  \centering
  \caption{Slope stability: Comparison with the exact solution using adaptive mesh.}
  \begin{tabular}{llcp{9em}ccccc}
    \toprule
    \multicolumn{2}{l}{\multirow{6}[12]{*}{$ \varphi = 20^{\circ} $}} & \multicolumn{1}{c}{\multirow{3}[6]{*}{\parbox{3.15cm}{Present approach (Adaptive bL-Q4)}}} & $N_{var}$                        & 1492   & 1837   & 2707   & 5332   & 13212  \\
    \cmidrule{4-9}    \multicolumn{2}{l}{}                            &                                                                                            & \multicolumn{1}{l}{$\alpha^{+}$} & 9.496  & 8.791  & 8.449  & 8.313  & 8.266  \\
    \cmidrule{4-9}    \multicolumn{2}{l}{}                            &                                                                                            & Relative error (\%)              & 15.66  & 7.08   & 2.91   & 1.25   & 0.68   \\
    \cmidrule{3-9}    \multicolumn{2}{l}{}                            & \multicolumn{1}{c}{\multirow{3}[6]{*}{\parbox{2.5cm}{Uniform mesh (bL-Q4)}}}               & $N_{var}$                        & 1492   & 5702   & 22282  & 88082  &        \\
    \cmidrule{4-9}    \multicolumn{2}{l}{}                            &                                                                                            & \multicolumn{1}{l}{$\alpha^{+}$} & 9.496  & 8.83   & 8.543  & 8.393  &        \\
    \cmidrule{4-9}    \multicolumn{2}{l}{}                            &                                                                                            & Relative error (\%)              & 15.66  & 7.55   & 4.06   & 2.23   &        \\
    \midrule
    \multicolumn{2}{l}{\multirow{6}[12]{*}{$ \varphi = 35^{\circ} $}} & \multicolumn{1}{c}{\multirow{3}[6]{*}{\parbox{3.15cm}{Present approach (Adaptive bL-Q4)}}} & $N_{var}$                        & 1492   & 1772   & 2507   & 4287   & 9027   \\
    \cmidrule{4-9}    \multicolumn{2}{l}{}                            &                                                                                            & \multicolumn{1}{l}{$\alpha^{+}$} & 20.075 & 16.38  & 14.842 & 14.195 & 13.984 \\
    \cmidrule{4-9}    \multicolumn{2}{l}{}                            &                                                                                            & Relative error (\%)              & 46.00  & 19.13  & 7.94   & 3.24   & 1.70   \\
    \cmidrule{3-9}    \multicolumn{2}{l}{}                            & \multicolumn{1}{c}{\multirow{3}[6]{*}{\parbox{2.5cm}{Uniform mesh (bL-Q4)}}}               & $N_{var}$                        & 1492   & 5702   & 22282  & 88082  &        \\
    \cmidrule{4-9}    \multicolumn{2}{l}{}                            &                                                                                            & \multicolumn{1}{l}{$\alpha^{+}$} & 20.075 & 16.457 & 15.028 & 14.407 &        \\
    \cmidrule{4-9}    \multicolumn{2}{l}{}                            &                                                                                            & Relative error (\%)              & 46.00  & 19.69  & 9.29   & 4.78   &        \\
    \bottomrule
  \end{tabular}%
  \label{tab:table5}%
\end{table}%

\begin{table}[H]
  \centering
  \caption{Slope stability: Comparison with other methods in the literature.}
  \begin{tabular}{llcc}
    \toprule
    \multicolumn{1}{c}{Approach}   & \multicolumn{1}{c}{Authors}                        & $ \varphi = 20^{\circ} $ & $ \varphi = 35^{\circ} $ \\
    \midrule
    \multirow{5}[14]{*}{Kinematic} & Krabbenhoft \textit{et al.}\cite{19Krabben}        & 8.440                    & --                       \\
    \cmidrule{2-4}                 & Lyamin \& Sloan \cite{21Lyamin}                    & 8.440                    & --                       \\
    \cmidrule{2-4}                 & Makrodimopoulos \& Martin \cite{23Makrodimopoulos} & 8.366                    & 14.19                    \\
    \cmidrule{2-4}                 & Chen \cite{97Chen}                                 & 8.300                    & 13.86                    \\
    \cmidrule{2-4}                 & Present (Adaptive bL-Q4)                           & 8.266                    & 13.984                   \\
    \midrule
    Static                         & Makrodimopoulos \& Martin \cite{22Makrodimopoulos} & 8.210                    & 13.75                    \\
    \bottomrule
  \end{tabular}%
  \label{tab:table6}%
\end{table}%

\end{document}